\documentclass[aps,prd,twocolumn,nofootinbib,showpacs,preprintnumbers,floatfix,longbibliography]{revtex4-2}
\usepackage{natbib}
\usepackage{longtable}
\usepackage[pdftex]{graphicx}
\usepackage[usenames]{color}
\usepackage[dvips]{epsfig}

\usepackage{hyperref}
\usepackage{bbm}
\usepackage{marvosym}
\usepackage{verbatim}
\usepackage{psfrag}
\usepackage{slashed}

\usepackage{amsmath}
\usepackage{amssymb}
\usepackage{wasysym}
\usepackage{multirow}
\usepackage{diagbox}%used in tables to display slash in a box
\usepackage{siunitx} % for 'S' column type
\usepackage{array} % for '\extrarowheight' macro
\usepackage{amssymb}  % Mathe
\usepackage{booktabs,amsmath}
\usepackage{bbold}
\usepackage[table,x11names]{xcolor}
\definecolor{maroon}{cmyk}{0,0.87,0.68,0.32}
\usepackage{amsfonts}
\definecolor{boxcolor}{HTML}{ffe6a1}

\newcommand{\eref}[1]{Eq.~(\ref{#1})}
\newcommand{\fref}[1]{Fig.~\ref{#1}}
\newcommand{\sref}[1]{Section~\ref{#1}}

\def \Nt {{N_{\tau}}}
\def \Nc {{N_{c}}}

\newcommand{\expval}[1]{\langle #1 \rangle}

\newcommand{\pbp}{\bar{\psi}\psi}

\newcommand{\Id}{{\mathbbm{1}}}
\newcommand{\lr}[1]{\left( #1 \right)}

\newcommand{\beqn} {\begin{equation}}
\newcommand{\eqn} {\end{equation}}

\def \beq{\begin{equation}}
\def \eeq{\end{equation}}
\def \bea{\begin{eqnarray}}
\def \eea{\end{eqnarray}}

\def \bet0{\beta_0}
\def \bet1{\beta_1}
\def \simgt{\,\rlap{\lower 7.5 pt\hbox{$\mathchar \sim$}}\raise 3 pt \hbox{$>$}\,}
\def \simlt{\,\rlap{\lower 7.5 pt\hbox{$\mathchar \sim$}}\raise 3 pt \hbox{$<$}\,}

\def\lsim{\raise0.3ex\hbox{$<$\kern-0.75em\raise-1.1ex\hbox{$\sim$}}}
\def\gsim{\raise0.3ex\hbox{$>$\kern-0.75em\raise-1.1ex\hbox{$\sim$}}}

\newcommand{\SU}{{\rm SU}}
\newcommand{\U}{{\rm U}}

\def \diag {\rm diag}
\newcommand{\Zero}{\mathbb{0}}
\newcommand{\deltaT}{\delta_{\hat{0},\hat{\mu}}}

\DeclareMathAlphabet{\mathpzc}{OT1}{pzc}{m} {it}
\begin{document}
\title{$U(N)$ gauge theory in the strong coupling limit on a quantum annealer}

\author{Jangho Kim$^{\rm a}$}
\email{j.kim@fz-fuelich.de}
\author{Thomas Luu$^{\rm a,b}$}
\email{t.luu@fz-fuelich.de}
\author{Wolfgang Unger$^{\rm c}$}
\email{wunger@physik.uni-bielefeld.de}
\affiliation{$^{\rm a}$ Institute for Advanced Simulation (IAS-4), Forschungszentrum J\"ulich}
\affiliation{$^{\rm b}$ Institut f\"ur Kernphysik (IKP-3), Forschungszentrum J\"ulich, 54245 J\"ulich, Germany}
\affiliation{$^{\rm c}$ Fakult\"at f\"ur Physik, Bielefeld University, D-33615 Bielefeld, Germany}

\begin{abstract}
Lattice QCD in the strong coupling regime can be formulated in dual variables which are integer-valued. It can be efficiently simulated for modest finite temperatures and finite densities via the worm algorithm, circumventing the finite density sign problem in this regime. However, the low temperature regime is more expensive to address. 
As the partition function is solely expressed in terms of integers, it can be cast as a combinatorial optimization problem that can be solved on a quantum annealer. We will first explain the setup of the system we want to study, and then present its reformulation suitable for a quantum annealer, and in particular the D-Wave.
As a proof of concept, we present first results obtained on D-Wave for gauge group $\U(1)$ and $\U(3)$, and outline the next steps towards gauge groups $\SU(3)$. We find that in addition, histogram reweighting greatly improves the accuracy of our observables when compared to analytic results.
\end{abstract}

\maketitle
\tableofcontents

\section{Introduction}

The goal of applying quantum computing to physically interesting quantum theories to obtain results that are too expensive on a classical computer is one of the great challenges of our time, and many groups worldwide are contributing to this endeavour. While some groups target gate-based multi-purpose quantum computers, many interesting physical questions can as well be reformulated as a optimization problem that can be addressed on a quantum annealer. For example, the lattice gauge theory for the non-Abelian dihedral groups $D_3$, $D_4$  \cite{Fromm:2022vaj} and $\SU(2)$ pure gauge theory \cite{ARahman:2021ktn} have recently been studied on the D-Wave quantum annealer. In this work we add another interesting lattice model which is an effective theory inspired by QCD and includes fermions.

This paper is organized as follows:
We first introduce the effective theory of QCD and identify the regime that is computationally expensive and requires alternative strategies such as quantum computing. We introduce the D-Wave quantum annealer as our method of choice to address the physics of our theory. We then derive the formalism for the general gauge group $\U(\Nc)$ in terms of a Quadratic Unconstrained Binary Optimization (QUBO) matrix.
Finally we present first results obtained on the D-Wave machine at FZ J\"ulich for gauge group $U(1)$ and $U(3)$ on various small lattices. We also outline the next steps towards larger lattices and gauge group $SU(3)$. 

\subsection{Lattice Gauge Theory at Strong Coupling}
The theory of strong interactions between quarks and gluons can be studied non-perturbatively via lattice QCD. One of the fundamental questions that after many decades of research is still unanswered is whether the QCD phase diagram has a chiral critical endpoint. 
To answer this question requires investigating lattice QCD at non-zero baryon density $\mu_B$~\cite{Philipsen2010}.
Unfortunately lattice QCD at $\mu_B \neq 0$  cannot be directly simulated via conventional hybrid Monte Carlo because of the infamous numerical sign problem \cite{deForcrand2010,Gattringer2016b}: the fermion determinant becomes complex-valued, prohibiting importance sampling.

A possible solution is to re-express the partition function in terms of dual variables. This results in a much milder sign problem and in some specific limits makes it sign problem-free. This is the case in the strong coupling limit of lattice QCD with staggered fermions (SC-LQCD), and it is by now a well-established effective theory of QCD. It has been studied in the last decades by many authors, both via mean-field theory \cite{KlubergStern1983,Faldt:1985ec,Bilic1992a,Nishida:2003fb,Miura2016} and Monte Carlo simulations \cite{Karsch:1988zx}, with the first study establishing the mean-field theory using the $1/d$ expansion \cite{Kawamoto1981} and the first Monte Carlo study \cite{Rossi:1984cv}, which introduced dimers as dual variables for mesons. It took however another 20 years before this effective theory could be studied efficiently, evaluated numerically via the worm algorithm \cite{Prokofev2001,Adams:2003cca,Fromm,deForcrand:2009dh}.\\ 

We will briefly describe the dual representation of SC-LQCD first - see \cite{Gagliardi2019,Kim2023} for more details - before we will address the shortcomings that might require alternative strategies such as quantum computing. 
In the strong coupling limit, the inverse gauge coupling $\beta=\frac{2}{N_c}\rightarrow 0$, hence the gauge action is absent. The essential idea is then to change the order of integration (first gauge links, then Grassmann variables), and use the fact that the integration over the gauge links factorizes.  
The final degrees of freedom are mesons and baryons, which are the dual variables of this approach. 
In this representation, the sign problem is essentially solved.
Despite the fact that in this limit the lattice is maximally coarse, this effective theory shares important features with full QCD, such as chiral symmetry breaking and confinement. 

When generalizing lattice QCD in the strong coupling limit with gauge groups $\U(\Nc)$ or $\SU(\Nc)$, the action for staggered fermions $\bar{\chi}$, $\chi$  becomes \cite{Susskind1976}
\begin{widetext}
\begin{align}
\label{eq:L}
S_F =
\sum_{x,\nu} \frac{\gamma^{\delta_{\nu 0}}}{2}\eta_{\nu}(x)
\big(
    e^{\mu\delta_{\nu 0}} \overline{\chi}(x) U_{\nu}(x) \chi(x+\hat{\nu})
  - e^{-\mu \delta_{\nu 0}} \overline{\chi}(x+\hat{\nu})U^{\dagger}_{\nu}(x) \chi(x)
\big)+am_q\sum_x{\bar{\chi}_x\chi_x}.
\end{align}
\end{widetext}
Here, $am_q$ is the bare quark mass and $\gamma$ is the bare anisotropy, which favors temporal gauge links in order to have a temperature that can be continuously varied.
From $Z=\int d\bar{\chi}\, d\chi\, dU\, e^{-S_F[\bar{\chi},\chi,U]}$, one can derive the partition function in the dual representation by integrating out the gauge links $U$ and Grassmann variables $\bar{\chi}$, $\chi$:
\begin{widetext}
\begin{align}
\label{eq:dual}
Z=\sum_{\{k,n,\ell\}}
\underbrace{\prod_{b=(x,\hat{\mu})}\frac{(N_c-k_b)!}{N_c!k_b!}\gamma^{2k_b\delta_{\hat{0},\hat{\mu}}}}_{\text{meson hoppings}}
\underbrace{\prod_{x}\frac{N_c!}{n_x!}(2am_q)^{n_x}}_{\text{chiral condensate}}
\underbrace{\prod_{\ell}w(\ell,\mu)}_{\text{baryon hoppings}}\,.
\end{align}
\end{widetext}%
This partition function describes a system of mesons and baryons.
The mesons live on the bonds\linebreak $b\equiv (x,\hat{\mu})$, where they hop to a
nearest neighbor $y=x+\hat{\mu}$, and the hopping multiplicities are given by
so-called dimers $k_b\in \{0,\ldots,N_c\}$.
Each baryon must form self-avoiding loops.
We have
\begin{widetext}
\begin{align}
w(\ell,\mu)=\dfrac{1}{\prod_{x\in\ell}N_c!}\sigma(\ell)\gamma^{\Nc N_{\hat{0}}} \exp{(\Nc N_t \omega_{\ell}a_t\mu)}\,,
\quad \sigma(\ell)=(-1)^{\omega_{\ell}+N_{-}(\ell)+1}\prod_{b=(x,\hat{\mu})\in\ell}\eta_{\hat{\mu}}(x)\,,
\end{align}
\end{widetext}
where $\ell$ denotes a baryon loop and $N_{\hat{0}}$ is the number of baryons in the
temporal direction. $N_t$ and $\omega_{\ell}$ are the number of temporal lattice sites and baryon winding number, respectively, in the temporal direction.
Finally $\sigma(\ell)=\pm 1$ is the sign which depends on the 
geometry of the baryon loop $\ell$.
By the Grassmann constraint, the summation over configurations
$\sum_{\{k,n,\ell\}}$ in \eref{eq:L} is constrained by the following
condition:
\begin{align}
n_x + \sum_{\mu=\pm 0,\cdots,\pm d}\bigg( k_{\mu}(x)+\frac{N_c}{2}|\ell_{\mu}(x)|\bigg) = \Nc.
\label{eq:GC}
\end{align}
The dual variable $n_x$ is the number of monomers that gives an explicit contribution to the chiral condensate.
In the chiral limit, monomers are absent, but as the quark mass increases, more monomers become present.

Two observables are particularly of interest, the chiral condensate $\expval{\pbp}$ and energy density $\langle \epsilon \rangle$: 
\begin{align}
   &a^{d-1}\expval{\pbp}= a^{d-1}\frac{T}{V}\frac{\partial \log Z}{\partial m_q}=\frac{1}{\Omega}\frac{1}{2am_q} \langle M \rangle  \\
   &a^d\expval{\epsilon}=- \frac{a^d}{V}\frac{\partial \log Z}{\partial T^{-1}}\nonumber
   %\\ 
   %&
   =\frac{1}{\Omega}\lr{ \frac{\xi}{\gamma}\frac{d\gamma}{d\xi}\langle 2D_t \rangle-\langle M \rangle}\nonumber\\
   &\phantom{a^d\expval{\epsilon}}=
   %\frac{1}{\Omega}( 2\frac{\kappa\gamma^2}{\gamma}(2\kappa\gamma)^{-1}\langle D_t \rangle-\langle M \rangle)
   \frac{1}{\Omega}(\langle D_t \rangle-\langle M \rangle)
\end{align}
with 
\begin{align}
M&=\sum_{x\in \Omega} n_x, & D_t=\sum_{x\in \Omega} k_0(x)
\label{eq:obs}
\end{align}
the total monomer number and temporal dimers, respectively, of a configuration on the d-dimensional lattice volume $\Omega=N_s^{d-1}\times N_t$, and $aT=\xi(\gamma)/N_t$, corresponding to a physical volume $V\times T^{-1}$.
In the last equation for $\expval{\epsilon}$
we used that $\xi(\gamma)=\kappa\gamma^2$ at strong coupling (see \cite{deForcrand2017} for details).
The lattice spacing $a$ does not take a specific value: the lattice is coarse at strong coupling. Instead of setting a scale, we consider dimensionless ratios. We will drop the lattice spacing when denoting $\expval{\pbp}$ and $\expval{\epsilon}$ in the following, always implying that our observables are expressed in lattice units.

Some of us contributed to this effective theory by extending it in various ways: by extending the phase diagram from the chiral limit to large quark masses \cite{Kim2016}, by including gauge corrections to the strong coupling limit \cite{deForcrand2014, Gagliardi2019}, i.e.~by a strong coupling expansion of the Wilson gauge action, and by formulating the theory as a quantum Hamiltonian system based on the continuous time limit \cite{Unger2011a,Klegrewe2020}.
To obtain the full $\mu_B$-$T$ phase diagram unambiguously, and study its properties, we had to overcome various technical issues such as the non-perturbative determination the temperature at fixed lattice spacing \cite{deForcrand2017,Kim2020a}. 

The finite density sign problem will get re-introduced as the lattice becomes finer, away from the strong coupling limit. 
Our findings show that in a regime where the sign problem is still manageable, the nuclear transition depends weakly on the inverse gauge coupling, but strongly on the quark mass \cite{Kim2019,Kim2023}.

The dual representation sampled by the worm algorithm works well to unravel the phase diagram, 
but the worm algorithm is highly inefficient at low temperatures: as the worm algorithm is based on a high temperature expansion, it requires longer simulation times as one goes to lower temperatures, which in turn requires larger temporal lattice extents $N_\tau$. This ultimately becomes too prohibitive numerically.

Thus the low-temperature regime in which the first order nuclear transition occurs remains particularly challenging,
%But in particular the low-temperature regime across the first order nuclear transition remains challenging, 
as well as determining the zero temperature observables such as the hadron spectrum. 
Our goal is to use our novel algorithm on the D-wave quantum computer to simulate at these low temperatures and on large volume, allowing us to investigate the first order phase chiral and nuclear transition with heavy quark masses. 
The algorithm we propose to use in conjunction with D-Wave is immune to this temperature constraint,
and parallelization is intrinsically built in, which cannot be achieved in the worm algorithm.
The dual representation of the strong coupling partition function \eref{eq:dual} is an exact rewriting in terms of integer valued dual variables, which resembles a monomer-dimer system for the meson dynamics, with additional baryonic world lines. All these dual degrees of freedom can be cast into binary numbers, which makes this effective theory of QCD suitable to be studied on a quantum annealer, and the D-Wave in particular.

\subsection{D-Wave quantum annealer}
\label{subsec:DWAVE_Description}
\begin{figure}
    \centering
    \includegraphics[width=0.48\textwidth]{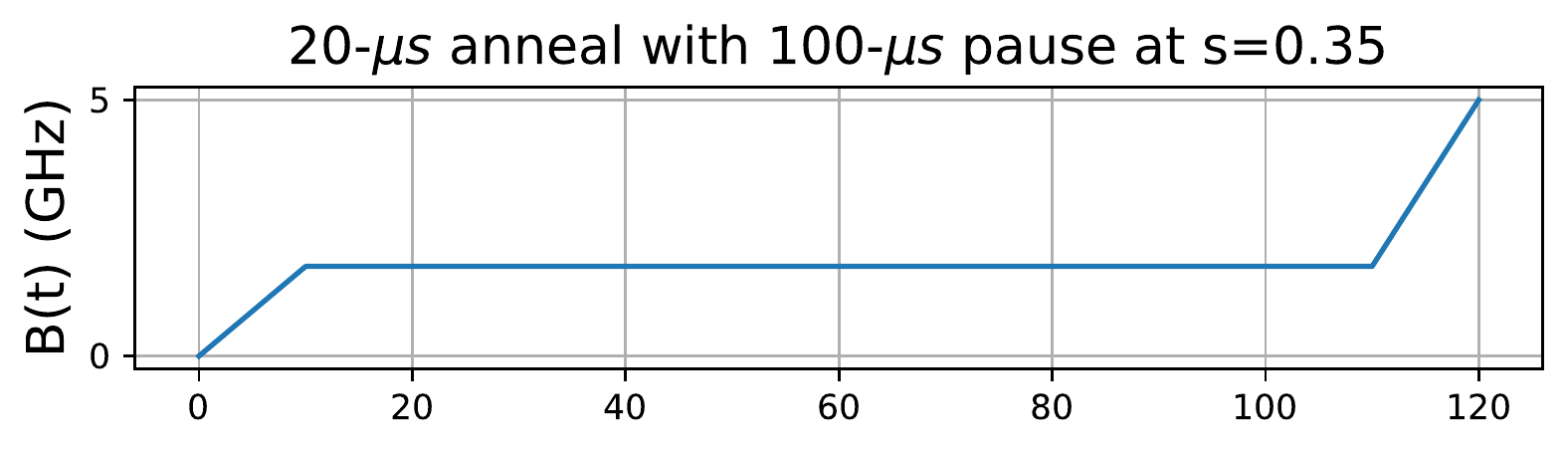}
    \includegraphics[width=0.48\textwidth]{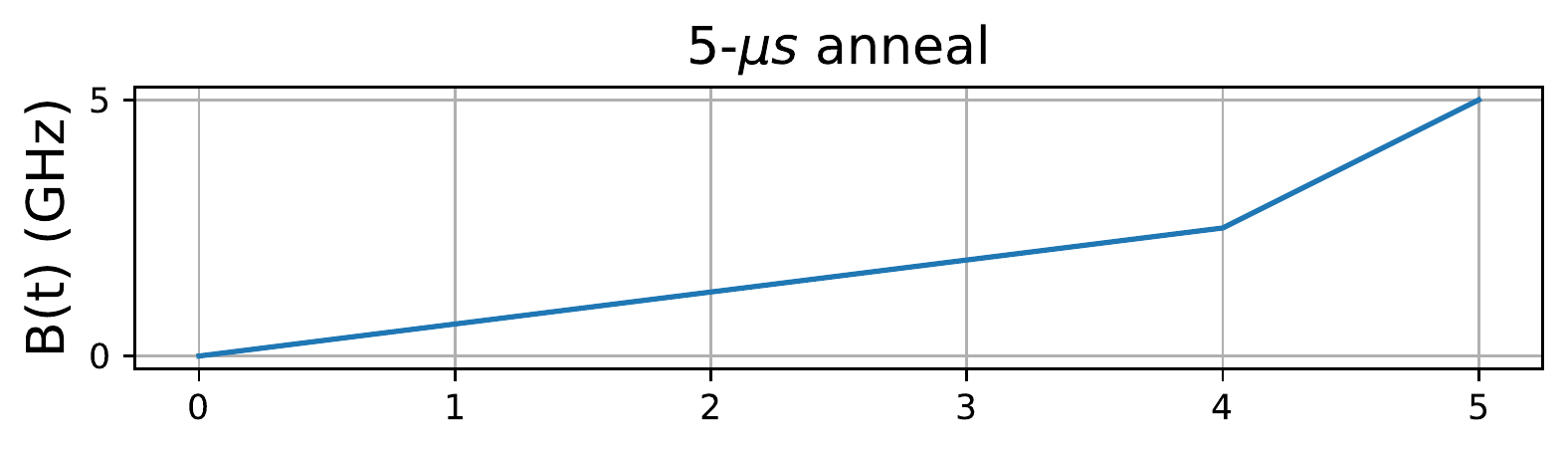}
    \caption{Our choice of annealing profiles~\cite{anneal}. The behavior of $A(t)$ (described in text) is roughly inversely proportional to $B(t)$.  The x-axis represents time in $\mu$-secs. Top profile was used for $U(1)$ simulations, bottom was used for $U(3)$ simulations.}
    \label{fig:anneal}
\end{figure}

The D-Wave quantum annealer consists of an array of metal loops with Josephson junctions that are cooled to a point that the loops are superconducting.  The two-state level scheme of each superconducting loop constitutes a single \emph{superconducting flux} qubit~\cite{PhysRevB.81.134510}.  The qubits are then placed in an array and coupled pairwise inductively.  This forms D-Wave's quantum processing unit (QPU).  The inductances between qubits, as well as the intrinsic inductance within each qubit, can be tuned \emph{in situ}~\cite{PhysRevB.82.024511}.  The array of qubits can then be modeled as an Ising spin glass,
\begin{align}
    H_{Ising}=-\sum_{i<j}K_{ij} \sigma^i_z\sigma^j_z+ \sum_i h_i \sigma^i_z\ ,
\end{align}  
where $\sigma_z$ is a Pauli spin matrix.
An application of an external magnetic field equates to introducing a non-commuting transverse field $\sigma_x$ at each site $i$, while diminishing the strength of the original `target' Ising Hamiltonian.  This competition between the transverse field and the Ising Hamiltonian is expressed by time-dependent coefficients $A(t)$ and $B(t)$,
\begin{align}
    H(s)=-A(t)\sum_i\sigma^i_x+B(t)H_{Ising}\ .
\end{align}
Thus the full Hamiltonian is that of the transverse (quantum) Ising model, but with coefficients $A(t)$ and $B(t)$ that depend on the strength and time-dependence $t$ of the external magnetic field.  The profile of the magnetic field is such that when $t=0$ one has $A/B\gg 1$ and the transverse field is dominant, while at some later `anneal time' $t_f$, $A/B\approx 0$ and one recovers only the Ising Hamiltonian. Fig.~\ref{fig:anneal} shows annealing profiles for $B(t)$ (and ones which we actually use in our calculations). D-Wave prepares the system to be in the (nearly) ground state of the transverse field at $t=0$ and by the process of quantum annealing, smoothly evolves the system to the (nearly) ground state of the Ising model at $t_f$.

The D-Wave Advantage system at FZJ has 5000 \emph{physical} qubits, each of which can be coupled to a maximum of 15 other \emph{physical} qubits.  Typically problems will require qubits with connectivities greater than 15, and in this case \emph{physical} qubits are `chained' together to form \emph{logical} qubits that have the requisite connectivity.  The physical qubits that constitute a logical qubit must act synchronously.  The degree in which they act in unison is controlled by a user-defined parameter \texttt{chain\_strength}.  Though this parameter does not change the physics of the system that we investigate, it does influence the level of quality of solutions that D-Wave obtains. We will return to this topic in later sections.

\section{Numerical Methods and Technical Developments}
\label{sec:tech_dev}

In this section, we first explain the general structure of the Quadratic Unconstrained Binary Optimization (QUBO) matrix and then proceed to map the lattice model in mind, i.e. strong coupling lattice QCD and modifications thereof with gauge group $\U(\Nc)$, to the QUBO matrix. We consider two parts: the action \eref{eq:L} and the Grassmann constraint \eref{eq:GC}. 
As we demonstrate in greater detail below, the constraints are incorporated into the QUBO matrix $Q$ by introducing penalty terms $p$.  This matrix $Q$ then serves as an objective function that is minimized, or \emph{optimized}.  The resulting optimization is therefore \emph{unconstrained}. 

\subsection{How to construct the QUBO matrix}

For our lattice model, we want to make use of the following minimization procedure: 
\begin{align}
	\chi^2 &= x^{T} W x + p ||Ax+b||^2\ ,
	\label{eq:QUBOminimization}
\end{align}
which combines the action $S=x^{T} Wx$ of the underlying lattice theory and the constraint of its variables $Ax+b=0$, with the penalty factor $p$. This factor needs to be sufficiently large. The aim is to find the solution vector $x$ which minimizes $\chi^2$.  Note that under the optimization, if the constraints are satisfied, then the terms proportional to the penalty terms $p$ vanish, resulting in a lower value for $\chi^2$.  Thus finding the absolute minimum of $\chi^2$ requires satisfying the original constraints.

The matrix formulation required by the D-Wave API is
\begin{align}
	\chi^2 &= x^T Q x + C\ .
 \label{eq:chi2}
\end{align}
Hence, the QUBO matrix $Q$ and the constant $C$ is
\begin{align}
	Q &=W+p\, \lr{A^{T}A + \diag\lr{2b^{T}A}}, & C &=p\, b^{T}b\ .
  \label{eq:QUBOdef}
\end{align}
It remains to determine the weight $W$ and the constraint $(A,b)$ for our lattice model, which in terms of integer valued dual variables can always be cast into a vector of binary numbers $x$ whose properties satisfy $x_i^2=x_i$.

For this exploratory study, we restrict ourselves to the gauge groups $\U(\Nc)$, $N_c=1,2,3$, where the effective degrees of freedom correspond to only mesons since baryons are absent (they only occur for gauge group $\SU(N_c)$).
Based on \eref{eq:dual}, the action takes the form of a logarithm:
\begin{widetext}
\begin{align}
	Z &=\sum_{\{k,n\}} \exp[-S],\nonumber\\
	S &=-\sum_{b=(x,\hat{\mu})}\log(\frac{(\Nc-k_b)!}{\Nc!k_b!}) - 2k_b\delta_{\hat{0},\hat{\mu}} \log(\gamma)	- \sum_{x} \log(\frac{N_c!}{n_x!}) - n_x \log(2am_q).
\end{align}
\end{widetext}
We will denote the contributions from the dimers $D_\Nc$ and 
from the monomers $M_\Nc$:
\begin{align}
 D_\Nc(k_b)&=-\log\lr{\frac{(N_c-k_b)!}{N_c!k_b!}} - 2k_b\delta_{\hat{0},\hat{\mu}} \log(\gamma),\nonumber\\	
 M_\Nc(n_x)&=-\log\lr{\frac{N_c!}{n_x!}} - n_x \log(2am_q)+c\ .
\label{WeightMatrix}
\end{align}
The action can be shifted by a constant $c$ without 
loss of generality,
\begin{align}
S &=\sum_{b=(x,\hat{\mu})} D_\Nc (k_b) + \sum_{x} M_\Nc(n_x) \ .
\end{align}
The action $S$, so far expressed as a function of only integers $k_b$ and $n_x$, can also be expressed in bilinear form by defining the binary vector $\vec{x}=(\vec{\tilde{k}}_b,\vec{\tilde{n}}_x)$.  This gives the following block structure:
\begin{align}
	S&= x^{T} W x \nonumber\\
 &= (\vec{\tilde{k}}_b^T,  \vec{\tilde{n}}_x^T) \left(	\begin{array}{cc}
		\tilde{D} \Id_{E\times E} & \Zero_{\Omega\times E}                  \\
		\Zero_{E\times \Omega}                                         & \tilde{M} \Id_{\Omega\times \Omega}        \\
	\end{array}
	\right)
	\left(  \begin{array}{r}
		 \vec{\tilde{k}}_b  \\
		 \vec{\tilde{n}}_x  \\
  \end{array}
	\right) \ \,,
\label{ActionBinary}
\end{align}
where $\Omega$ is the number of lattice sites and $E=\Omega\times d$ the number of bonds on a $d$-dimensional hypercubic lattice.
The weight matrix $W$ is the one that shows up in the QUBO matrix \eref{eq:QUBOdef}. It remains to specify $\tilde{D}$ and $\tilde{M}$ in the binary basis for the specific gauge group under consideration.

\subsubsection{Gauge group $\U(1)$}
Let us first consider $\Nc=1$. 
Since $k_b(x)\in {0,1}$ and $n(x)\in {0,1}$, we already have a binary format from the start.  
The entries in the weight matrix $W$ are,
\begin{align}
 D_1(0)&=0,& D_1(1)&=-2\deltaT\log(\gamma), \nonumber\\
 M_1(0)&=0,& M_1(1)&=-\log(2 am_q), \nonumber\\
 c&=0.
\end{align}

This implies that $\tilde{D}$ and $\tilde{M}$ in \eref{ActionBinary} are just real numbers:  $\tilde{D}=D_1(1)$, $\tilde{M}=M_1(1)$, as $D_1(0)=0$ and when $k_b=0$, no weight is picked up.  The same occurs when $m_x=0$.\\

\begin{figure}[h!]
\vspace{2mm}
	\center
	\includegraphics[width=0.42\textwidth]{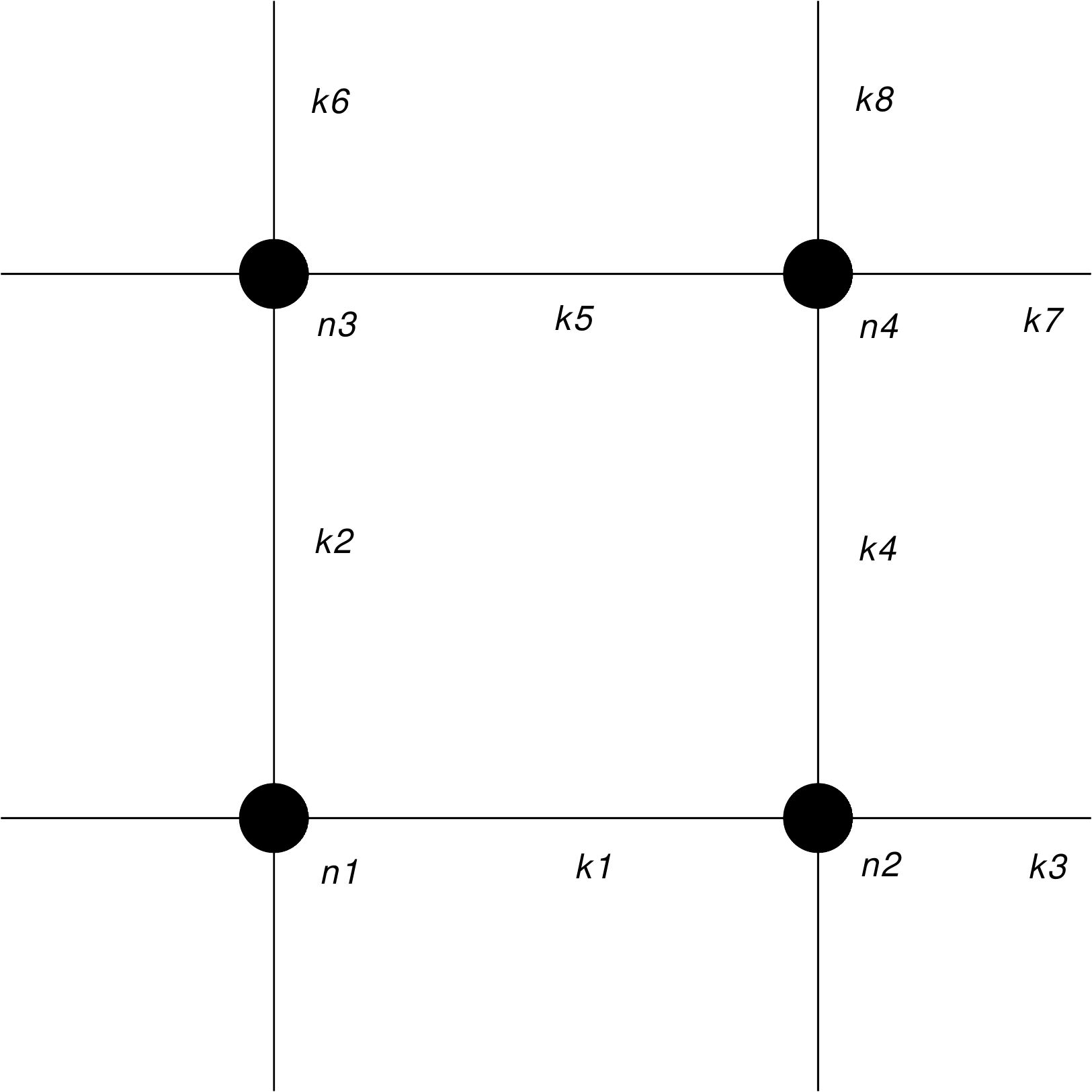}
	\caption{\label{fig:U1_2x2} $2 \times 2$ lattice with indexing the sites and bonds as used in the QUBO matrix
 \\[1mm]
 }
\end{figure}

In order to discuss the constraint $(A,b)$ in \eref{eq:QUBOminimization} explicitly, we now restrict ourselves to a small $2\times 2$ lattice, where
\begin{align}
\vec{k}_b^T&=(k_1,k_2,k_3,k_4,k_5,k_6,k_7,k_8), \nonumber\\ 
\vec{n}_x^T&=(n_1,n_2,n_3,n_4),
\end{align}
The Grassmann constraint \eref{eq:GC} for $\Nc=1$ reads 
\begin{align}
\sum_{\mu=\pm 0,\cdots,\pm d} k_{\mu}(x) +n_x &= 1,
\end{align}
where the sum is over all bonds attached to a lattice site.  This gives four equations, one for each lattice site, with the labels of dimers and monomers as in Fig.~\ref{fig:U1_2x2}.

The matrix form of this constraint is:
\begin{widetext}
\begin{align}
 A \cdot x + b &= 
 \left(\begin{array}{rrrrrrrr|rrrr}
		1 & 1 & 1 & 0 & 0 & 1 & 0 & 0 & 1 & 0 & 0 & 0 \\
		1 & 0 & 1 & 1 & 0 & 0 & 0 & 1 & 0 & 1 & 0 & 0 \\
		0 & 1 & 0 & 0 & 1 & 1 & 1 & 0 & 0 & 0 & 1 & 0 \\
		0 & 0 & 0 & 1 & 1 & 0 & 1 & 1 & 0 & 0 & 0 & 1 \\
	\end{array}
	\right)
	 \left(  \begin{array}{r}
  %\vec{k}_b_{8 \times 1}\\
     k_1  \\
     \vdots \\
     k_8  \\
     \hline
     n_1  \\
     \vdots\\
     n_4  \\
  \end{array}
  \right)
	+  
	\left(  \begin{array}{r}
     -1  \\
     -1  \\
     -1  \\
     -1  \\
  \end{array}
  \right)=0 
  \label{U1_GC}
\end{align}
\end{widetext}
The binary solution vector $x$ has dimension \linebreak $E+\Omega=12$ on this $2\times 2$ lattice. 
Clearly, the computation of $(A,b)$ generalizes to arbitrary hypercubic lattices. The resulting QUBO matrix $Q$ in \eref{eq:QUBOdef} is specified by 
\begin{align}
2b^{T}A&=-(4,4,4,4,4,4,4,4,2,2,2,2), & C &=4p.
  \label{QUBOU1}
\end{align}

The weight matrix $W$ and the constraint $(A,b)$ can be similarly constructed for any lattice volumes. In particular, the lattices $4\times 4$ and $2\times 2\times 2$ have been already simulated on D-Wave, as we will discuss in \sref{sec:result}.

\subsubsection{Gauge group $\U(2)$ and $\U(3)$}
\label{subsec:u3}
In the case of $N_c>1$, we have $k_b\in\{0,1,\ldots \Nc\}$ and $n_x\in\{0,1,\ldots \Nc\}$. 
In order to express these as binary vectors, we define each component of $x$ in terms of a vector of binary numbers
 \begin{align}
 &\tilde{k}_b = (k_b^{(1)},\ldots  k_b^{(r)}), & 	&\tilde{n}_x = (n_x^{(1)},\ldots n_x^{(r)}), \\
  &k_b^{(i)} \in\{0,1\}, &  &n_x^{(i)} \in\{0,1\},
 \end{align}
with $r=\lceil\log(\Nc+1)/\log(2)\rceil$, i.e.~$r=2$ for $\Nc=2$ and $\Nc=3$, $r=3$ for $\Nc\in \{4,5,6,7\}$.
We restrict here to the physically interesting cases of gauge groups $\U(2)$ and $\U(3)$, where the assignments are  
\begin{align}
0 &\mapsto (0,0),&	1 &\mapsto  (0,1),&	2 &\mapsto (1,0) & 3 &\mapsto (1,1)
\end{align}
where for $\Nc=2$ the last assignment must vanish. 
The entries in the weight matrix $W$ for $\Nc=2$ and $\Nc=3$ are
\begin{widetext}
\begin{align}
 D_2(0)&=0,& D_2(1)&=\log(2)-2\deltaT\log(\gamma),& D_2(2)&=\log(4)-4\deltaT\log(\gamma),\nonumber\\ 
 M_2(0)&=-\log(2)+c,& M_2(1)&=-\log(2)-\log(2 am_q)+c,& M_2(2)&=-2\log(2 am_q)+c,\nonumber\\
  c&=\log(2),\\
 D_3(0)&=0,& D_3(1)&=\log(3)-2\deltaT\log(\gamma),& D_3(2)&=\log(12)-4\deltaT\log(\gamma),\nonumber\\
&&&& D_3(3)&=\log(36)-6\deltaT\log(\gamma),\nonumber\\ 
 M_3(0)&=-\log(6)+c,& M_3(1)&=-\log(6)-\log(2 am_q)+c,& M_3(2)&=-\log(3)-2\log(2 am_q)+c,\nonumber\\
 &&&& M_3(3)&=-3\log(2 am_q)+c,\nonumber\\
  c&=\log(6).
\end{align}
\end{widetext}
Hence, for $\Nc=2,3$ both $\tilde{D}$ and $\tilde{M}$ in \eref{ActionBinary} are $2\times2$ matrices 
given by
\begin{align}
	\tilde{D}_{\Nc=2}&=
	\left( 
  \begin{array}{cc}
		D_2(2) & -D_2(1)-D_2(2) \\
		0 & D_2(1) \\
  \end{array}
  \right), \\
	\tilde{D}_{\Nc=3}&=
	\left( 
  \begin{array}{cc}
		D_3(2) & 0 \\
		0 & D_3(1) \\
  \end{array}
  \right) \ .
\end{align}
For $\Nc=3$ it happens that $D_3(3)=D_3(1)+D_3(2)$, and $\tilde{k}_b^T=(1,1)$ picks up this sum, whereas for $\Nc=2$, the off-diagonal term is introduced in order to cancel the unphysical contribution $\tilde{k}_b^T=(1,1)$.

In the same fashion:
\begin{align}
	\tilde{M}_{\Nc=2}&=
	\left( 
  \begin{array}{cc}
		M_2(2) & -M_2(1)-M_2(2) \\
		0 & M_2(1) \\
  \end{array}
  \right), \\
	\tilde{M}_{\Nc=3}&=
	\left( 
  \begin{array}{cc}
		M_3(2) & M_3(3)-M_3(1)-M_3(2) \\
		0 & M_3(1) \\
  \end{array}
  \right) 
\end{align}
where the off-diagonal term in $\tilde{M}$ does not vanish as $ M_3(3)\neq M_3(1)+M_3(2)$

Again, we determine the constraint $(A,b)$ from the Grassmann constraint \eref{eq:GC}, which for illustration is given for U(3) on a $2\times2$ lattice:
\begin{widetext}
{
\begin{align}
  & A \cdot x + b &= \left(\begin{array}{rrrrrrrrrrrrrrrr|rrrrrrrr}
    2 & 1 & 2 & 1 & 2 & 1 & 0 & 0 & 0 & 0 & 2 & 1 & 0 & 0 & 0 & 0 & 2 & 1 & 0 & 0 & 0 & 0 & 0 & 0 \\
    2 & 1 & 0 & 0 & 2 & 1 & 2 & 1 & 0 & 0 & 0 & 0 & 0 & 0 & 2 & 1 & 0 & 0 & 2 & 1 & 0 & 0 & 0 & 0 \\
    0 & 0 & 2 & 1 & 0 & 0 & 0 & 0 & 2 & 1 & 2 & 1 & 2 & 1 & 0 & 0 & 0 & 0 & 0 & 0 & 2 & 1 & 0 & 0 \\
    0 & 0 & 0 & 0 & 0 & 0 & 2 & 1 & 2 & 1 & 0 & 0 & 2 & 1 & 2 & 1 & 0 & 0 & 0 & 0 & 0 & 0 & 2 & 1 \\
  \end{array}
  \right)
   \left(  \begin{array}{r}
		 k_1^{(1)}  \\
		 k_1^{(2)}  \\
   \vdots \\
		 k_8^{(1)}  \\
		 k_8^{(2)}  \\
     \hline
     n_1^{(1)}  \\
     n_1^{(2)}  \\
     \vdots \\
     n_4^{(1)}  \\
     n_4^{(2)}  \\
  \end{array}
  \right)
  -
  \left(  \begin{array}{r}
     3  \\
     3  \\
     3  \\
     3  \\
  \end{array}
  \right)=0
\end{align}
}
\end{widetext}
The size of the solution vector $x$ for $\U(2)$ and  $\U(3)$ is twice of that of $\U(1)$.

The QUBO matrix $Q$ in \eref{eq:QUBOdef} is now specified by 
\begin{widetext}
\begin{align}
2b^{T}A&=-(24,12,24,12,24,12,24,12,24,12,24,12,24,12,24,12,12,6,12,6,12,6,12,6), & C &=36p.
  \label{QUBOU3}
\end{align}
\end{widetext}

For the gauge group $\U(3)$ we have already obtained preliminary results, but as the analysis is still ongoing, we will not address it in this paper, but see \ref{Outlook}.

\subsection{\label{subsec:QUBO} How physical parameters enter the QUBO matrix}

The QUBO matrix Q discussed above depends on the bare quark mass $am_q$ and the bare anisotropy $\gamma$, 
which at strong coupling is related to the temperature via $aT=\kappa\gamma^2/N_t$. 
In the previous section we restricted ourselves to gauge groups $\U(\Nc)$ which do not depend on the baryon chemical potential $\mu_B$. 
Also, as we are in the strong coupling limit where $\beta=0$, we have no dependence on the lattice spacing $a(\beta)$. 
Since both bare  parameters $\gamma$ and $am_q$ enter $Q$ logarithmically, the special choice $\gamma=1$ and \linebreak $am_q=1/2$ simplifies $Q$ considerably. For the gauge group $\U(1)$ the weight matrix $W$ in \eref{WeightMatrix} is zero, as all configurations have equal weight and are purely governed by the constraint $(A,b)$, which is independent of the parameters and remains non-trivial. 

Other choices of $am_q$ and $\gamma$ will be also interesting to study.  For example, towards the chiral limit $am_q\rightarrow0$ the weights in $\tilde{M}$ of \eref{ActionBinary} will diverge logarithmically. Also, in the low-temperature limit $\gamma\rightarrow 0$, the minimization will disfavor temporal meson hoppings over spatial meson hoppings, resulting in a chirally broken phase at low temperatures. 

The zero temperature limit is usually addressed on isotropic lattices $\gamma=1$ by taking $\Nt$ large. In this regime it is interesting to consider the free energy density of the vacuum and related properties. 
In Monte Carlo simulations that use the Worm algorithm, this limit is quite expensive, with simulations using $\Nt=16$ taking days to obtain sufficient statistics. 
On the other hand, we expect that on D-Wave such low temperature simulations should be readily accessible.
In the next section, we however fix the lattice volume and choose a low temperature by setting $\gamma=0.5$.

\section{Results \label{sec:result}}

In this section, we provide an explicit example on an extremely small problem to make our description cogent and where we can compare with analytic results.
Our example generalizes to more degrees of freedom (larger volumes, larger dimensions,  larger gauge groups) where D-Wave is applicable as well, and may outperform the classical Monte Carlo simulations.

We have tested our setup for the gauge group $U(1)$ on $2 \times 2$, $4 \times 4$, and $2 \times 2 \times 2$ lattices.
The physical parameters are:
\begin{align}
    am_q&=0.3, 0.6,&  \gamma&=0.5, 1.0 \quad \text{for $U(1)$}, \\
    am_q&=0.5, 1.0,&  \gamma&=0.5, 1.0 \quad \text{for $U(3)$}.
\end{align}
which correspond to intermediate quark masses and rather low temperatures. 

\subsection{Parameter tuning}

\begin{figure}
\includegraphics[width=0.5\textwidth]{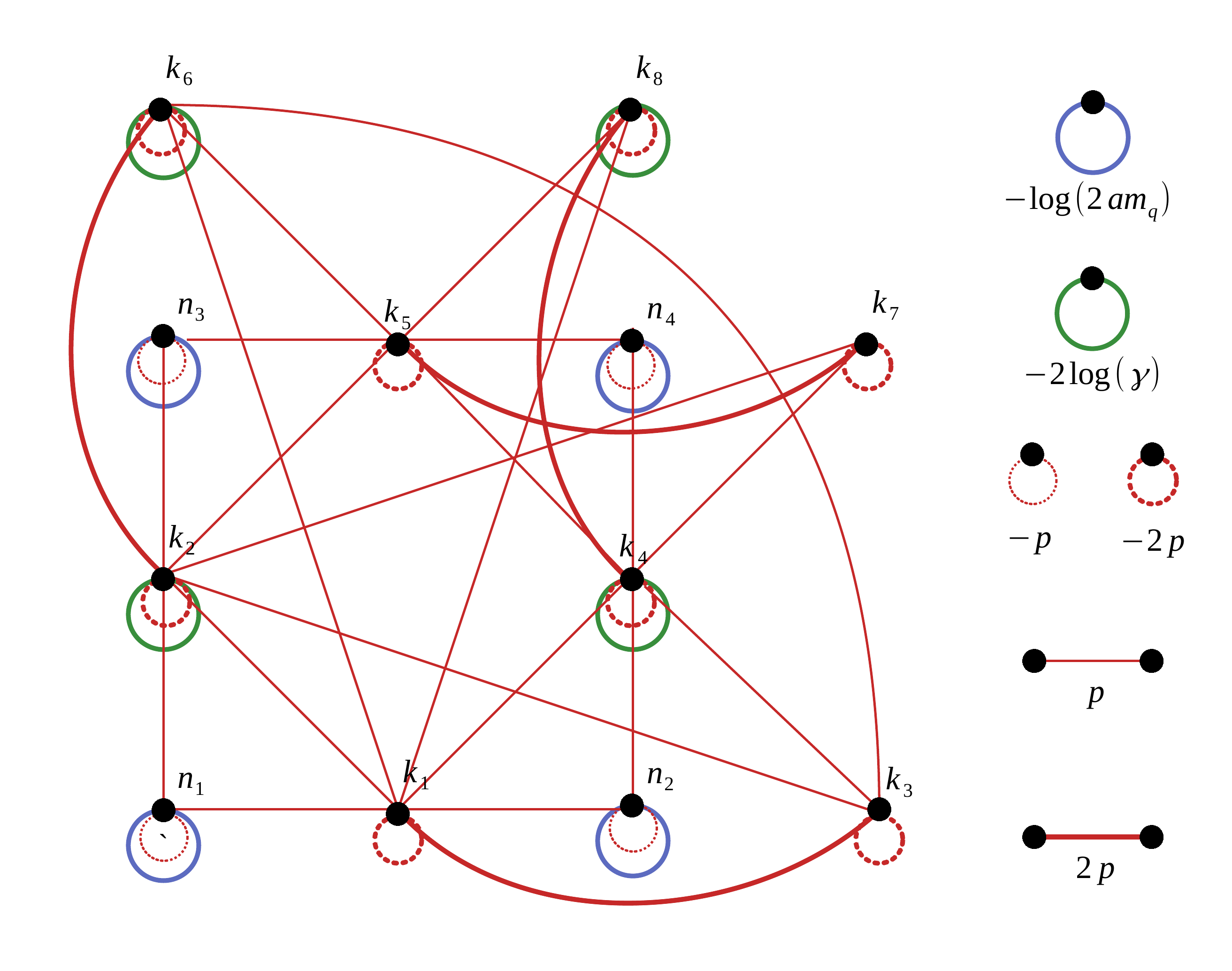}
\caption{The problem graph for gauge group $\U(1)$ on the $2\times 2$ lattice, every site is connected to itself and 2 dimers, every bond is connected itself, to two sites and 5 other bonds.}
\label{problemgraph}
\end{figure}
One tunable parameter is the penalty factor $p$ given in~\eref{eq:QUBOdef}.  It controls the balance between the action \eref{ActionBinary} and the Grassmann constraint \eref{U1_GC} in the QUBO matrix $Q$.
Another free parameter is the \texttt{chain\_strength} introduced in~\ref{subsec:DWAVE_Description}.  As its name suggests, it controls the strengths of chains used to build physical qubits into logical qubits, ensuring that physical qubits act in unison~\cite{chain}.  Varying this parameter also changes the relative size of the elements of $Q$ in relation to the \texttt{chain\_strength}~\cite{chain}. 
Since our problem graph, corresponding to the adjacency graph of our QUBO matrix, see \fref{problemgraph}, does not necessarily have the same topology as the QPU, such that no one-to-one embedding can be found, 
a non-trivial \texttt{chain\_strength} is required in order to retain the logical qubits.  Too weak a value results in `chain breaks', meaning the physical qubits within a logical qubit do not act in unison and our topology is broken.  On the other hand, too strong a value overwhelms the QUBO matrix and essentially makes its contribution to the minimization negligible.  In this case one obtains purely random solutions with no connection to the system under investigation.

Therefore it is important to find the optimal \texttt{chain\_strength} that ensures the correct topology dictated by the adjacency matrix of our system while providing the maximum rate of valid solutions.
%which should be not too small and not too large: not too small in order not to break the chain, not too large to account for the QUBO matrix in order to generate valid configurations more efficiently. 
Since the run-time on D-wave to generate 500 samples is constantly about 600 msec when we use the annealing schedule of the top panel in Fig.~\ref{fig:anneal}, a higher rate of valid solution vectors corresponds to more statistics.
After a single simulation, the D-wave system provides the \texttt{chain\_break\_fraction}, which is the ratio of the number of broken chains over the total number of chains. If this value is non-zero, then the solution vector encodes mostly invalid configurations or sub-optimal solutions. 
We define the \texttt{chain\_break\_rate} by the percentage of the broken chains to the solutions vectors returned from D-wave.
The \texttt{unbroken\_chain\_rate} is $1-\texttt{chain\_break\_rate}$.

\begin{figure*}[btp]
    \centerline{
    \includegraphics[width=0.49\textwidth]{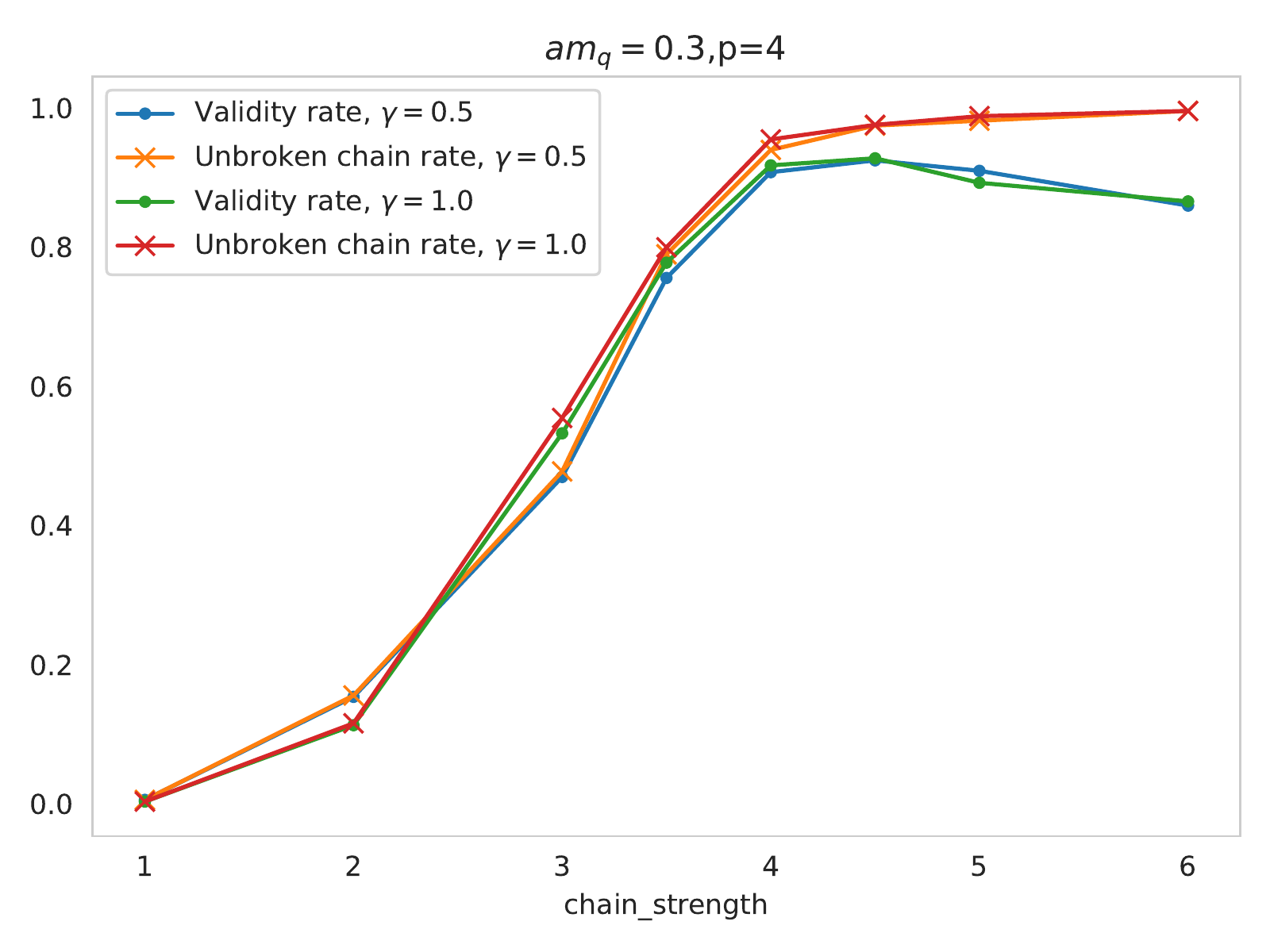}
    \includegraphics[width=0.49\textwidth]{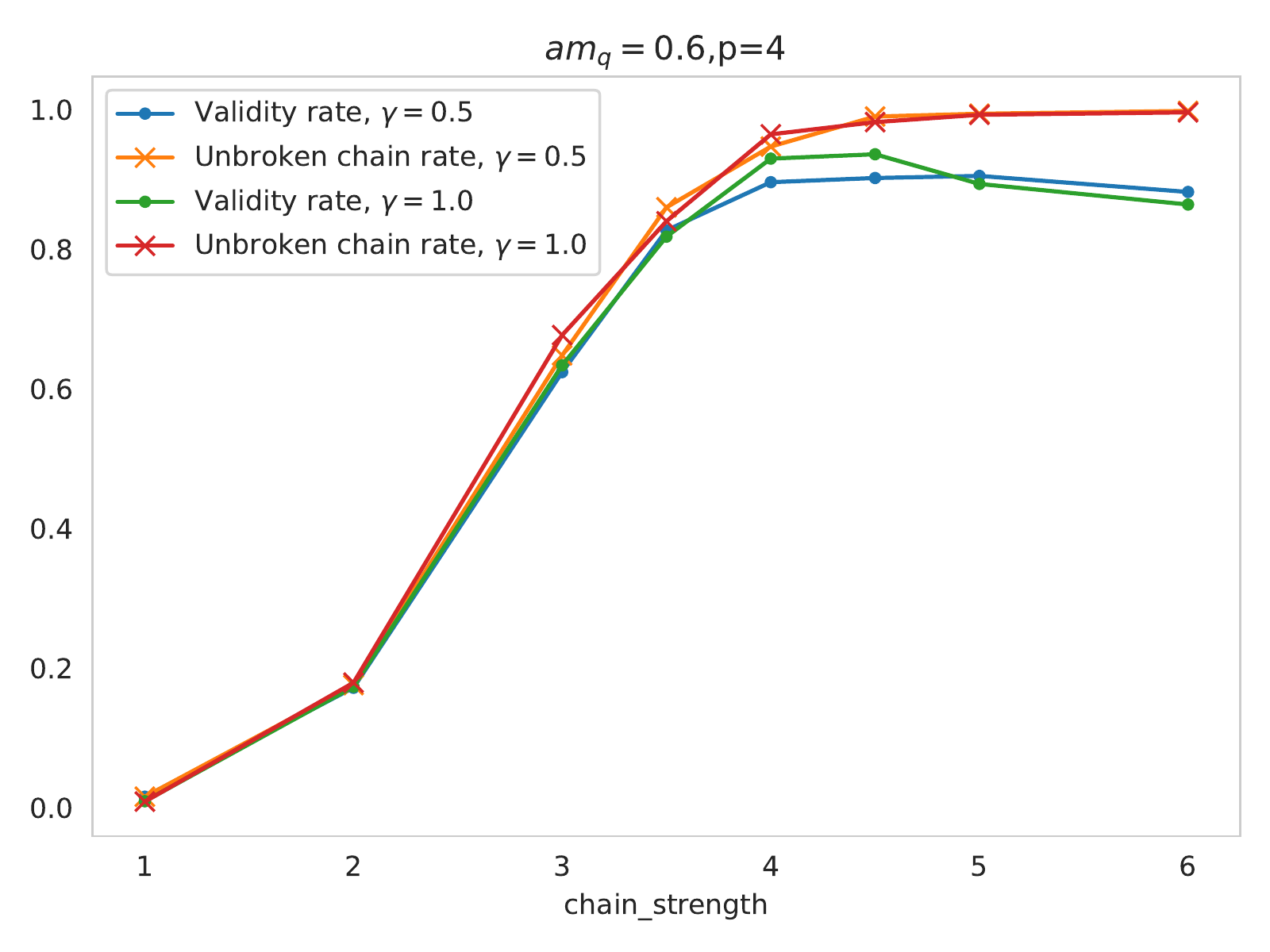}
    }
    \caption{Comparing the valid configuration rate and unbroken chain rate at $am_q=0.3,0.6$, $\gamma=0.5,1.0$, $p=4$ on $2 \times 2$.
    The \texttt{unbroken\_chain\_rate} is the percentage of unbroken chains to the solutions from D-wave. The validity rate and \texttt{unbroken\_chain\_rate} are the same for small values of \texttt{chain\_strength} and starts deviating with increasing \texttt{chain\_strength}. We select the optimal value of \texttt{chain\_strength} that corresponds to the validity rate being maximal.  }
    \label{fig:chain_tuning}
\end{figure*}

For small values of \texttt{chain\_strength}, we find that the validity rate, i.e.~the percentage of valid solution vectors, is identical to the \texttt{unbroken\_chain\_rate}.
This behavior persists up to some specific value of \texttt{chain\_strength}, after which the validity rate drops again and hence starts deviating from \texttt{unbroken\_chain\_rate}. This correlation is shown in Fig.~\ref{fig:chain_tuning}, where we compare the validity rate and \texttt{unbroken\_chain\_rate} for $am_q=0.3,0.6$ and $\gamma=0.5,1.0$ at $p=4$. We find that the correlation is almost independent of the physical parameters.
The \texttt{unbroken\_chain\_rate} approaches 1 as the \texttt{chain\_strength} becomes larger than $\simeq p$, after which
the validity rate drops. The optimal value of \texttt{chain\_strength} can thus be selected as the point of maximal validity rate, which is typically slightly below $p$.

\begin{figure*}
    \centering
    \includegraphics[width=0.24\textwidth]{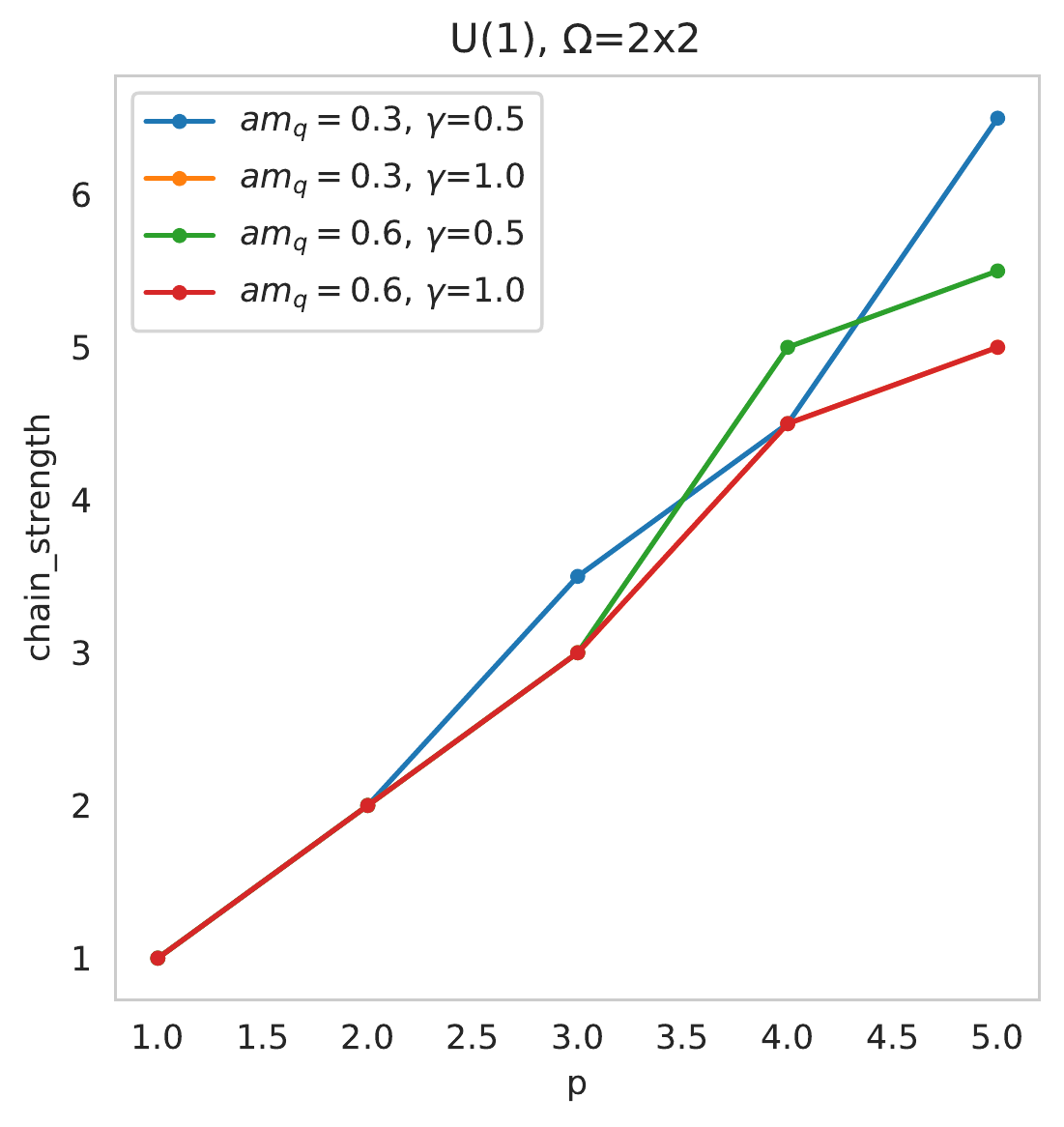}
    \includegraphics[width=0.24\textwidth]{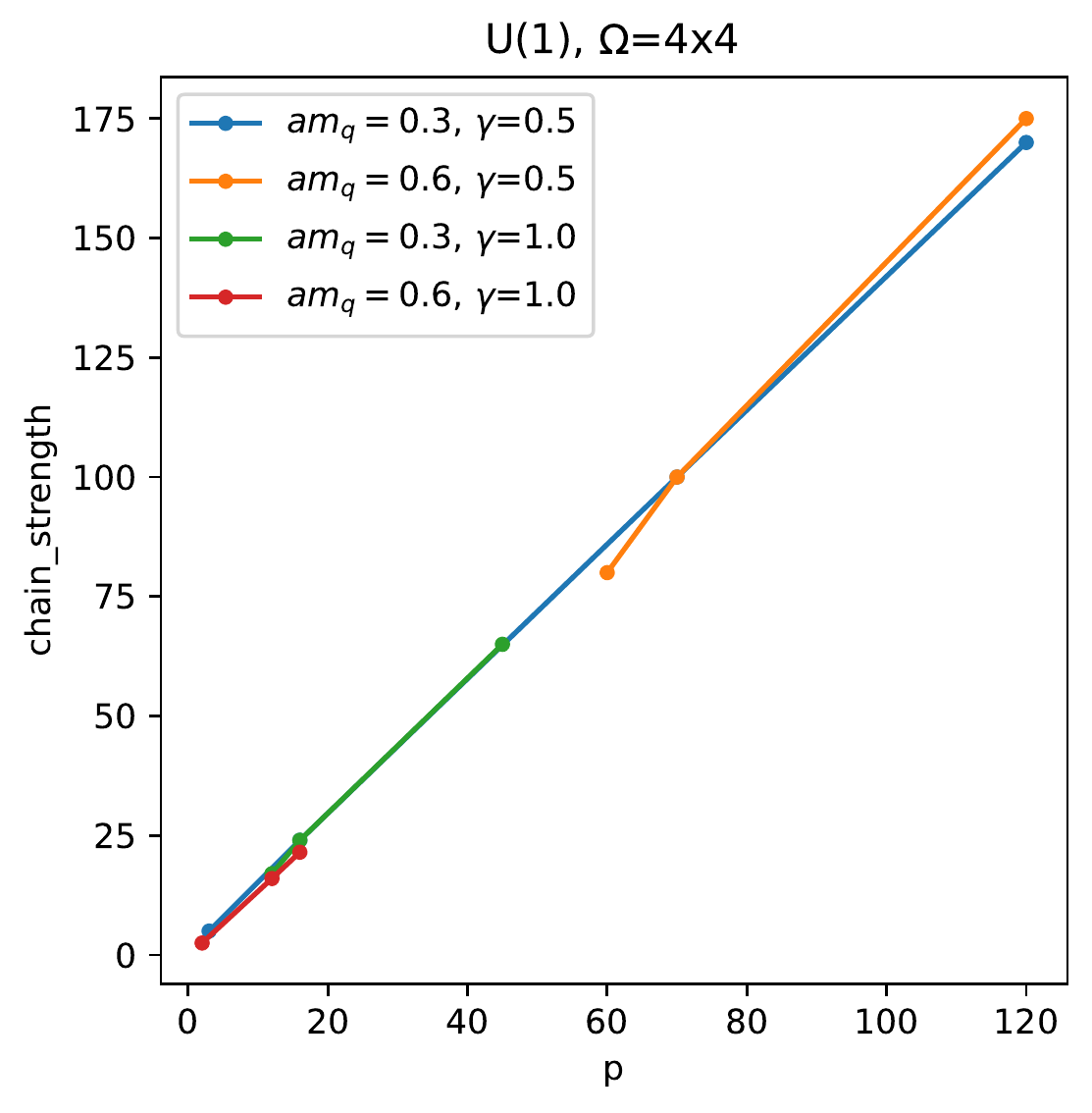}
    \includegraphics[width=0.24\textwidth]{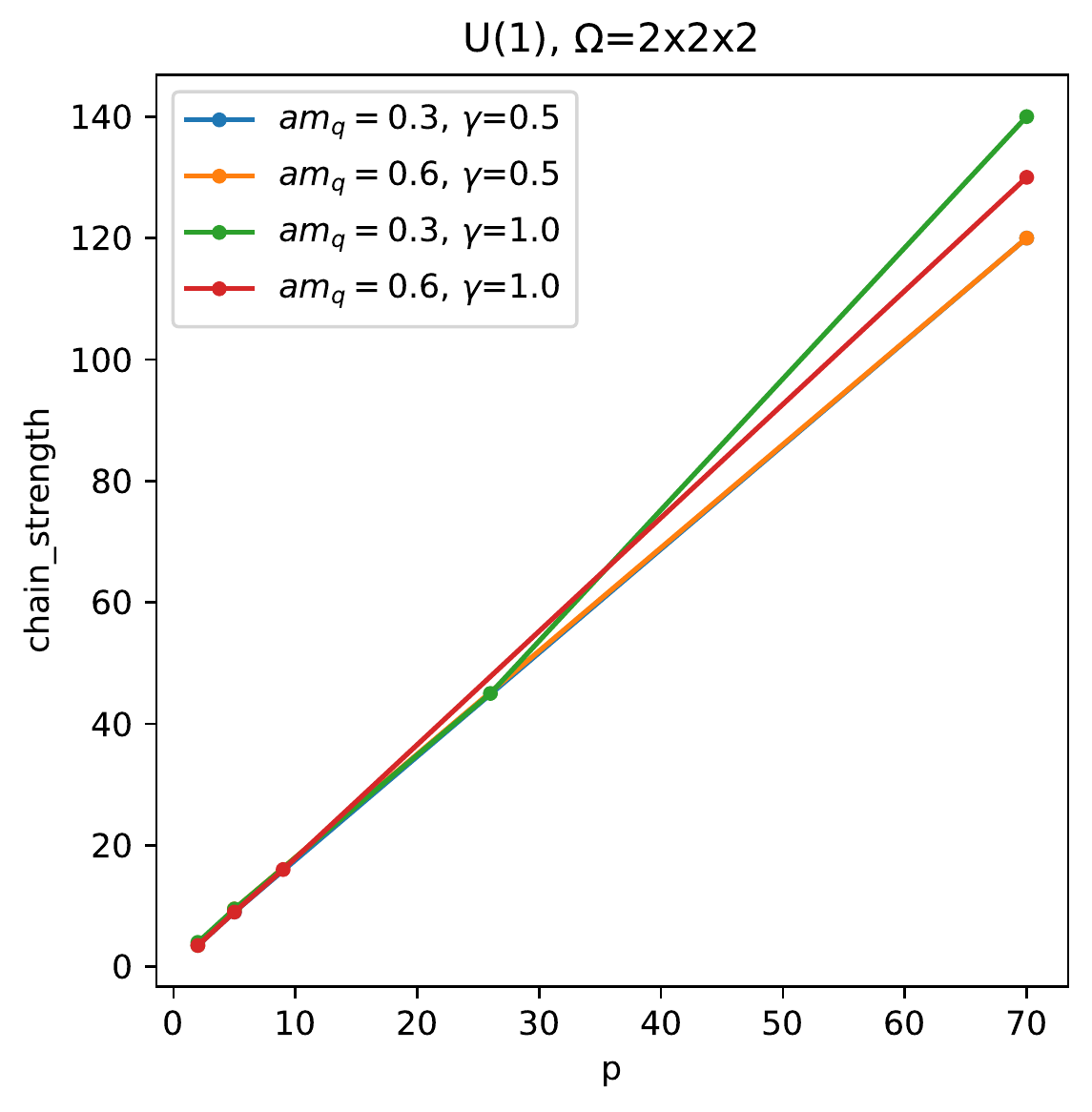}
    \includegraphics[width=0.24\textwidth]{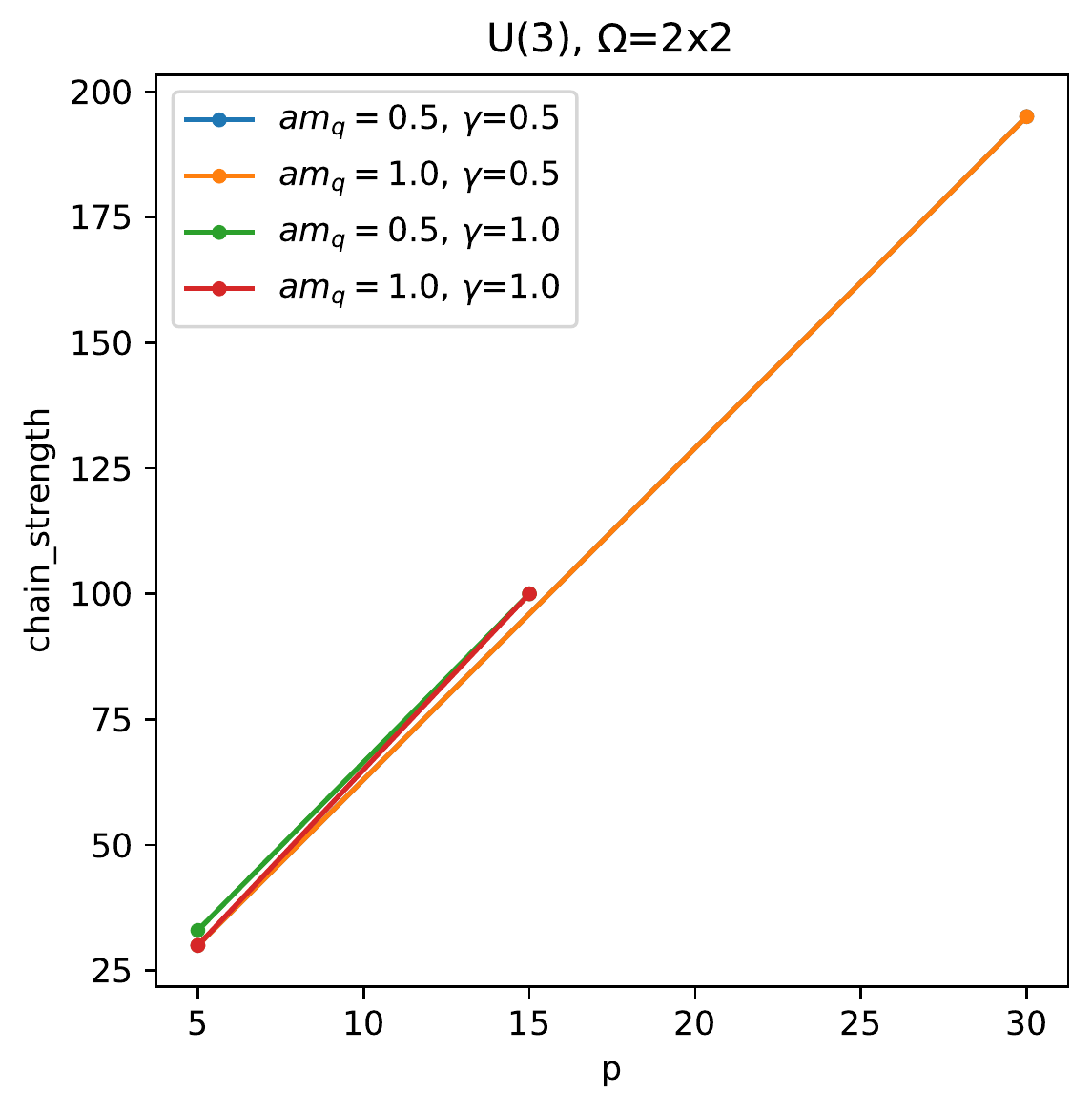}
    \caption{$p$ dependence of \texttt{chain\_strength} for $U(1)$ on $2 \times 2$, $4 \times 4$ and $2 \times 2 \times 2$ and $U(3)$ on $2 \times 2$. The optimal \texttt{chain\_strength} is proportional to $p$ because the maximum absolute element of our QUBO matrix for the selected physical parameters in the paper is itself proportional to $p$. However, we need stronger \texttt{chain\_strength} for larger volumes and larger $N_c$ to find the best validity rate.}
    \label{fig:chain_vs_p}
\end{figure*}

By determining the optimal \texttt{chain\_strength} for various values of $p$, we find their correlation to be proportional. For our physical parameters, $am_q=0.3,0.6$ and $\gamma=0.5,1.0$, the maximum absolute value of the weight matrix $W$ \eref{ActionBinary} is smaller than 2. Thus the maximum absolute value of QUBO matrix is always $2p$ since we use $p$ values greater than 1. Once the QUBO matrix is submitted to the QPU, D-Wave automatically scales the elements of $Q$ into the range of $[-1,1]$ using the maximum absolute value of all $Q$ elements. If \texttt{chain\_strength} is larger than the largest weight in $Q$, \texttt{chain\_strength} is used for this scaling. Hence, if \texttt{chain\_strength} is too large, the weight of $Q$ shrinks to near zero. Ideally, \texttt{chain\_strength} should be $2p$ in our system. We found that when \texttt{unbroken\_chain\_rate} is smaller than 1, the validity rate is maximal and the corresponding \texttt{chain\_strength} is smaller than $2p$. This explains the data shown in Fig.~\ref{fig:chain_vs_p} for the three different volumes under consideration.

\begin{figure*}
    \centering
    \includegraphics[width=0.24\textwidth]{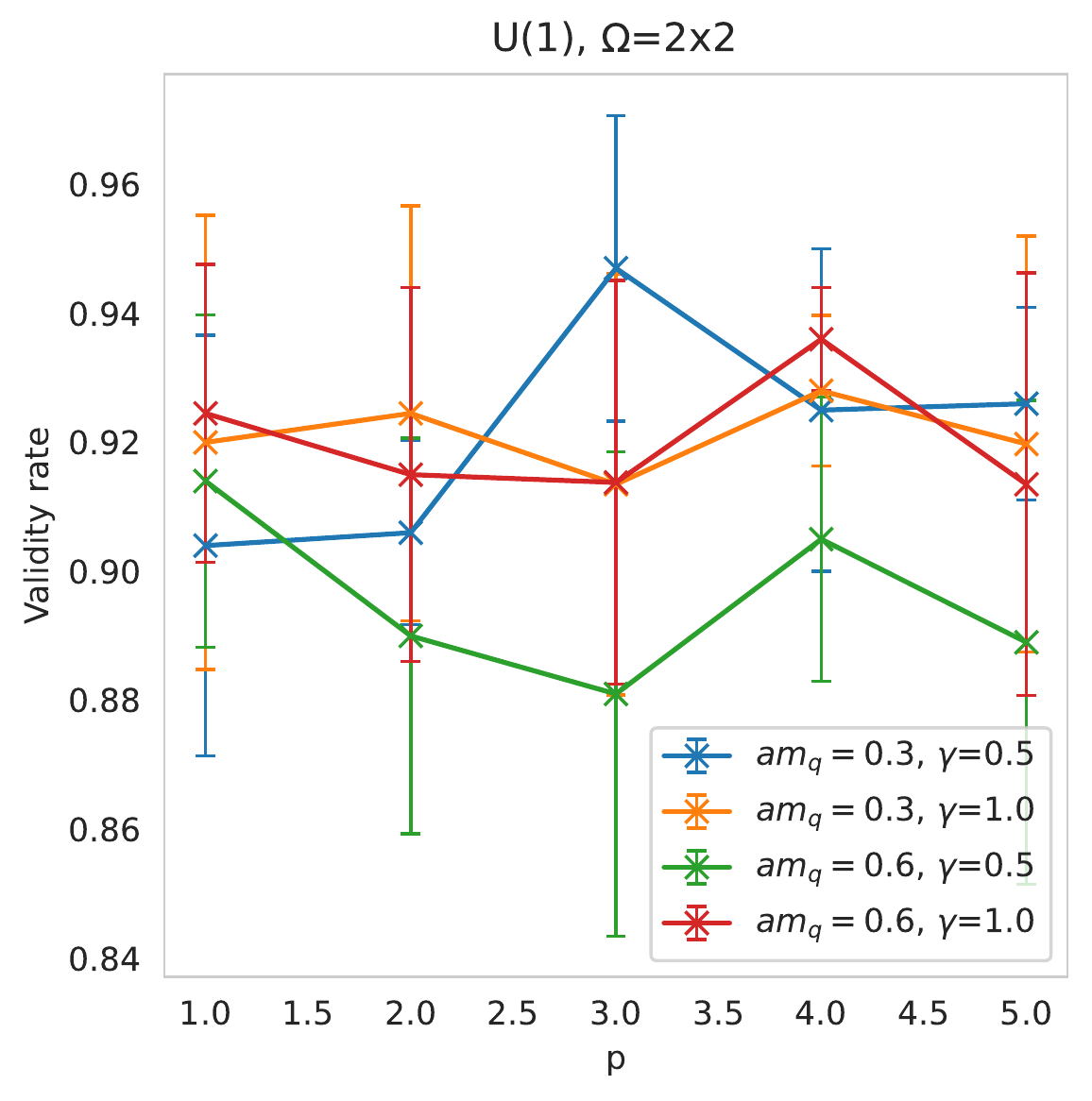}
    \includegraphics[width=0.24\textwidth]{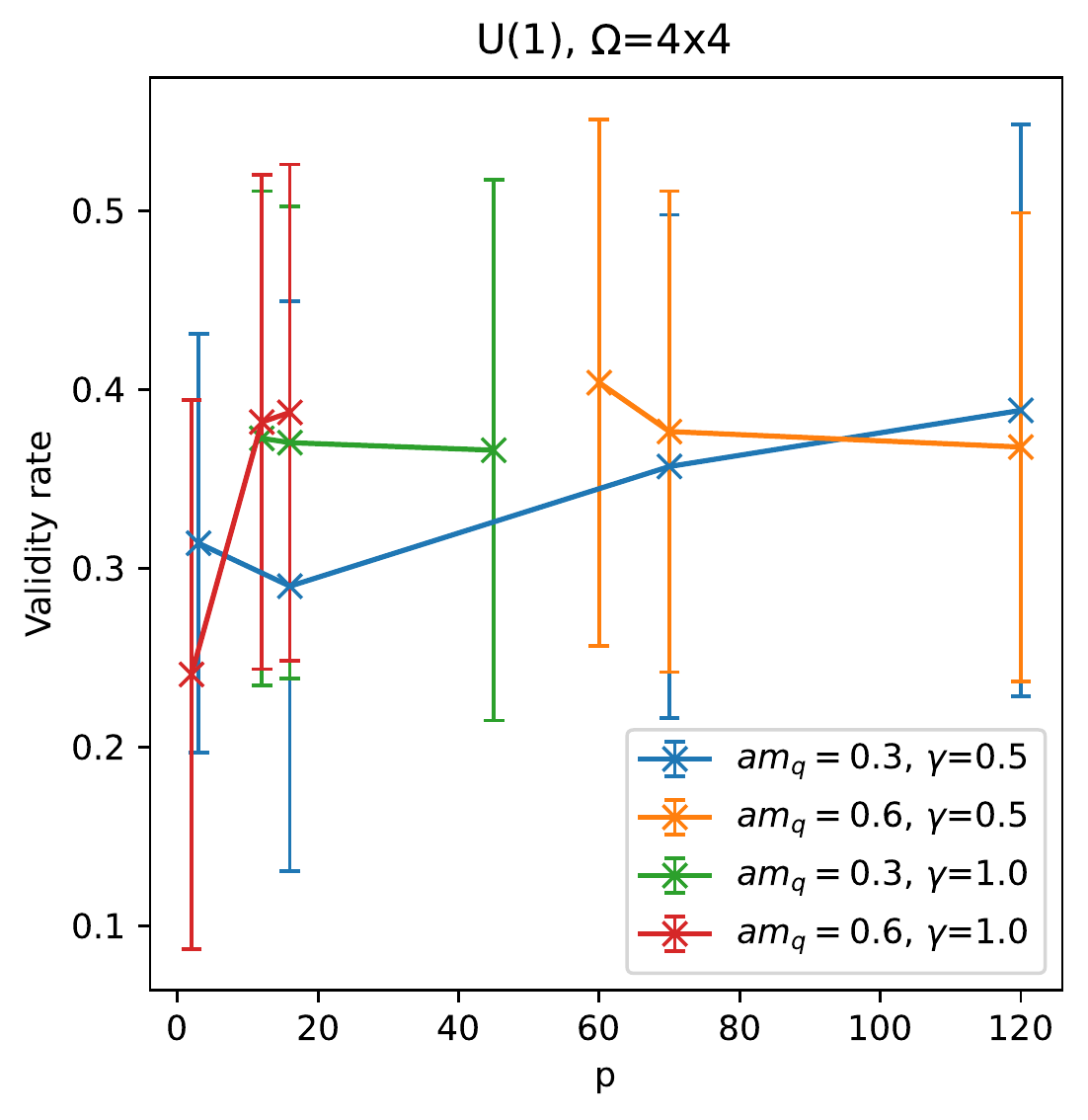}
    \includegraphics[width=0.24\textwidth]{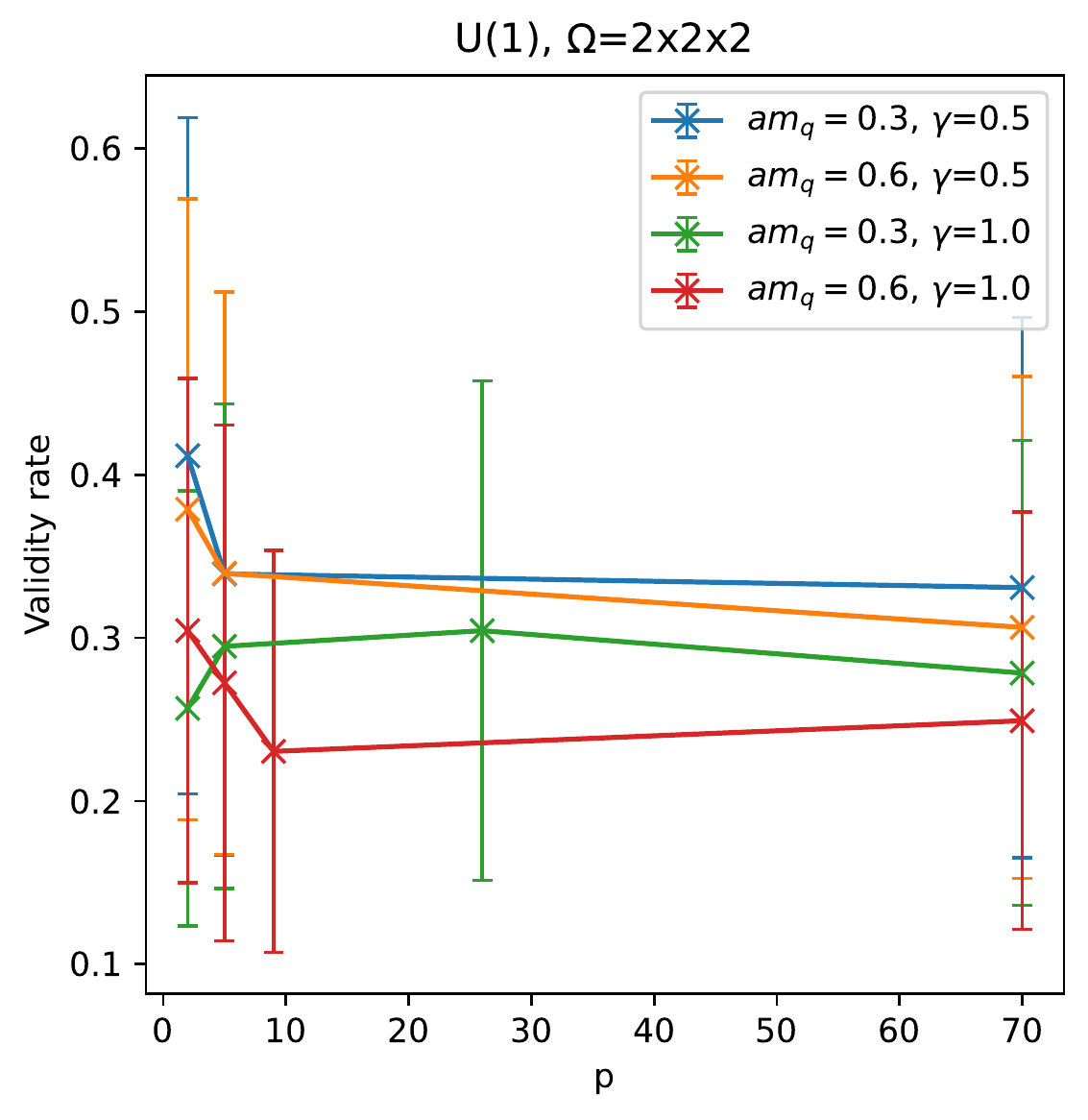}
    \includegraphics[width=0.24\textwidth]{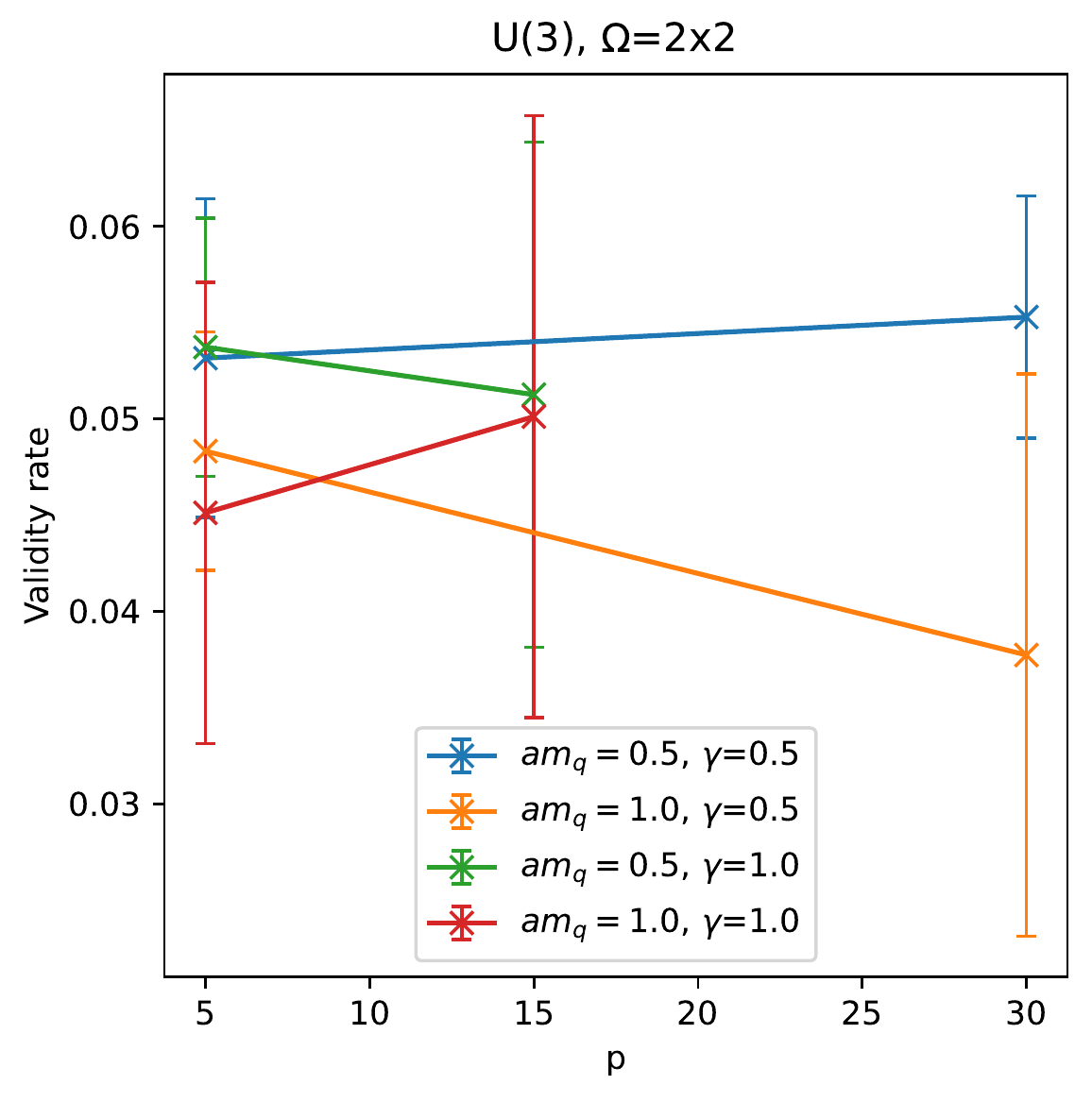}
    \caption{$p$ dependence of the validity rate on $2 \times 2$, $4 \times 4$ and $2 \times 2 \times 2$ systems. The validity rate is almost 90\% for $2 \times 2$ volume, but for the larger volumes this decreases to about 40\%.  The lower validation rates for the larger volumes is due to the fact that the ratio of the total number of valid configurations to total number of binary-vector solutions that D-Wave can generate is very small:   $0.000016042\%$ for $2 \times 2 \times 2$ and $41025/2^{48}$ for $4 \times 4$.  In contrast, in the smallest volume, $2\times 2$, the valid solutions span 0.42\% of all possible solutions. We find that the validity rate is small for $U(3)$, about 5\%.}
    \label{fig:valid_vs_p}
\end{figure*}

In Fig.~\ref{fig:valid_vs_p}, we present the rate of produced valid configurations at the optimal \texttt{chain\_strength} at each $p$ for three volumes. There is no strong dependence on $p$. For the $2 \times 2$ lattice, there are 17 valid configurations and the total configuration space is $2^{E+\Omega}=4096$. The probability to find the valid configuration randomly is $17/4096=0.42\%$. Other cases have, for example, $689/2^{32}=0.000016042\%$ for the $2 \times 2 \times 2$ lattice and $41025/2^{48}$ for the $4 \times 4$ volume. Hence, finding valid configurations is much more difficult on larger volumes. 
In the case of $U(3)$, it requires much larger \texttt{chain\_strength} since the structure of $Q$ matrix is much more complicate. Even with stronger \texttt{chain\_strength} and higher probability to find valid configurations $695/2^{24}$, the validity rate is about $5\%$. 

%However, the validity rate remains consistently over 0.3 and does not decrease drastically as the volume increases. This is a strong indication that D-wave can also address larger volumes and could potentially outperform classical computers. \jangho{we need to modify the above paragraph.}

The finding that this is roughly independent of the physical parameters suggests that some sort of importance sampling, in a manner similar to Monte Carlo algorithms, is being done by the D-Wave.
The number of valid configurations for other gauge groups and volumes are given in Table:~\ref{tab:N_valid}. 

\subsection{Observables and histogram reweighting}
\begin{figure}
    \centering
    \includegraphics[width=0.5\textwidth]{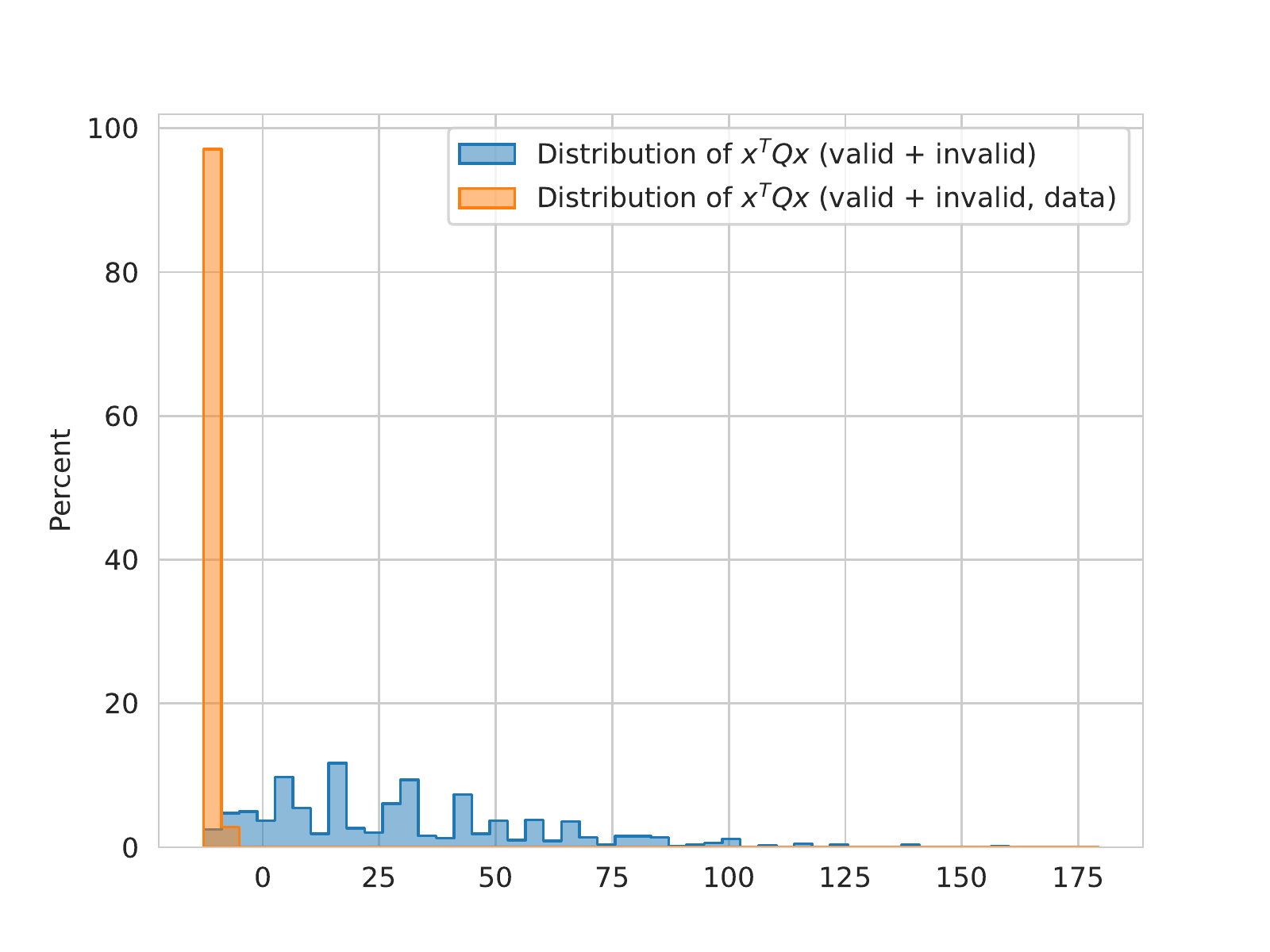}
    \caption{Histogram of $x^T Q x$ at $am_q=0.6$, $\gamma=1$, $p=3$, $\texttt{chain\_strength}=3$ on the $2 \times 2$ lattice. Blue histogram shows all possible 4096 configurations and orange comes from the D-wave samples. The D-wave samples show solutions $x$ which minimize  $x^T Q x$. This indicates that D-wave generates samples that take into account the QUBO matrix $Q$.}
    \label{fig:Q_histo}
\end{figure}

\begin{figure*}[h]
    \includegraphics[width=0.45\textwidth]{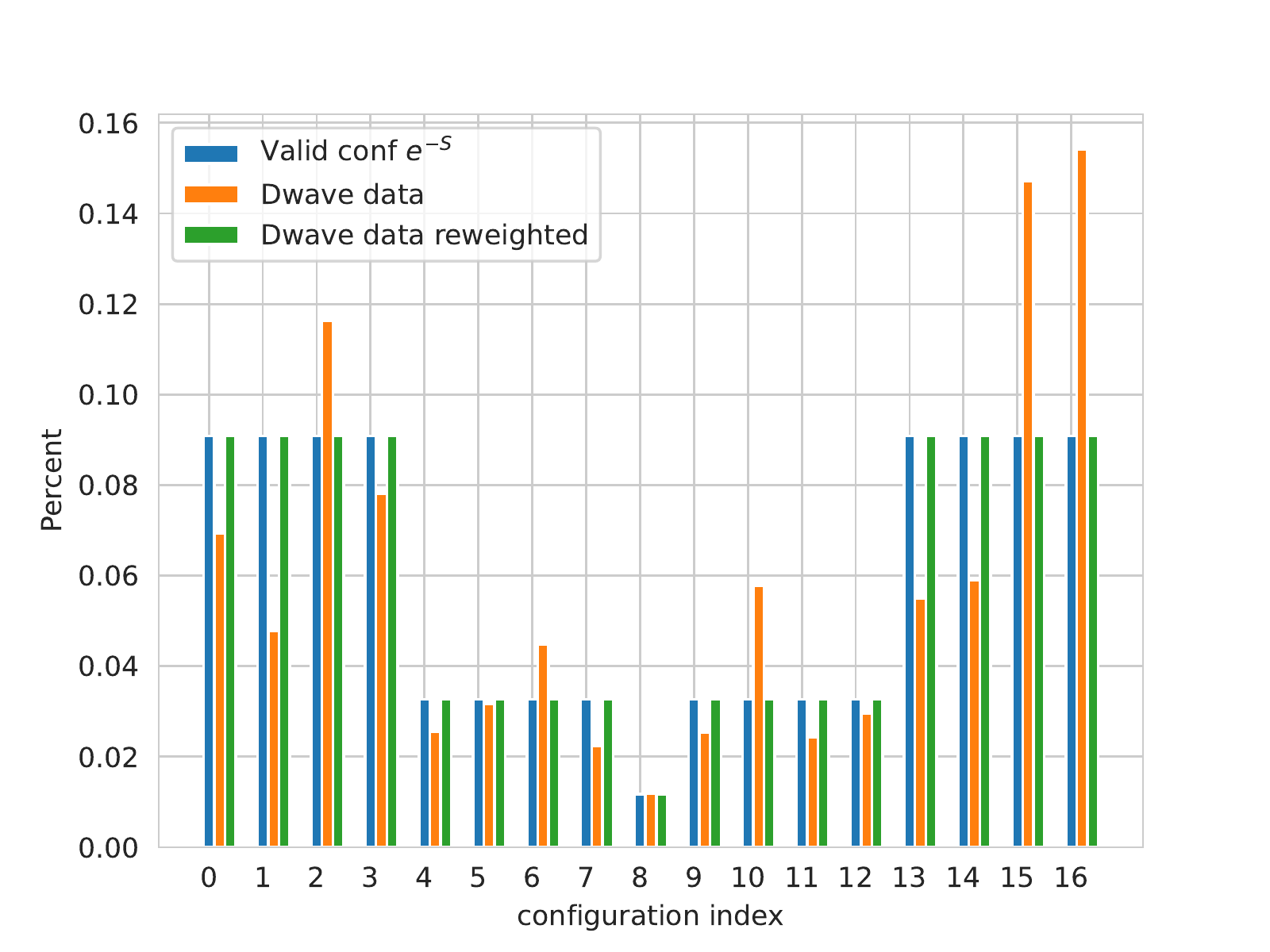}
    \includegraphics[width=0.45\textwidth]{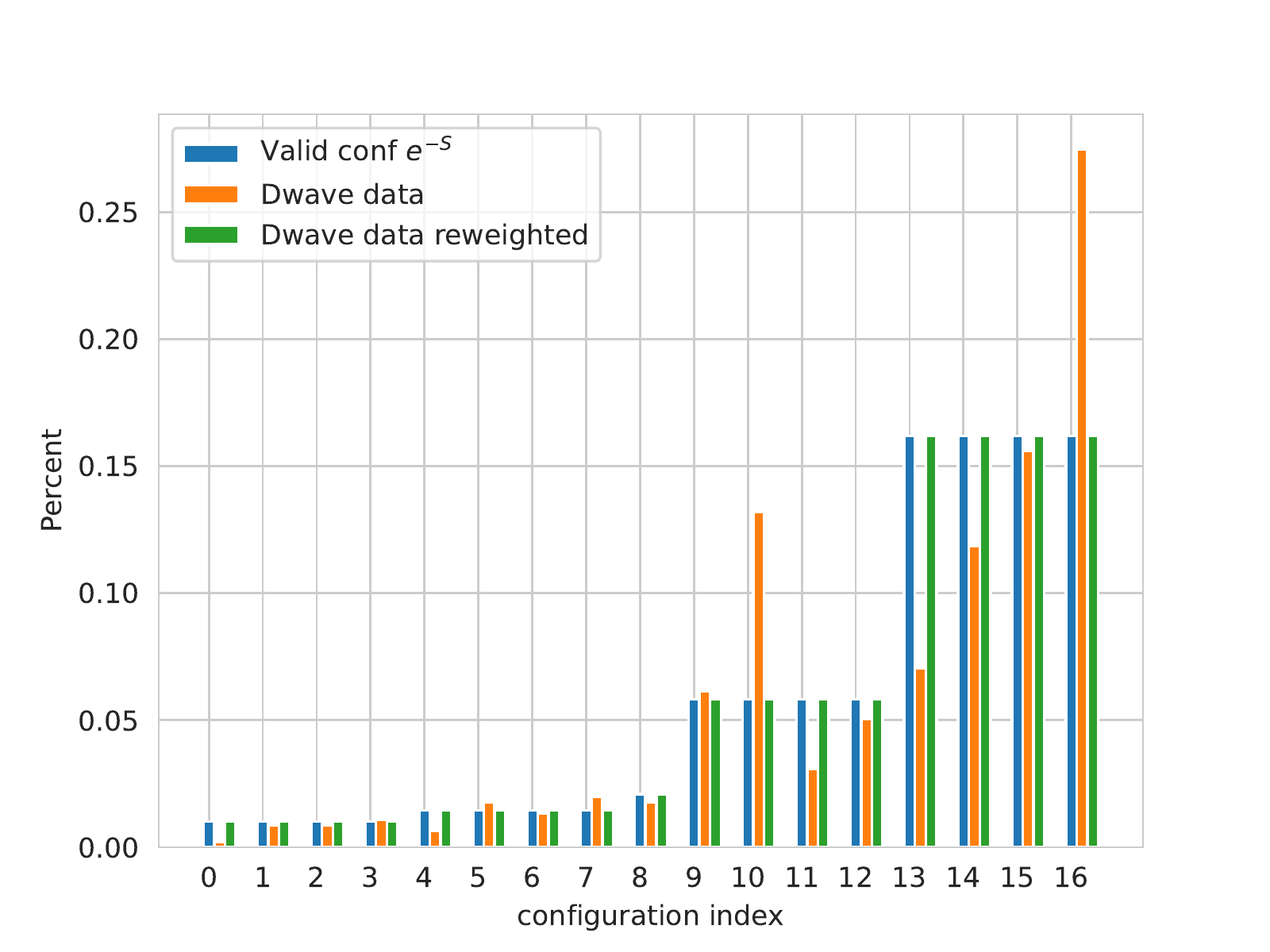}
    \caption{Histogram of D-wave output for the known valid configurations at $am_q=0.3$, $\gamma=1$, $p=3$(left), $\texttt{chain\_strength}=3$ and $am_q=0.3$, $\gamma=0.5$, $p=4$, $\texttt{chain\_strength}=5$(right) on $2 \times 2$ lattice. The orange distribution is not exactly same with the blue distribution, but similar by considering the statistical fluctuation. Using the reweighting method, the orange distribution can be reconstructed to green distribution without knowing the analytic solutions.}
    \label{fig:histo_conf}
\end{figure*}

D-wave attempts to find solutions which minimize the $\chi^2=x^TQx$ \eref{eq:chi2}. In Fig.~\ref{fig:Q_histo}, we show the distribution of $\chi^2$ for all possible $x$ (blue histogram) for $\U(1)$ on $2 \times 2$ volume. The number of possible $x$ is $2^{E+\Omega}=2^{12}=4096$ and the distribution depends on the physical parameters $\gamma$, $am_q$, and the penalty term $p$. The number of data generated by D-wave (orange histogram) is 14000. Since the orange peak does not follow the blue distribution and forms around the minimum, the QUBO matrix we implement is being taken into account for D-wave sampling. The detailed distribution of the D-wave samples are presented in Fig.~\ref{fig:histo_conf}.
%From the Fig.~\ref{fig:Q_histo}, D-wave generates the configuration which minimize $x^T Q x$, not only the absolute minimum but also the distribution near minimum. 
Since the output of D-wave also includes invalid configurations, a post-process check of valid configurations that satisfy the Grassmann constraint, $A\cdot x + b=0$, is necessary.  This check is efficient and simple to perform.

With the coordinate convention in Fig.~\ref{fig:U1_2x2}, we provide a list of all valid configurations in Table~\ref{tab:valid_conf}.
\begin{table}
\label{tab:conf_2x2}
    \centering
    \resizebox{0.5\textwidth}{!}{
    \begin{tabular}{c|cccccccccccc|ccc}
     Index & ($k_1$, &$k_2$, &$k_3$, &$k_4$, &$k_5$, &$k_6$, &$k_7$, &$k_8$, &$n_1$, &$n_2$, &$n_3$, &$n_4$) & $D_t$& $D_s$ & $M$\\
     \hline
     0 & (0,&0,&1,&1,&0,&0,&0,&0,&0,&0,&0,&0) & 2 & 0 & 0\\ %2 temporal dimers
     1 & (0,&0,&1,&0,&0,&0,&0,&1,&0,&0,&0,&0) & 2 & 0 & 0 \\
     2 & (0,&0,&0,&1,&0,&0,&1,&0,&0,&0,&0,&0) & 2 & 0 & 0 \\
     3 & (0,&0,&0,&0,&0,&0,&1,&1,&0,&0,&0,&0) & 2 & 0 & 0 \\
     \hline
     4 & (0,&0,&1,&0,&0,&0,&0,&0,&0,&1,&0,&1) & 1 & 0 & 2 \\ %1 temporal dimer and 2 monomers
     5 & (0,&0,&0,&1,&0,&0,&0,&0,&1,&0,&1,&0) & 1 & 0 & 2 \\
     6 & (0,&0,&0,&0,&0,&0,&1,&0,&0,&1,&0,&1) & 1 & 0 & 2 \\
     7 & (0,&0,&0,&0,&0,&0,&0,&1,&1,&0,&1,&0) & 1 & 0 & 2 \\
     \hline
     8 & (0,&0,&0,&0,&0,&0,&0,&0,&1,&1,&1,&1) & 0 & 0 & 4 \\ % 4 monomers
     \hline
     9 & (1,&0,&0,&0,&0,&0,&0,&0,&0,&0,&1,&1) & 0 & 1 & 2 \\ % 1 spatial dimer and 2 monomers
     10 & (0,&1,&0,&0,&0,&0,&0,&0,&0,&0,&1,&1) & 0 & 1 & 2 \\
     11 & (0,&0,&0,&0,&1,&0,&0,&0,&1,&1,&0,&0) & 0 & 1 & 2 \\
     12 & (0,&0,&0,&0,&0,&1,&0,&0,&1,&1,&0,&0) & 0 & 1 & 2 \\
     \hline
     13 & (1,&0,&0,&0,&1,&0,&0,&0,&0,&0,&0,&0) & 0 & 2 & 0 \\ % 2 spatial dimers
     14 & (1,&0,&0,&0,&0,&1,&0,&0,&0,&0,&0,&0) & 0 & 2 & 0 \\
     15 & (0,&1,&0,&0,&1,&0,&0,&0,&0,&0,&0,&0) & 0 & 2 & 0 \\
     16 & (0,&1,&0,&0,&0,&1,&0,&0,&0,&0,&0,&0) & 0 & 2 & 0 \\     
 \end{tabular}
 }
    \caption{List of valid configurations for $U(1)$ on $2 \times 2$. $D_t$: the number of temporal direction dimers, $D_s$: the number of spatial direction dimers, $M$: the number of monomers.}
    \label{tab:valid_conf}
\end{table}    
In this Table, the configuration with index (0-3) have two temporal dimers and index (4-7) have one temporal dimer and two monomers.
four monomer configuration is index (8) and index (9-12) represent one spatial dimer and two monomers configurations.
Index (13-16) have two spatial dimers. 
For example, at large quark mass, the configurations which have many monomers have larger weight.
This configuration index convention is applied to Fig.~\ref{fig:histo_conf}.

We compare the distribution of the valid configurations obtained from D-Wave with its corresponding weight $e^{-S}$ in Fig.~\ref{fig:histo_conf}. Because of low statistics and sub-optimal tuned parameters, \texttt{chain\_strength} and $p$, this distribution has some fluctuations from the analytic solution $e^{-S}$, though there is qualitative agreement. Once we find all valid configurations, reconstructing the exact same distribution with the analytic solution is possible by the single histogram reweighting method~\cite{Ferrenberg:1988yz}. The quark mass $am_q=0.3$ is small so the configurations which do not contain monomers are preferred and in the case of $\gamma=1$, spatial and temporal dimers have the same weight (left in Fig.~\ref{fig:histo_conf}). If $\gamma=0.5$, spatial dimers are more preferred than temporal dimers (right in Fig.~\ref{fig:histo_conf}). 

\begin{figure*}
    \centerline{
    \includegraphics[width=0.49\textwidth]{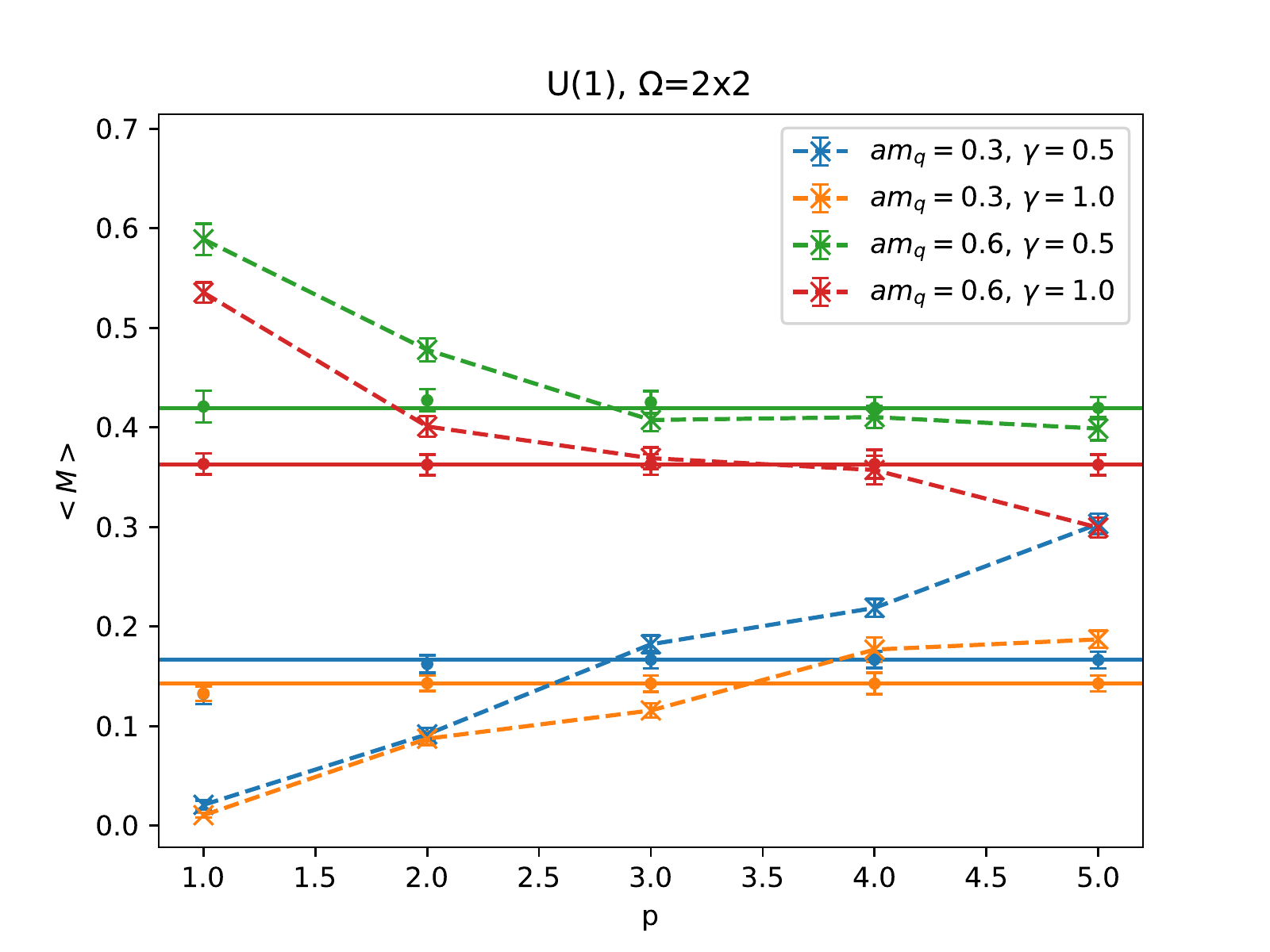}
    \includegraphics[width=0.49\textwidth]{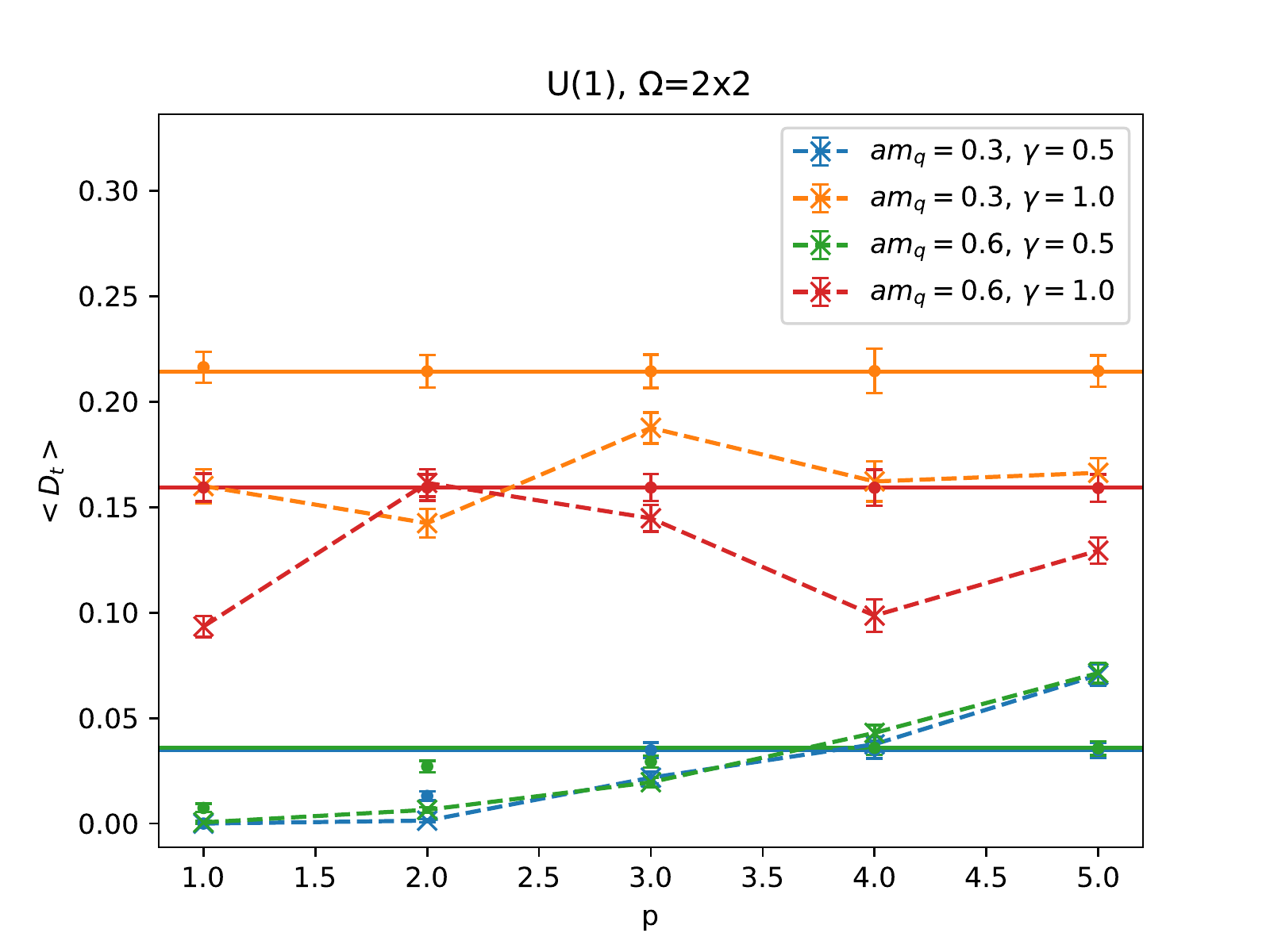}
    }
    \caption{$p$ dependence of monomer density and temporal dimer density. The cross points are the raw data of D-wave and circle points are reweighted data. The raw data are connected by dotted line to guide the eyes and the solid line is the analytic solutions on $2 \times 2$. When $p$ is greater than 2, D-wave find all valid configurations, so reweighting method can reconstruct the distribution exactly same with the ideal distribution of the analytic solution.}
    \label{fig:M_E_2x2}
\end{figure*}

\begin{figure*}
    \centerline{
    \includegraphics[width=0.49\textwidth]{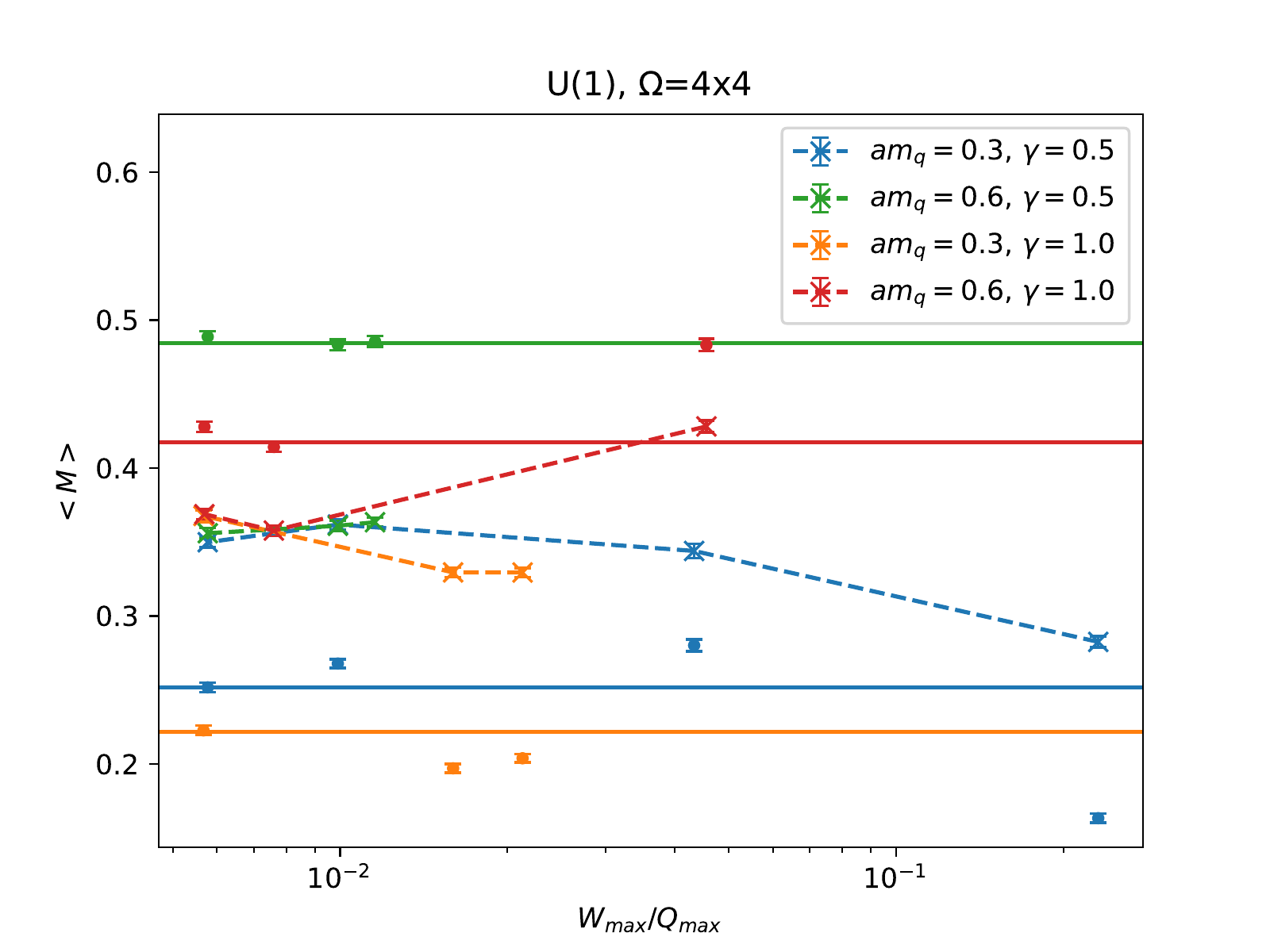}
    \includegraphics[width=0.49\textwidth]{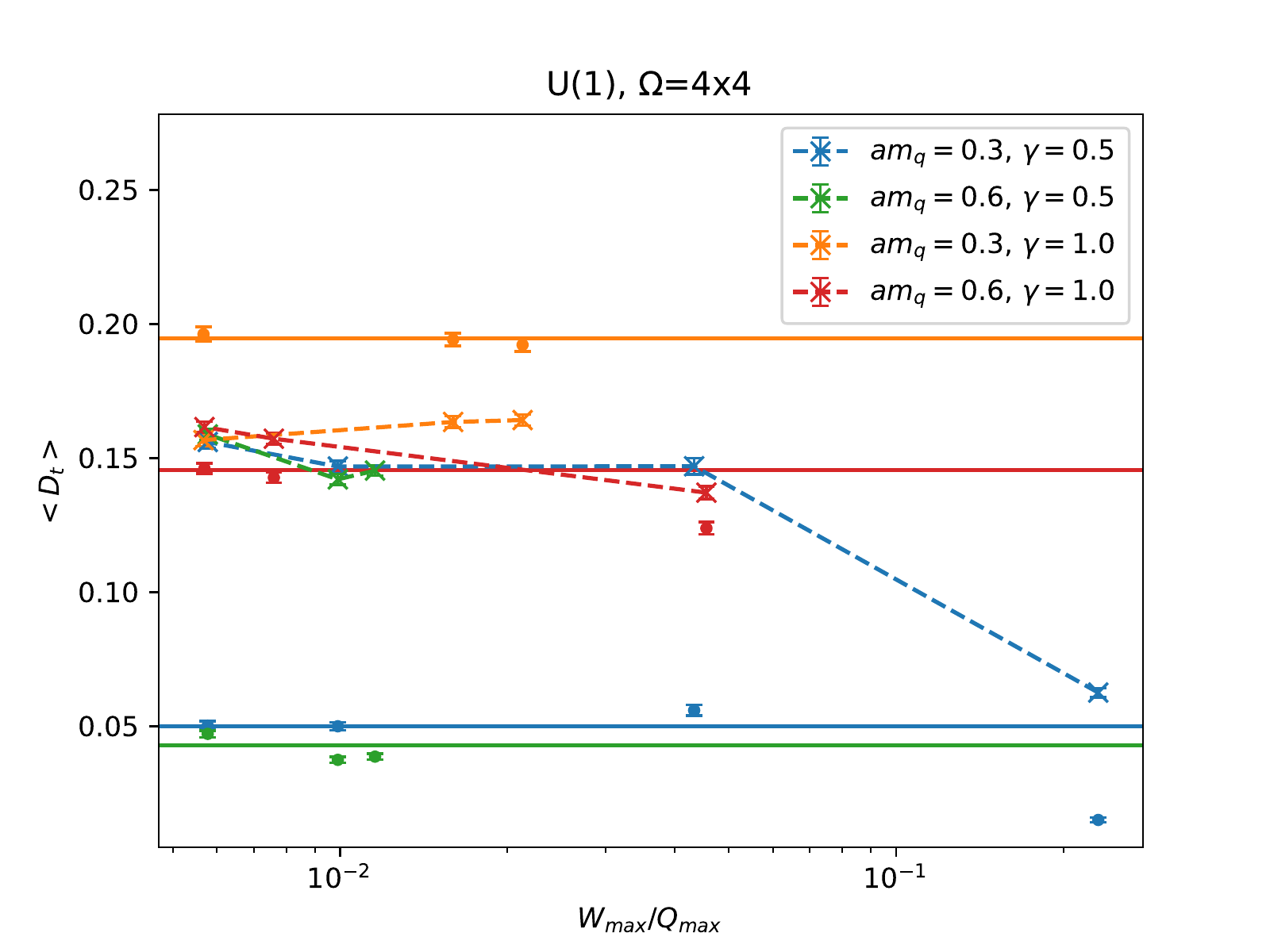}
    }
    \caption{$W_{max}/Q_{max}$ dependence of monomer density and temporal dimer density. The cross points are the raw data of D-wave and circle points are reweighted data. The raw data are connected by dotted line to guide the eyes and the solid line is the analytic solutions on $4 \times 4$. Here we have the statistics about $O(700-1800)$ valid configurations in 40125 total which is $1.7-4.3\%$.}
    \label{fig:M_E_4x4}
\end{figure*}

\begin{figure*}
    \centerline{   
    \includegraphics[width=0.49\textwidth]{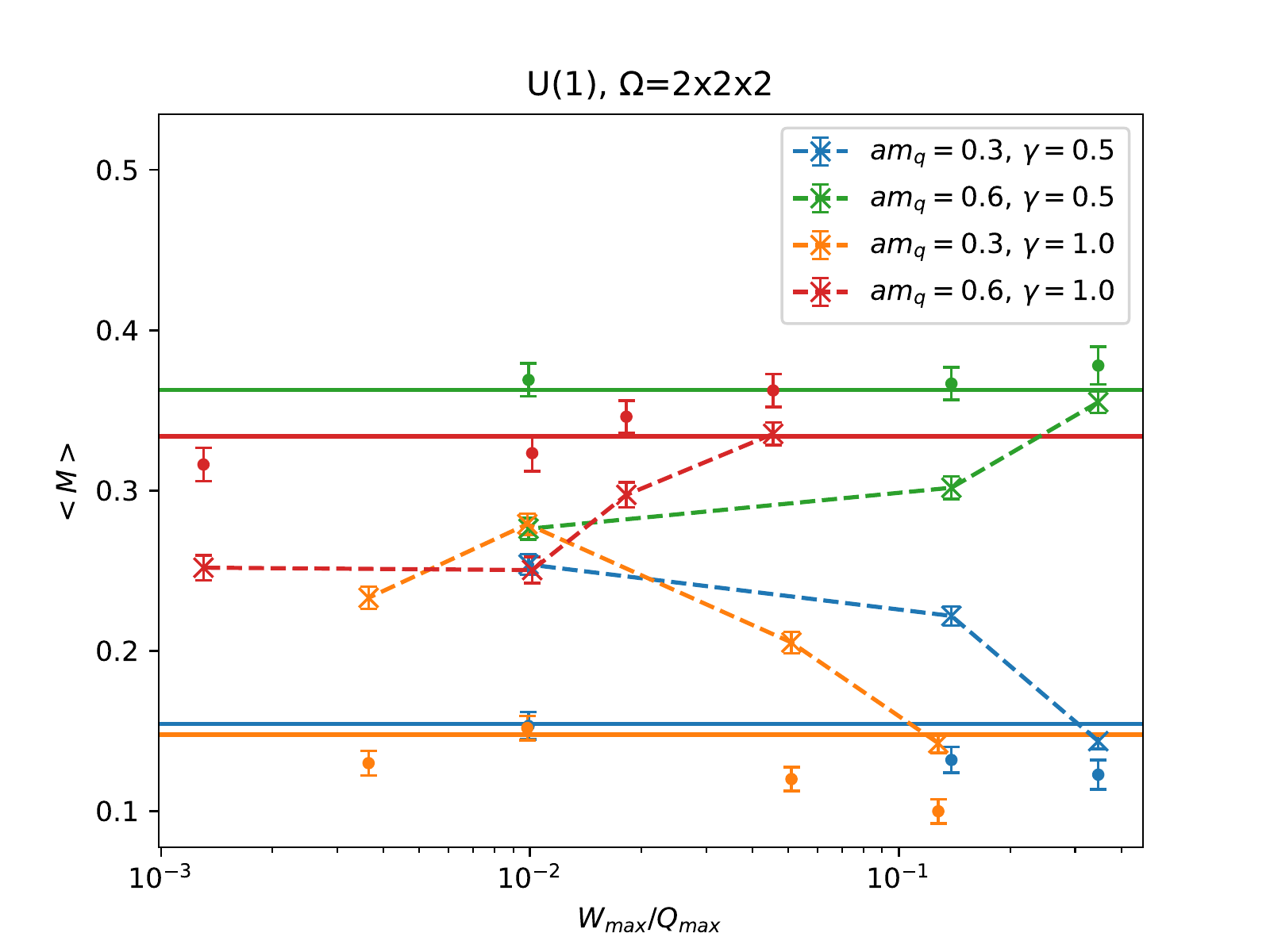}
    \includegraphics[width=0.49\textwidth]{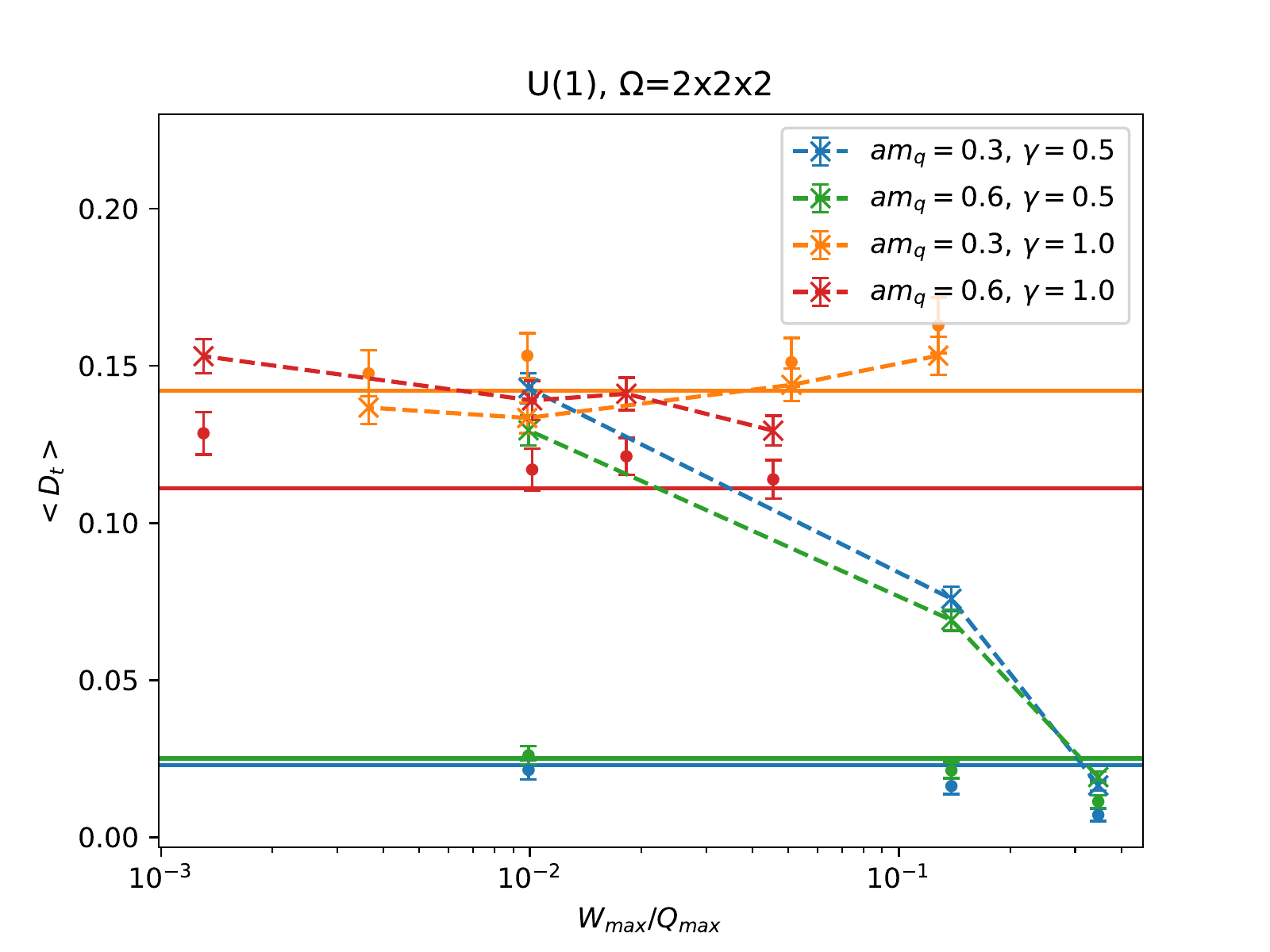}
    }
    \caption{$W_{max}/Q_{max}$ dependence of monomer density and temporal dimer density. The cross points are the raw data of D-wave and circle points are reweighted data. The raw data are connected by dotted line to guide the eyes and the solid line is the analytic solutions on $2 \times 2 \times 2$. We have about $40-65\%$ of valid configurations of 689 total, so the reweighting method guides the observables toward the analytic solution.}
    \label{fig:M_E_2x2x2}
\end{figure*}

\begin{figure*}
    \centerline{   
    \includegraphics[width=0.49\textwidth]{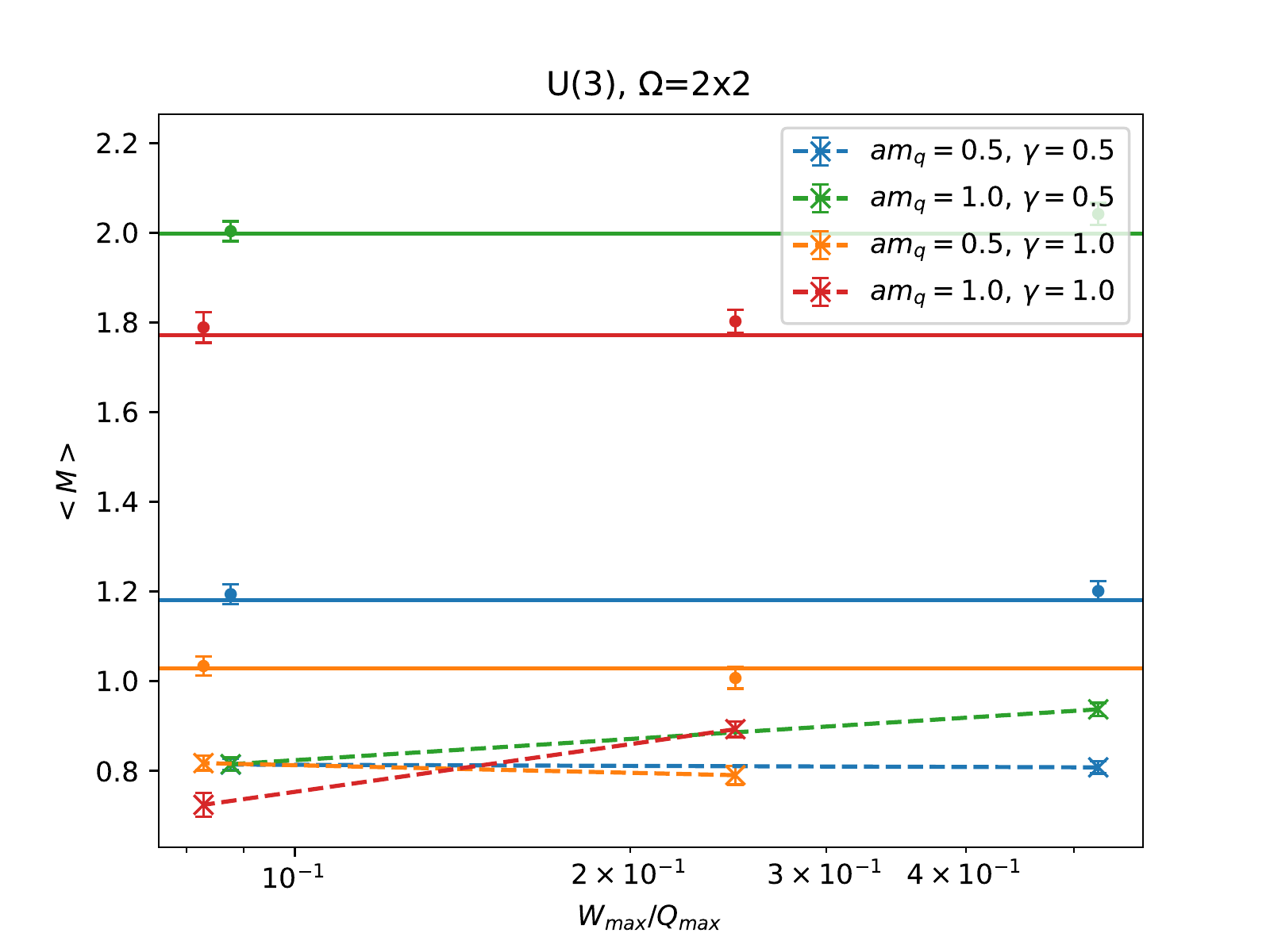}
    \includegraphics[width=0.49\textwidth]{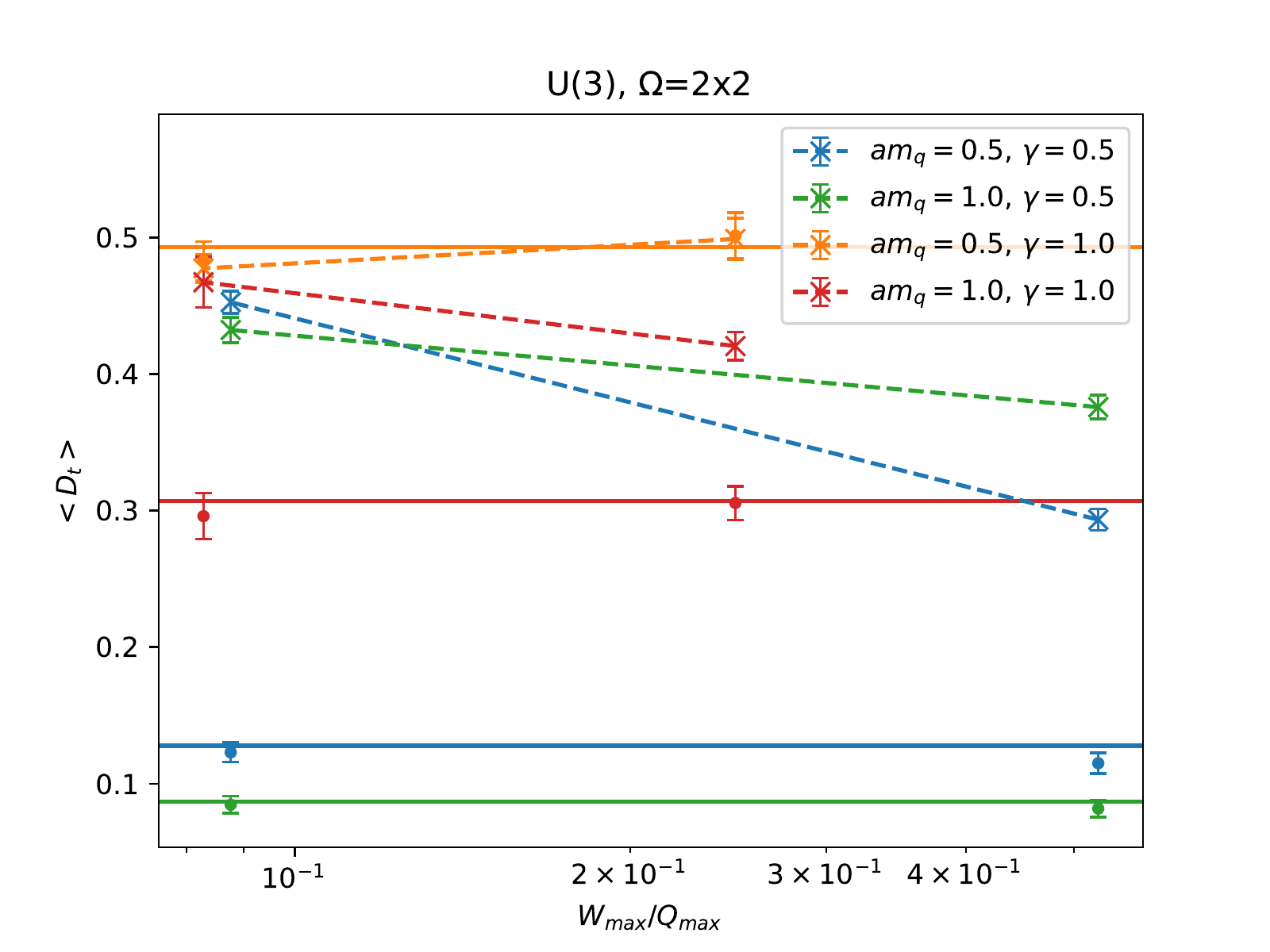}
    }
    \caption{$W_{max}/Q_{max}$ dependence of monomer density and temporal dimer density. The cross points are the raw data of D-wave and circle points are reweighted data. The raw data are connected by dotted line to guide the eyes and the solid line is the analytic solutions for $U(3)$ on $2 \times 2$. We have about $38-86\%$ of valid configurations of 695 total, so the reweighting method guides the observables toward the analytic solution.}
    \label{fig:U3_M_E_2x2}
\end{figure*}

In Fig.~\ref{fig:M_E_2x2},~\ref{fig:M_E_4x4},~\ref{fig:M_E_2x2x2} and ~\ref{fig:U3_M_E_2x2}, we present two independent observables \eref{eq:obs} $\langle M \rangle$ and $\langle D_t \rangle$ for several $p$ values with optimal \texttt{chain\_strength}. If $p$ is very small, the action in the QUBO matrix is emphasized. Hence, D-wave samples the distribution very near to the global minimum. However, D-wave does not find all 17 valid configurations and the reweighting method works poorly here. For large enough $p$, D-wave finds all 17 valid configurations and reweighting method adequately produces the  correct distribution. Since the $2 \times 2$ lattice is very small, finding all valid configuration is not difficult. For large volumes, the number of valid configurations is much larger and therefore finding all valid configurations with finite number of samplings is not possible. For example, the $4 \times 4$ volume has 41025 valid configurations, of which $O(700-1800)$ valid configurations ($1.7-4.3\%$) are produced during our D-Wave simulations thus far.  As such, it is not enough to evaluate the correct values of observables even with the reweighting method. This situation will improve, however, as we perform repeated runs of this system with D-Wave.  On the other hand, the $2 \times 2 \times 2$ reweighted results have better agreement with exact results.  Here the total number of valid configurations is 689 and $O(280-450)$ configurations are produced during each D-Wave simulation, representing between $40-65\%$ of the total. We stress, however, that in all cases considered so far, reweighted data gives better estimations, as shown in Fig.~\ref{fig:M_E_2x2x2}. In Fig.~\ref{fig:U3_M_E_2x2}, we show that our formalism for $N_c=3$ ~\ref{subsec:u3} works as well.

In Fig.~\ref{fig:M_E_4x4} and Fig.~\ref{fig:M_E_2x2x2} we show the dependence of our observables as a function of the ratio $W_{max}/Q_{max}$, where $W_{max}$ is the maximum absolute element of the weight matrix $W$ and $Q_{max}=2p$ is the maximum absolute element of $Q$ matrix. 
These plots show how $W$ is scaled by D-wave. We find that when $0.005 \lesssim W_{max}/Q_{max} \lesssim 0.01$, importance sampling works well.
For both $2 \times 2 \times 2$ and $4 \times 4$ volumes, where $0.005 \lesssim W_{max}/Q_{max} \lesssim 0.01$, the reweighted data points agree with the analytic solution. $W_{max}$ is 1.38,1.38,0.51,0.18 for $(am_q, \gamma)$ is $(0.3,0.5)$, $(0.6,0.5)$, $(0.3,1.0)$ and $(0.6,1.0)$. $W_{max}/Q_{max} \sim 0.01$ corresponds to $p=70,70,25,9$. In this manner we can find the proper $p$ value for any physical parameters. 

\begin{figure*}
    \centering
    \includegraphics[width=0.24\textwidth]{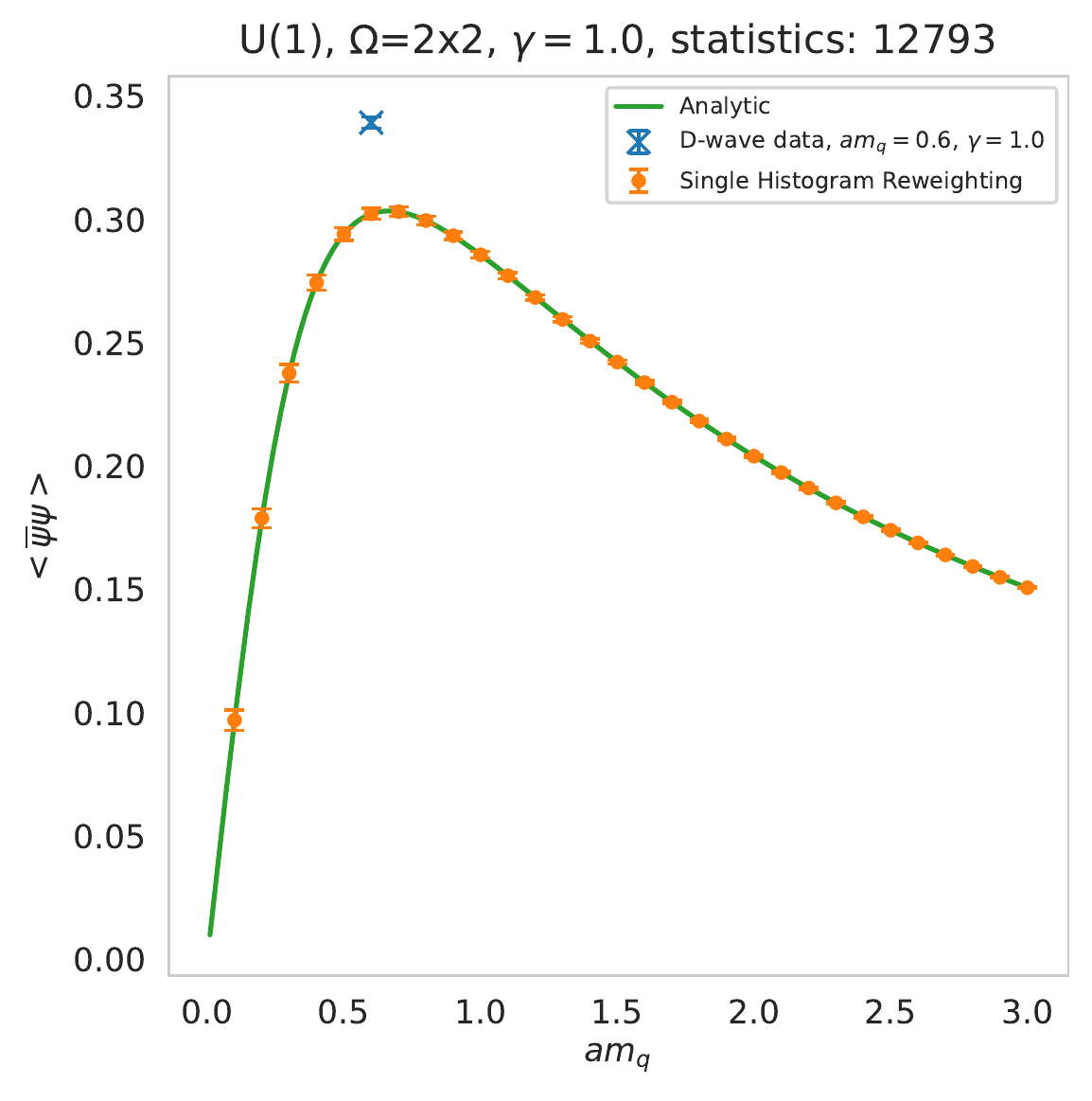}
    \includegraphics[width=0.24\textwidth]{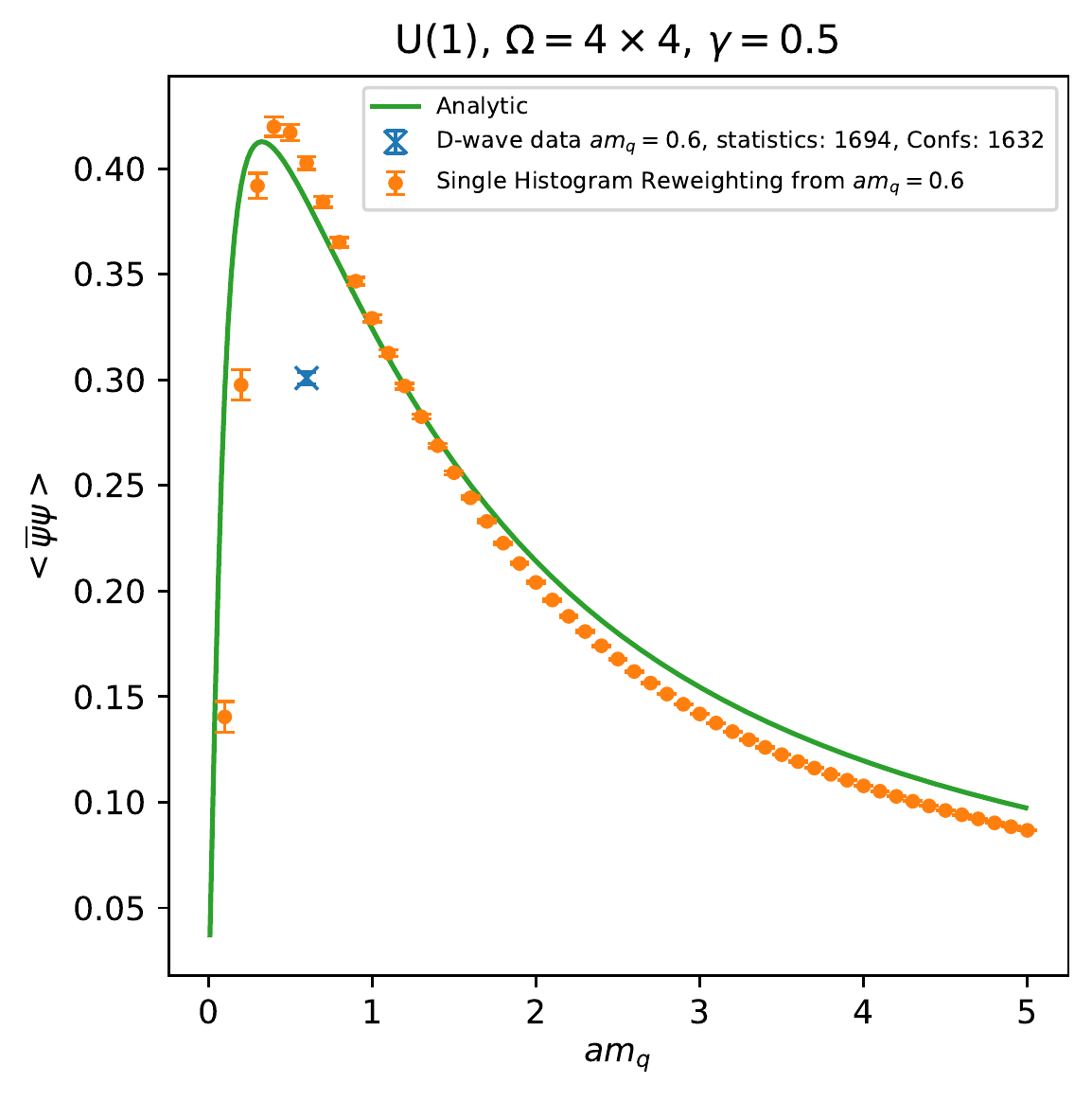}
    \includegraphics[width=0.24\textwidth]{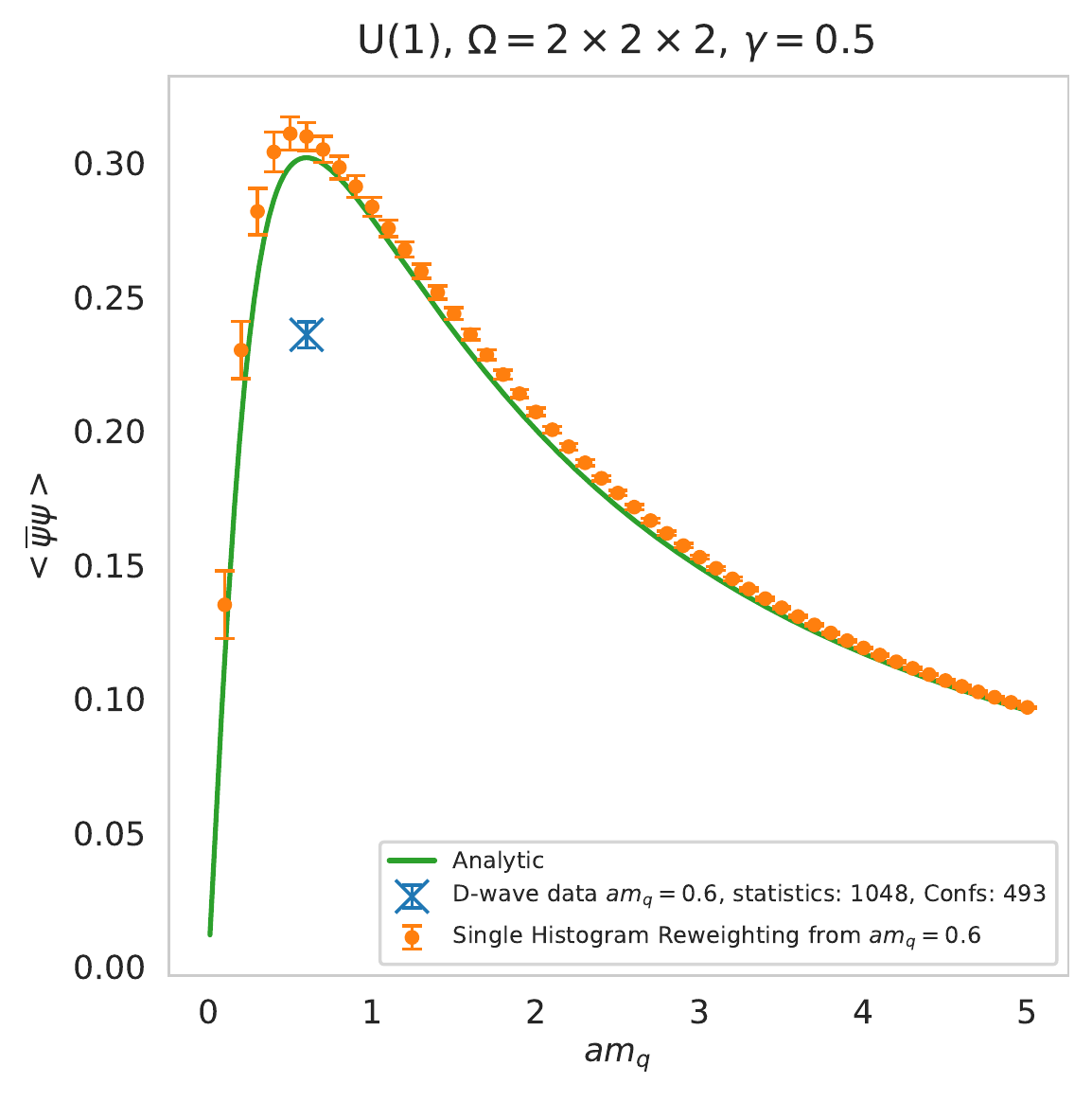}
    \includegraphics[width=0.24\textwidth]{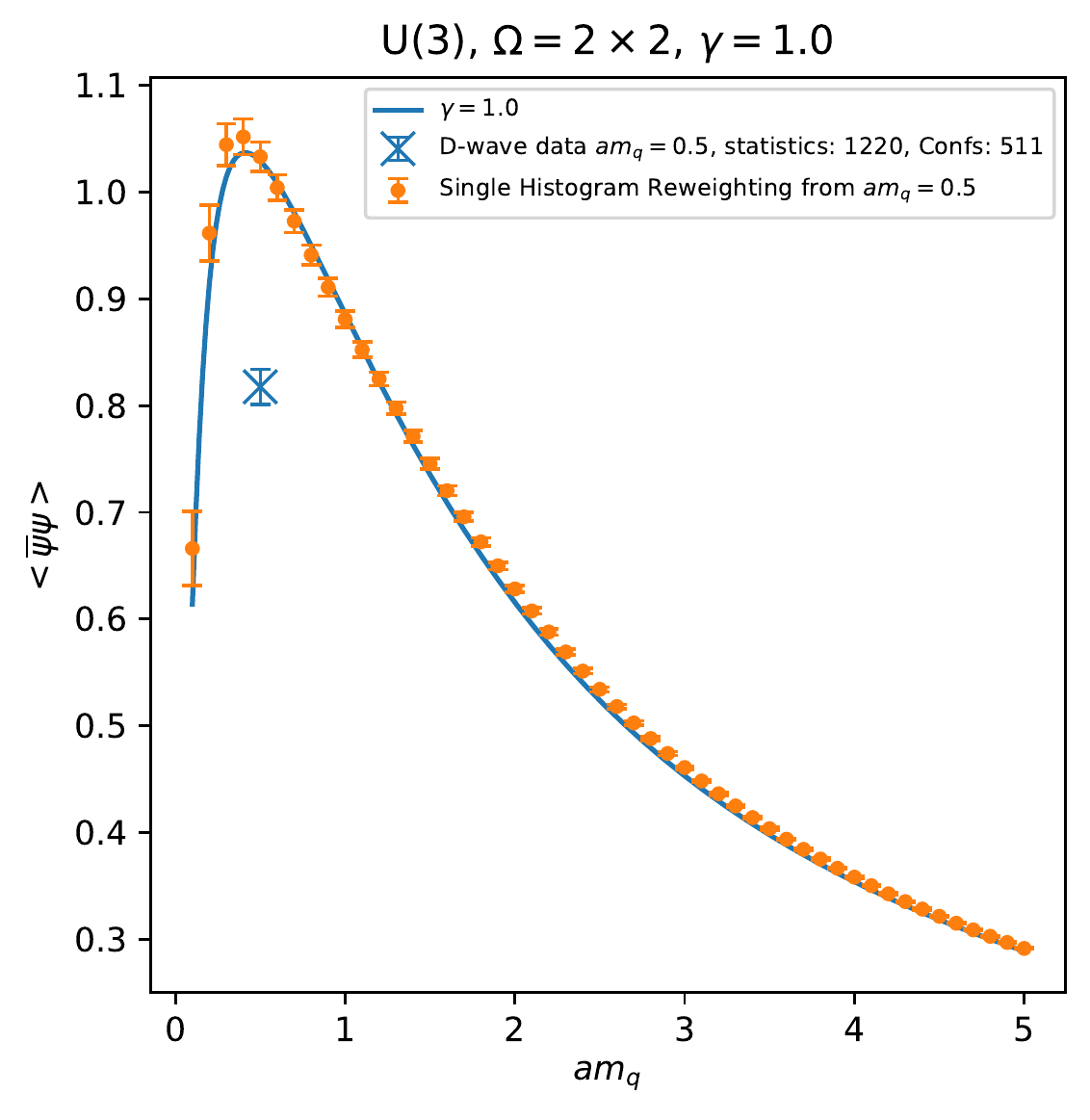}
    \caption{Chiral condensate by single histogram reweighting method from D-wave data points. Since we have all 17 valid configurations for the $2 \times 2$ lattice, the reweighting method can calculate this observable in any range of physical parameters. For $U(1)$ larger volumes, we present the reweighting results from physical parameters $am_q=0.6$ with fixed $\gamma=0.5$. For the $4 \times 4$ lattice, we have $\sim 4\%$ of the total number of configurations, and the 16 monomer configuration is missing. Therefore the reweighting method does not describe well the large quark mass region. We have $\sim 72\%$ of the configurations in the $U(1)$ $2 \times 2 \times 2$ and $38-86\%$ for $U(3)$ $2 \times 2$ lattices, so the reweighting works for a much longer range. The errorbars are purely statistical, so it does not explain the systematic uncertainty coming from the overlap problem.}
    \label{fig:reweighting}
\end{figure*}
Originally the single histogram reweighting method was developed as a means to change the physical parameters of the system without the need to perform new simulations.  %to change the physical parameters. 
Thus we also apply the reweighting method to change the quark mass and present these results in Fig.~\ref{fig:reweighting}.
For the $2 \times 2$ lattice, because D-wave finds all 17 valid configurations, the exact values of observables can be calculated for any choice of physical parameters by simply reweighting. 
For $4 \times 4$ and $2 \times 2 \times 2$ lattices, we fix $\gamma=0.5$ and quark masses $am_q=0.6$ for reweighted data point (orange dots). The blue cross points are D-wave raw data. We generate a very limited number of samples for the $4 \times 4$ system, and so the reweighting method works only for a range of the quark mass which still have overlap with the distribution that has been generated on D-Wave. In particular, the configuration with the maximal number of monomer $M=16$ is not sampled.  This further enhances the mismatch with the analytic solutions in the large quark mass region. The $2 \times 2 \times 2$ system has a larger percentage of sampled valid configurations compared to the $4 \times 4$ case, and therefore the reweighted data agree with the analytic solutions for a much broader range of quark masses. We also confirm that our $U(3)$ formulation works as well but the validity rate of this case is about $5\%$. 

\section{Conclusion}
\label{Conclusion}
\label{Outlook}
In this paper, we have demonstrated that lattice gauge theory in the strong coupling limit, and for now restricting to gauge group U(1) as a proof of principle, can be successfully simulated by the D-Wave quantum annealer. In particular, we have demonstrated that importance sampling is feasible on the D-wave, and with quantum annealing in general. With some fine tuning of the D-Wave simulation parameters and using the raw data distribution from D-wave, we find that the distribution of valid configurations is roughly produced according to the weight matrix, which in turn is dictated by the physical parameters of the system. The accuracy is greatly enhanced by the histogram reweighting method. In that case, the tuning of D-Wave parameters is less crucial.  We have generated and compared results for the lattice volumes $2\times2$, $4\times 4$ and $2\times 2 \times 2$ for $U(1)$ and $U(3)$ on $2 \times 2$.

In order to deal with a more realistic, and thus physical problem, we need to address larger volumes. To do this we propose an iterative scheme by decomposing local updates on even and odd sites. We expect that the solution vectors will converge to the equilibrium distributions after some number of iterations, and this can be thought of as a thermalization time.
Such a procedure would involve coupling the D-Wave system with classical computing, providing a means for an hybrid classical/quantum computing approach for simulation.
We are currently preparing this strategy to be tested on a $8 \times 8$ lattice.

We optimize the annealing schedule (bottom panel in Fig.~\ref{fig:anneal}) and use this annealing schedule for our $\U(3)$ runs. For the other runs we used the annealing schedule shown in the top panel in Fig.~\ref{fig:anneal}.  Here we generated 500 samples within 600 msec, corresponding to 1.2 msec per sample. For the optimized annealing schedule, we generated 1500 samples within 1 second, corresponding to~0.6 msec per sample. 
%A further possible optimization is the modification of the annealing schedule. In this study we have fixed the annealing schedule, however, further customization may be advantageous for the validity rate. For our annealing schedule described in \sref{subsec:DWAVE_Description}, we generated 500 samples within 600 msec, i.e.~1.2msec per one sample. 
By reducing the sampling time, we increased statistics while using the same compute resources.
%We test the shorter annealing schedule ~\ref{fig:anneal} which takes $5\mu s$ per sample. This allows to generate 1500 samples in $1$ sec with no difference from a longer annealing schedule.

From a computational aspect, we will also make comparisons for our lattice model concerning the performance on D-Wave, as compared to a classical Monte Carlo algorithms. Questions we want to address are: how fast D-wave equilibrate in comparison to the worm-algorithm, in particular on large volumes and low temperatures? How do the autocorrelation times compare?

It is then straightforward to go to higher dimensions, and the strategy can also be extended to the more interesting gauge group $\SU(\Nc)$, in particular strong coupling lattice QCD, where baryons are present and the baryon chemical potential enters as an additional physical parameter. In this paper, dealing with the sign problem was not included, but as 
we will extend our study to $\SU(\Nc)$ gauge theory, the sign problem due to non-vanishing baryon loops will be investigated. The corresponding QUBO matrix has already been worked out and will be presented in a forthcoming publication.

\begin{acknowledgments}
We would like to thank Christopher K\"ober and Arthur Witt for helpful discussions and helping the initial setting.
The authors gratefully acknowledge the J\"ulich Supercomputing Centre (https://www.fzjuelich.de/ias/jsc) for funding this project by providing computing time on the D-Wave Advantage™ System JUPSI through the J\"ulich UNified Infrastructure for Quantum computing (JUNIQ).
J.K. and T. L. were supported by the Deutsche Forschungsgemeinschaft (DFG, German Research Foundation) through the funds provided to the Sino-German Collaborative Research Center TRR110 "Symmetries and the Emergence of Structure in QCD" (DFG Project-ID 196253076 - TRR 110)
W.U. is supported by the Deutsche Forschungsgemeinschaft (DFG) through the CRC-TR 211 'Strong-interaction matter under extreme conditions'– project number 315477589 – TRR 211. 
\end{acknowledgments}

\section{Appendix}

\subsection{Exact enumeration}

We provide the results from exact enumeration on the lattices $2\times2$, $4\times4$ and  $2\times 2 \times 2$ for gauge group $\U(1)$ and $2\times 2$ for gauge group $\U(3)$, by listing the number of configurations $C$. 
Since for the U(1) gauge group, the combinatorical factor for every configuration has weight 1 (this is not the case for other gauge groups), the partition function in a finite volume is a polynomial: 
\begin{align}
Z&=\sum_M \sum_{D_t} w_{M,D_t}\gamma^{2D_t} \sum_M (2am_q)^M,
\end{align}
%\jangho{we need this formula for $U(3)$ and it need $z$ instead of $C$ in the table. @Wolfgang, could you modify the table and add equation?}
%\wolfgang{same formula, $w$ instead of $z$ is better, as these are weights}
where the coefficients $w_{M,D_t}$ are given by the sum of weights of configurations $z_{M,D_t}$ in each sector $(M,D_t)$. For gauge group $\U(1)$, $w_{M,D_t}=z_{M,D_t}$, for other gauge groups that is not the case. They are given for the various lattice volumes in the Tables \ref{tab:Enum2}, \ref{tab:Enum4} \ref{tab:Enum222} and \ref{tab:Enum_U3_2x2}. The number of valid configurations including other volumes and models, obtained from exact enumeration, is given in Table \ref{tab:N_valid}.\\

\begin{table}[ptb]
\begin{tabular}{rrrr}
\hline
& $M$ & $D_t$ & $z=w$\\ 
\hline
& 0 & 0 & 4\\ 
& 0 & 2 & 4\\
& 2 & 0 & 4\\ 
& 2 & 1 & 4\\ 
& 4 & 0 & 1\\ 
\hline
$\Sigma$ & & & 17\\
\hline
\end{tabular}
\caption{Contributions to the partition function of $\U(1)$ on a periodic $2\times2$ lattice.}
\label{tab:Enum2}
\end{table}   

\begin{table}[ptb]
\begin{tabular}{rrrr}
\hline
& $M$ & $D_t$ & $z=w$\\ 
\hline
&0 & 0 &  16\\ 
&0 & 2 &  64\\ 
&0 & 4 & 112\\ 
&0 & 6 &  64\\ 
&0 & 8 &  16\\ 
&2 & 0 & 128\\
&2 & 1 & 256\\
&2 & 2 & 640\\
&2 & 3 & 832\\
&2 & 4 & 832\\
&2 & 5 & 640\\
&2 & 6 & 256\\
&2 & 7 & 128\\
&4 & 0 & 416\\
&4 & 1 &1280\\
&4 & 2 &2592\\
&4 & 3 &3072\\
&4 & 4 &2592\\
&4 & 5 &1280\\
&4 & 6 & 416\\
&6 & 0 & 704\\
&6 & 1 &2368\\
&6 & 2 &4032\\
&6 & 3 &4032\\
&6 & 4 &2368\\
&6 & 5 & 704\\
&8 & 0 & 664\\
&8 & 1 &2048\\
&8 & 2 &2832\\
&8 & 3 &2048\\
&8 & 4 & 664\\
&10 & 0&  352\\
&10 & 1&  896\\
&10 & 2&  896\\
&10 & 3&  352\\
&12 & 0&  104\\
&12 & 1&  192\\
&12 & 2&  104\\
&14 & 0&   16\\
&14 & 1&   16\\
&16 & 0&    1\\
\hline
$\Sigma$ & & & 41025\\
\hline
\end{tabular}
\caption{Contributions to the partition function of $\U(1)$ on a periodic $4\times4$ lattice.}
\label{tab:Enum4}
\end{table}   

\begin{table}[ptb]
\begin{tabular}{rrrr}
\hline
& $M$ & $D_t$ & $z=w$\\ 
\hline
& 0 & 0 & 64\\ 
& 0 & 2 & 64\\ 
& 0 & 4 & 16\\ 
& 2 & 0 & 128\\
& 2 & 1 & 128\\ 
& 2 & 2 & 64\\
& 2 & 3 & 32\\ 
& 4 & 0 & 80\\
& 4 & 1 & 64\\
& 4 & 2 & 24\\ 
& 6 & 0 & 16\\ 
& 6 & 1 & 8\\ 
& 8 & 0 & 1\\ 
\hline
$\Sigma$ & & & 689\\
\hline
\end{tabular} 
\caption{Contributions to the partition function of $\U(1)$ on a periodic $2\times2 \times 2$ lattice.}
\label{tab:Enum222}
\end{table}  

\begin{table}[ptb]
\begin{tabular}{rrrrr}
\hline
& $M$ & $D_t$  & $z$&  $w$\\ 
\hline
&0 &0 & 16 & 16  \\
&0 &2 & 36 & 44.44\\
&0 &4 & 36 & 44.44\\
&0 &6& 16 & 16\\
&2 &0 & 24 & 80\\
&2 &1 & 36 & 133.33\\
&2 &2 & 48 & 213.33\\
&2 &3 & 48 & 213.33\\
&2 &4 & 36 & 133.33\\
&2 &5& 24 & 80\\
&4 &0 & 25 & 148\\
&4 &1 & 48 & 320\\
&4 &2 & 64 & 416\\
&4 &3 & 48 & 320\\
&4 &4 & 25 & 148\\
&6 &0 & 20 & 128\\
&6 &1 & 40 & 272\\
&6 &2 & 40 & 272\\
&6 &3 & 20 & 128\\
&8 &0 & 10 & 56\\
&8 &1 & 16 & 96\\
&8 &2 & 10 & 56\\
&10 &0 & 4 & 12\\
&10 &1 & 4 & 12\\
&12 &0 & 1 & 1\\
\hline
$\Sigma$ & & & 695\\
\hline
\end{tabular} 
\caption{Contributions to the partition function of $\U(3)$ on a periodic $2\times2$ lattice.}
\label{tab:Enum_U3_2x2}
\end{table}  

\begin{table}[ptb]
\begin{tabular}{rrr}
\hline
gauge group & lattice & configurations \\
\hline
U(1) & $2\times 2$ & 17  \\ 
U(1) & $4\times 4$ & 41025  \\
U(1) & $6\times 6$ & 23079663560  \\ 
U(1) & $2\times 2\times 2$ & 689  \\
U(1) & $2\times 2\times 2\times 2$ & 1898625  \\
\hline
U(2) & $2\times 2$ & 135  \\ 
\hline
U(3) & $2\times 2$ & 695 \\ 
\hline 
\end{tabular}
\caption{Table of total number of configurations for various gauge groups and periodic lattices.\label{tab:N_valid}}
\end{table}

\subsection{Statistics of solution vectors}
We provide the information of the runs with the optimal \texttt{chain\_strength}, validity rate and the number of independent configurations for each physical parameters $\gamma$, $am_q$ and penalty term $p$ for three volumes in the Tables~\ref{tab:runs_enum_2x2},\ref{tab:runs_enum_4x4}, \ref{tab:runs_enum_2x2x2} and \ref{tab:runs_enum_U3_2x2}.

\begin{table*}[bt]
\begin{tabular}{cc|cccccccccc}
\hline
$\U(1)$ &$2 \times 2$ & \multicolumn{5}{c|}{$am_q=0.3$, $\gamma=0.5$} & \multicolumn{5}{c}{$am_q=0.6$, $\gamma=0.5$} \\   
\hline
\multicolumn{2}{c|}{$p$}                              & 1    & 2    & 3    & 4    & 5    & 1    & 2    & 3    & 4    & 5    \\
\multicolumn{2}{c|}{optimal \texttt{chain\_strength}} & 1    & 2    & 3.5  & 4.5  & 6.5  & 1    & 2    & 3    & 5    & 5.5  \\
\multicolumn{2}{c|}{validity rate}                    & 0.90 & 0.91 & 0.95 & 0.93 & 0.93 & 0.91 & 0.89 & 0.88 & 0.91 & 0.89 \\
\multicolumn{2}{c|}{independent configurations}        & 8    & 12   & 17   & 17   & 17   & 10   & 13   & 14   & 17   & 17   \\
\multicolumn{2}{c|}{runs}                             & 1    &    2 &    2 &    2 &    2 &    1 &    2 &    2  &    2 &  2   \\
\hline
\hline
 & & \multicolumn{5}{c|}{$am_q=0.3$, $\gamma=1$} & \multicolumn{5}{c}{$am_q=0.6$, $\gamma=1$} \\ 
\hline
\multicolumn{2}{c|}{$p$}                              & 1    & 2    & 3    &   4  &  5   &   1  &    2 &    3 &    4 &    5 \\
\multicolumn{2}{c|}{optimal \texttt{chain\_strength}} & 1    & 2    & 3    & 4.5  &  5   &    1 &    2 &    3 &  4.5 & 5    \\
\multicolumn{2}{c|}{validity rate}                    & 0.92 & 0.92 & 0.91 & 0.93 & 0.92 & 0.92 & 0.92 & 0.91 & 0.94 & 0.91 \\
\multicolumn{2}{c|}{independent configurations}        & 16   & 17   & 17   & 17   & 17   & 17   & 17   & 17   & 17   & 17   \\
\multicolumn{2}{c|}{runs}                             & 8    & 8    & 28   & 1    &  8   &   8  &    8 &  28  &    1 &    8 \\
\hline
\end{tabular} \\
\caption{Table of total number of runs on $2 \times 2$. Each run generates 500 solutions. \label{tab:runs_enum_2x2}}
\end{table*}

\begin{table*}[bt]
\begin{tabular}{cc|ccccccc}
\hline
$\U(1)$ &$4 \times 4$ & \multicolumn{4}{c|}{$am_q=0.3$, $\gamma=0.5$} & \multicolumn{3}{c}{$am_q=0.6$, $\gamma=0.5$} \\   
\hline
\multicolumn{2}{c|}{$p$}                              & 3    & 16    & 70    & 120   & 60    & 70     & 120   \\
\multicolumn{2}{c|}{optimal \texttt{chain\_strength}} & 5    & 24    & 100   & 170   & 80    & 100    & 175   \\
\multicolumn{2}{c|}{validity rate}                    & 0.31 & 0.29  & 0.36  & 0.39  & 0.40  & 0.38   & 0.37  \\
\multicolumn{2}{c|}{independent configurations}       & 1230 & 712   & 1564  & 1313  & 1777  & 1632   & 1593  \\
\multicolumn{2}{c|}{runs}                             & 8    & 4     & 8     & 6     & 8     & 8      & 8     \\
\hline
\hline
 & & \multicolumn{3}{c|}{$am_q=0.3$, $\gamma=1$} & \multicolumn{3}{c}{$am_q=0.6$, $\gamma=1$} \\ 
\hline
\multicolumn{2}{c|}{$p$}                              & 12    &    16   &   45   &   2    &   12   &  16    \\
\multicolumn{2}{c|}{optimal \texttt{chain\_strength}} & 17    &    24   &   65   &   2.5  &   16   &  21.5  \\
\multicolumn{2}{c|}{validity rate}                    & 0.37  &    0.37 &   0.37 &   0.24 &   0.38 &  0.39  \\
\multicolumn{2}{c|}{independent configurations}       & 1635  & 1633    &   1264 &  1059  & 1679   & 1685   \\
\multicolumn{2}{c|}{runs}                             & 8     &    8    &   6    &   8    &   8    &  8     \\
\hline
\end{tabular} \\
\caption{Table of total number of runs on $4 \times 4$. Each run generates 500 solutions.\label{tab:runs_enum_4x4}}
\end{table*}

\begin{table*}[bt]
\begin{tabular}{cc|cccccccc}
\hline
$\U(1)$ &$2 \times 2 \times 2$                         & \multicolumn{3}{c|}{$am_q=0.3$, $\gamma=0.5$} & \multicolumn{3}{c}{$am_q=0.6$, $\gamma=0.5$} \\   
\hline
\multicolumn{2}{c|}{$p$}                               & 2    & 5    & 70    & 2    & 5    & 70     \\
\multicolumn{2}{c|}{optimal \texttt{chain\_strength}}  & 3.5  & 9    & 120   & 3.5  & 9.5  & 120    \\
\multicolumn{2}{c|}{validity rate}                     & 0.41 & 0.34 & 0.33  & 0.38 & 0.34 & 0.31   \\
\multicolumn{2}{c|}{independent configurations}        & 280  & 408  & 398   & 343  & 432  & 409    \\
\multicolumn{2}{c|}{runs}                              & 4    &    4 &    4  &    4 &    4 &    4   \\
\hline
\hline
 & & \multicolumn{4}{c|}{$am_q=0.3$, $\gamma=1$} & \multicolumn{4}{c}{$am_q=0.6$, $\gamma=1$} \\ 
\hline
\multicolumn{2}{c|}{$p$}                               & 2    & 5    & 26   &  70   &   2     &  5     &   9    &  70   \\
\multicolumn{2}{c|}{optimal \texttt{chain\_strength}}  & 4    & 9.5  & 45   &  140  &   3.5   &  9     &   16   &  130  \\
\multicolumn{2}{c|}{validity rate}                     & 0.26 & 0.29 & 0.30 & 0.28  &   0.30  &  0.27  &   0.23 &  0.25 \\
\multicolumn{2}{c|}{independent configurations}        & 317  & 386  & 443   & 396  &   410   & 399    &  361   &  373  \\
\multicolumn{2}{c|}{runs}                              & 4    & 4    & 4    & 4     &   4     &  4     &   4    &  4    \\
\hline
\end{tabular} \\
\caption{Table of total number of runs on $2 \times 2 \times 2$. Each run generates 500 solutions.\label{tab:runs_enum_2x2x2}}
\end{table*}

\begin{table*}[bt]
\begin{tabular}{cc|cccc}
\hline
$\U(3)$ &$2 \times 2 $                         & \multicolumn{2}{c|}{$am_q=0.5$, $\gamma=0.5$} & \multicolumn{2}{c}{$am_q=1.0$, $\gamma=0.5$} \\   
\hline
\multicolumn{2}{c|}{$p$}                               & 5     & 30    &   5     & 30     \\
\multicolumn{2}{c|}{optimal \texttt{chain\_strength}}  & 30    & 195   &   30    & 195    \\
\multicolumn{2}{c|}{validity rate}                     & 0.053 & 0.055 &   0.048 & 0.038  \\
\multicolumn{2}{c|}{independent configurations}        & 522   & 599   &   500   & 540    \\
\multicolumn{2}{c|}{runs}                              & 26    & 27    &   26    & 28     \\
\hline
\hline
 & & \multicolumn{2}{c|}{$am_q=0.5$, $\gamma=1$} & \multicolumn{2}{c}{$am_q=1$, $\gamma=1$} \\ 
\hline
\multicolumn{2}{c|}{$p$}                               & 5    & 15    & 5   &  15      \\
\multicolumn{2}{c|}{optimal \texttt{chain\_strength}}  & 33    & 100  & 30   &  100    \\
\multicolumn{2}{c|}{validity rate}                     & 0.054 & 0.051 & 0.045 & 0.050 \\
\multicolumn{2}{c|}{independent configurations}        & 383  & 511  & 457   & 263     \\
\multicolumn{2}{c|}{runs}                              & 9    & 17    & 18    & 6      \\
\hline
\end{tabular} \\
\caption{Table of total number of runs for $U(3)$ on $2 \times 2$. Each run generates 1500 solutions.\label{tab:runs_enum_U3_2x2}}
\end{table*}

%\clearpage

\bibliography{references}

\begin{thebibliography}{32}%
\makeatletter
\providecommand \@ifxundefined [1]{%
 \@ifx{#1\undefined}
}%
\providecommand \@ifnum [1]{%
 \ifnum #1\expandafter \@firstoftwo
 \else \expandafter \@secondoftwo
 \fi
}%
\providecommand \@ifx [1]{%
 \ifx #1\expandafter \@firstoftwo
 \else \expandafter \@secondoftwo
 \fi
}%
\providecommand \natexlab [1]{#1}%
\providecommand \enquote  [1]{``#1''}%
\providecommand \bibnamefont  [1]{#1}%
\providecommand \bibfnamefont [1]{#1}%
\providecommand \citenamefont [1]{#1}%
\providecommand \href@noop [0]{\@secondoftwo}%
\providecommand \href [0]{\begingroup \@sanitize@url \@href}%
\providecommand \@href[1]{\@@startlink{#1}\@@href}%
\providecommand \@@href[1]{\endgroup#1\@@endlink}%
\providecommand \@sanitize@url [0]{\catcode `\\12\catcode `\$12\catcode
  `\&12\catcode `\#12\catcode `\^12\catcode `\_12\catcode `\%12\relax}%
\providecommand \@@startlink[1]{}%
\providecommand \@@endlink[0]{}%
\providecommand \url  [0]{\begingroup\@sanitize@url \@url }%
\providecommand \@url [1]{\endgroup\@href {#1}{\urlprefix }}%
\providecommand \urlprefix  [0]{URL }%
\providecommand \Eprint [0]{\href }%
\providecommand \doibase [0]{https://doi.org/}%
\providecommand \selectlanguage [0]{\@gobble}%
\providecommand \bibinfo  [0]{\@secondoftwo}%
\providecommand \bibfield  [0]{\@secondoftwo}%
\providecommand \translation [1]{[#1]}%
\providecommand \BibitemOpen [0]{}%
\providecommand \bibitemStop [0]{}%
\providecommand \bibitemNoStop [0]{.\EOS\space}%
\providecommand \EOS [0]{\spacefactor3000\relax}%
\providecommand \BibitemShut  [1]{\csname bibitem#1\endcsname}%
\let\auto@bib@innerbib\@empty
%</preamble>
\bibitem [{\citenamefont {Fromm}\ \emph {et~al.}(2022)\citenamefont {Fromm},
  \citenamefont {Philipsen},\ and\ \citenamefont {Winterowd}}]{Fromm:2022vaj}%
  \BibitemOpen
  \bibfield  {author} {\bibinfo {author} {\bibfnamefont {M.}~\bibnamefont
  {Fromm}}, \bibinfo {author} {\bibfnamefont {O.}~\bibnamefont {Philipsen}},\
  and\ \bibinfo {author} {\bibfnamefont {C.}~\bibnamefont {Winterowd}},\
  }\bibfield  {title} {\bibinfo {title} {{Dihedral Lattice Gauge Theories on a
  Quantum Annealer}},\ }\href@noop {} {\  (\bibinfo {year} {2022})},\ \Eprint
  {https://arxiv.org/abs/2206.14679} {arXiv:2206.14679 [hep-lat]} \BibitemShut
  {NoStop}%
\bibitem [{\citenamefont {A~Rahman}\ \emph {et~al.}(2021)\citenamefont
  {A~Rahman}, \citenamefont {Lewis}, \citenamefont {Mendicelli},\ and\
  \citenamefont {Powell}}]{ARahman:2021ktn}%
  \BibitemOpen
  \bibfield  {author} {\bibinfo {author} {\bibfnamefont {S.}~\bibnamefont
  {A~Rahman}}, \bibinfo {author} {\bibfnamefont {R.}~\bibnamefont {Lewis}},
  \bibinfo {author} {\bibfnamefont {E.}~\bibnamefont {Mendicelli}},\ and\
  \bibinfo {author} {\bibfnamefont {S.}~\bibnamefont {Powell}},\ }\bibfield
  {title} {\bibinfo {title} {{SU(2) lattice gauge theory on a quantum
  annealer}},\ }\href {https://doi.org/10.1103/PhysRevD.104.034501} {\bibfield
  {journal} {\bibinfo  {journal} {Phys. Rev. D}\ }\textbf {\bibinfo {volume}
  {104}},\ \bibinfo {pages} {034501} (\bibinfo {year} {2021})},\ \Eprint
  {https://arxiv.org/abs/2103.08661} {arXiv:2103.08661 [hep-lat]} \BibitemShut
  {NoStop}%
\bibitem [{\citenamefont {Philipsen}(2010)}]{Philipsen2010}%
  \BibitemOpen
  \bibfield  {author} {\bibinfo {author} {\bibfnamefont {O.}~\bibnamefont
  {Philipsen}},\ }\bibfield  {title} {\bibinfo {title} {{Lattice QCD at
  non-zero temperature and baryon density}},\ }in\ \href@noop {} {\emph
  {\bibinfo {booktitle} {{Les Houches Summer School: Session 93: Modern
  perspectives in lattice QCD: Quantum field theory and high performance
  computing}}}}\ (\bibinfo {year} {2010})\ \Eprint
  {https://arxiv.org/abs/1009.4089} {arXiv:1009.4089 [hep-lat]} \BibitemShut
  {NoStop}%
\bibitem [{\citenamefont {de~Forcrand}(2009)}]{deForcrand2010}%
  \BibitemOpen
  \bibfield  {author} {\bibinfo {author} {\bibfnamefont {P.}~\bibnamefont
  {de~Forcrand}},\ }\bibfield  {title} {\bibinfo {title} {{Simulating QCD at
  finite density}},\ }\bibfield  {booktitle} {\emph {\bibinfo {booktitle}
  {{Proceedings, 27th International Symposium on Lattice field theory (Lattice
  2009): Beijing, P.R. China, July 26-31, 2009}}},\ }\href
  {https://doi.org/10.22323/1.091.0010} {\bibfield  {journal} {\bibinfo
  {journal} {PoS}\ }\textbf {\bibinfo {volume} {LAT2009}},\ \bibinfo {pages}
  {010} (\bibinfo {year} {2009})},\ \Eprint {https://arxiv.org/abs/1005.0539}
  {arXiv:1005.0539 [hep-lat]} \BibitemShut {NoStop}%
%%CITATION = ARXIV:1005.0539;%%
\bibitem [{\citenamefont {Gattringer}\ and\ \citenamefont
  {Langfeld}(2016)}]{Gattringer2016b}%
  \BibitemOpen
  \bibfield  {author} {\bibinfo {author} {\bibfnamefont {C.}~\bibnamefont
  {Gattringer}}\ and\ \bibinfo {author} {\bibfnamefont {K.}~\bibnamefont
  {Langfeld}},\ }\bibfield  {title} {\bibinfo {title} {{Approaches to the sign
  problem in lattice field theory}},\ }\href
  {https://doi.org/10.1142/S0217751X16430077} {\bibfield  {journal} {\bibinfo
  {journal} {Int. J. Mod. Phys.}\ }\textbf {\bibinfo {volume} {A31}},\ \bibinfo
  {pages} {1643007} (\bibinfo {year} {2016})},\ \Eprint
  {https://arxiv.org/abs/1603.09517} {arXiv:1603.09517 [hep-lat]} \BibitemShut
  {NoStop}%
%%CITATION = ARXIV:1603.09517;%%
\bibitem [{\citenamefont {Kluberg-Stern}\ \emph {et~al.}(1983)\citenamefont
  {Kluberg-Stern}, \citenamefont {Morel},\ and\ \citenamefont
  {Petersson}}]{KlubergStern1983}%
  \BibitemOpen
  \bibfield  {author} {\bibinfo {author} {\bibfnamefont {H.}~\bibnamefont
  {Kluberg-Stern}}, \bibinfo {author} {\bibfnamefont {A.}~\bibnamefont
  {Morel}},\ and\ \bibinfo {author} {\bibfnamefont {B.}~\bibnamefont
  {Petersson}},\ }\bibfield  {title} {\bibinfo {title} {{Spectrum of Lattice
  Gauge Theories with Fermions from a 1/D Expansion at Strong Coupling}},\
  }\href {https://doi.org/10.1016/0550-3213(83)90259-6} {\bibfield  {journal}
  {\bibinfo  {journal} {Nucl. Phys.}\ }\textbf {\bibinfo {volume} {B215}},\
  \bibinfo {pages} {527} (\bibinfo {year} {1983})}\BibitemShut {NoStop}%
%%CITATION = NUPHA,B215,527;%%
\bibitem [{\citenamefont {Faldt}\ and\ \citenamefont
  {Petersson}(1986)}]{Faldt:1985ec}%
  \BibitemOpen
  \bibfield  {author} {\bibinfo {author} {\bibfnamefont {G.}~\bibnamefont
  {Faldt}}\ and\ \bibinfo {author} {\bibfnamefont {B.}~\bibnamefont
  {Petersson}},\ }\bibfield  {title} {\bibinfo {title} {{Strong Coupling
  Expansion of Lattice Gauge Theories at Finite Temperature}},\ }\href
  {https://doi.org/10.1016/0550-3213(86)90414-1} {\bibfield  {journal}
  {\bibinfo  {journal} {Nucl. Phys.}\ }\textbf {\bibinfo {volume} {B265}},\
  \bibinfo {pages} {197} (\bibinfo {year} {1986})}\BibitemShut {NoStop}%
%%CITATION = NUPHA,B265,197;%%
\bibitem [{\citenamefont {Bilic}\ \emph {et~al.}(1992)\citenamefont {Bilic},
  \citenamefont {Karsch},\ and\ \citenamefont {Redlich}}]{Bilic1992a}%
  \BibitemOpen
  \bibfield  {author} {\bibinfo {author} {\bibfnamefont {N.}~\bibnamefont
  {Bilic}}, \bibinfo {author} {\bibfnamefont {F.}~\bibnamefont {Karsch}},\ and\
  \bibinfo {author} {\bibfnamefont {K.}~\bibnamefont {Redlich}},\ }\bibfield
  {title} {\bibinfo {title} {{Flavor dependence of the chiral phase transition
  in strong coupling QCD}},\ }\href {https://doi.org/10.1103/PhysRevD.45.3228}
  {\bibfield  {journal} {\bibinfo  {journal} {Phys. Rev.}\ }\textbf {\bibinfo
  {volume} {D45}},\ \bibinfo {pages} {3228} (\bibinfo {year}
  {1992})}\BibitemShut {NoStop}%
%%CITATION = PHRVA,D45,3228;%%
\bibitem [{\citenamefont {Nishida}(2004)}]{Nishida:2003fb}%
  \BibitemOpen
  \bibfield  {author} {\bibinfo {author} {\bibfnamefont {Y.}~\bibnamefont
  {Nishida}},\ }\bibfield  {title} {\bibinfo {title} {{Phase structures of
  strong coupling lattice QCD with finite baryon and isospin density}},\ }\href
  {https://doi.org/10.1103/PhysRevD.69.094501} {\bibfield  {journal} {\bibinfo
  {journal} {Phys. Rev.}\ }\textbf {\bibinfo {volume} {D69}},\ \bibinfo {pages}
  {094501} (\bibinfo {year} {2004})},\ \Eprint
  {https://arxiv.org/abs/hep-ph/0312371} {arXiv:hep-ph/0312371 [hep-ph]}
  \BibitemShut {NoStop}%
%%CITATION = HEP-PH/0312371;%%
\bibitem [{\citenamefont {Miura}\ \emph {et~al.}(2017)\citenamefont {Miura},
  \citenamefont {Kawamoto}, \citenamefont {Nakano},\ and\ \citenamefont
  {Ohnishi}}]{Miura2016}%
  \BibitemOpen
  \bibfield  {author} {\bibinfo {author} {\bibfnamefont {K.}~\bibnamefont
  {Miura}}, \bibinfo {author} {\bibfnamefont {N.}~\bibnamefont {Kawamoto}},
  \bibinfo {author} {\bibfnamefont {T.~Z.}\ \bibnamefont {Nakano}},\ and\
  \bibinfo {author} {\bibfnamefont {A.}~\bibnamefont {Ohnishi}},\ }\bibfield
  {title} {\bibinfo {title} {{Polyakov loop effects on the phase diagram in
  strong-coupling lattice QCD}},\ }\href
  {https://doi.org/10.1103/PhysRevD.95.114505} {\bibfield  {journal} {\bibinfo
  {journal} {Phys. Rev.}\ }\textbf {\bibinfo {volume} {D95}},\ \bibinfo {pages}
  {114505} (\bibinfo {year} {2017})},\ \Eprint
  {https://arxiv.org/abs/1610.09288} {arXiv:1610.09288 [hep-lat]} \BibitemShut
  {NoStop}%
%%CITATION = ARXIV:1610.09288;%%
\bibitem [{\citenamefont {Karsch}\ and\ \citenamefont
  {Mutter}(1989)}]{Karsch:1988zx}%
  \BibitemOpen
  \bibfield  {author} {\bibinfo {author} {\bibfnamefont {F.}~\bibnamefont
  {Karsch}}\ and\ \bibinfo {author} {\bibfnamefont {K.~H.}\ \bibnamefont
  {Mutter}},\ }\bibfield  {title} {\bibinfo {title} {{STRONG COUPLING QCD AT
  FINITE BARYON NUMBER DENSITY}},\ }\href
  {https://doi.org/10.1016/0550-3213(89)90396-9} {\bibfield  {journal}
  {\bibinfo  {journal} {Nucl. Phys.}\ }\textbf {\bibinfo {volume} {B313}},\
  \bibinfo {pages} {541} (\bibinfo {year} {1989})}\BibitemShut {NoStop}%
%%CITATION = NUPHA,B313,541;%%
\bibitem [{\citenamefont {Kawamoto}\ and\ \citenamefont
  {Smit}(1981)}]{Kawamoto1981}%
  \BibitemOpen
  \bibfield  {author} {\bibinfo {author} {\bibfnamefont {N.}~\bibnamefont
  {Kawamoto}}\ and\ \bibinfo {author} {\bibfnamefont {J.}~\bibnamefont
  {Smit}},\ }\bibfield  {title} {\bibinfo {title} {{Effective Lagrangian and
  Dynamical Symmetry Breaking in Strongly Coupled Lattice QCD}},\ }\href
  {https://doi.org/10.1016/0550-3213(81)90196-6} {\bibfield  {journal}
  {\bibinfo  {journal} {Nucl. Phys.}\ }\textbf {\bibinfo {volume} {B192}},\
  \bibinfo {pages} {100} (\bibinfo {year} {1981})},\ \bibinfo {note}
  {[,556(1981)]}\BibitemShut {NoStop}%
%%CITATION = NUPHA,B192,100;%%
\bibitem [{\citenamefont {Rossi}\ and\ \citenamefont
  {Wolff}(1984)}]{Rossi:1984cv}%
  \BibitemOpen
  \bibfield  {author} {\bibinfo {author} {\bibfnamefont {P.}~\bibnamefont
  {Rossi}}\ and\ \bibinfo {author} {\bibfnamefont {U.}~\bibnamefont {Wolff}},\
  }\bibfield  {title} {\bibinfo {title} {{Lattice {QCD} With Fermions at Strong
  Coupling: A Dimer System}},\ }\href
  {https://doi.org/10.1016/0550-3213(84)90589-3} {\bibfield  {journal}
  {\bibinfo  {journal} {Nucl. Phys.}\ }\textbf {\bibinfo {volume} {B248}},\
  \bibinfo {pages} {105} (\bibinfo {year} {1984})}\BibitemShut {NoStop}%
%%CITATION = NUPHA,B248,105;%%
\bibitem [{\citenamefont {Prokof'ev}\ and\ \citenamefont
  {Svistunov}(2001)}]{Prokofev2001}%
  \BibitemOpen
  \bibfield  {author} {\bibinfo {author} {\bibfnamefont {N.}~\bibnamefont
  {Prokof'ev}}\ and\ \bibinfo {author} {\bibfnamefont {B.}~\bibnamefont
  {Svistunov}},\ }\bibfield  {title} {\bibinfo {title} {{Worm Algorithms for
  Classical Statistical Models}},\ }\href
  {https://doi.org/10.1103/PhysRevLett.87.160601} {\bibfield  {journal}
  {\bibinfo  {journal} {Phys. Rev. Lett.}\ }\textbf {\bibinfo {volume} {87}},\
  \bibinfo {pages} {160601} (\bibinfo {year} {2001})},\ \Eprint
  {https://arxiv.org/abs/cond-mat/0103146} {arXiv:cond-mat/0103146 [cond-mat]}
  \BibitemShut {NoStop}%
%%CITATION = COND-MAT/0103146;%%
\bibitem [{\citenamefont {Adams}\ and\ \citenamefont
  {Chandrasekharan}(2003)}]{Adams:2003cca}%
  \BibitemOpen
  \bibfield  {author} {\bibinfo {author} {\bibfnamefont {D.~H.}\ \bibnamefont
  {Adams}}\ and\ \bibinfo {author} {\bibfnamefont {S.}~\bibnamefont
  {Chandrasekharan}},\ }\bibfield  {title} {\bibinfo {title} {{Chiral limit of
  strongly coupled lattice gauge theories}},\ }\href
  {https://doi.org/10.1016/S0550-3213(03)00350-X} {\bibfield  {journal}
  {\bibinfo  {journal} {Nucl. Phys.}\ }\textbf {\bibinfo {volume} {B662}},\
  \bibinfo {pages} {220} (\bibinfo {year} {2003})},\ \Eprint
  {https://arxiv.org/abs/hep-lat/0303003} {arXiv:hep-lat/0303003 [hep-lat]}
  \BibitemShut {NoStop}%
%%CITATION = HEP-LAT/0303003;%%
\bibitem [{\citenamefont {Fromm}(2010)}]{Fromm}%
  \BibitemOpen
  \bibfield  {author} {\bibinfo {author} {\bibfnamefont {M.}~\bibnamefont
  {Fromm}},\ }\bibfield  {title} {\bibinfo {title} {{Lattice QCD at string
  coupling: thermodynamics and nuclear physics}},\ }\href@noop {} {\bibfield
  {journal} {\bibinfo  {journal} {Thesis}\ } (\bibinfo {year}
  {2010})}\BibitemShut {NoStop}%
\bibitem [{\citenamefont {de~Forcrand}\ and\ \citenamefont
  {Fromm}(2010)}]{deForcrand:2009dh}%
  \BibitemOpen
  \bibfield  {author} {\bibinfo {author} {\bibfnamefont {P.}~\bibnamefont
  {de~Forcrand}}\ and\ \bibinfo {author} {\bibfnamefont {M.}~\bibnamefont
  {Fromm}},\ }\bibfield  {title} {\bibinfo {title} {{Nuclear Physics from
  lattice QCD at strong coupling}},\ }\href
  {https://doi.org/10.1103/PhysRevLett.104.112005} {\bibfield  {journal}
  {\bibinfo  {journal} {Phys. Rev. Lett.}\ }\textbf {\bibinfo {volume} {104}},\
  \bibinfo {pages} {112005} (\bibinfo {year} {2010})},\ \Eprint
  {https://arxiv.org/abs/0907.1915} {arXiv:0907.1915 [hep-lat]} \BibitemShut
  {NoStop}%
%%CITATION = ARXIV:0907.1915;%%
\bibitem [{\citenamefont {Gagliardi}\ and\ \citenamefont
  {Unger}(2020)}]{Gagliardi2019}%
  \BibitemOpen
  \bibfield  {author} {\bibinfo {author} {\bibfnamefont {G.}~\bibnamefont
  {Gagliardi}}\ and\ \bibinfo {author} {\bibfnamefont {W.}~\bibnamefont
  {Unger}},\ }\bibfield  {title} {\bibinfo {title} {{New dual representation
  for staggered lattice QCD}},\ }\href
  {https://doi.org/10.1103/PhysRevD.101.034509} {\bibfield  {journal} {\bibinfo
   {journal} {Phys. Rev. D}\ }\textbf {\bibinfo {volume} {101}},\ \bibinfo
  {pages} {034509} (\bibinfo {year} {2020})},\ \Eprint
  {https://arxiv.org/abs/1911.08389} {arXiv:1911.08389 [hep-lat]} \BibitemShut
  {NoStop}%
\bibitem [{\citenamefont {Kim}\ \emph {et~al.}(2023)\citenamefont {Kim},
  \citenamefont {Pattanaik},\ and\ \citenamefont {Unger}}]{Kim2023}%
  \BibitemOpen
  \bibfield  {author} {\bibinfo {author} {\bibfnamefont {J.}~\bibnamefont
  {Kim}}, \bibinfo {author} {\bibfnamefont {P.}~\bibnamefont {Pattanaik}},\
  and\ \bibinfo {author} {\bibfnamefont {W.}~\bibnamefont {Unger}},\ }\bibfield
   {title} {\bibinfo {title} {{Nuclear liquid-gas transition in the strong
  coupling regime of lattice QCD}},\ }\href
  {https://doi.org/10.1103/PhysRevD.107.094514} {\bibfield  {journal} {\bibinfo
   {journal} {Phys. Rev. D}\ }\textbf {\bibinfo {volume} {107}},\ \bibinfo
  {pages} {094514} (\bibinfo {year} {2023})},\ \Eprint
  {https://arxiv.org/abs/2303.01467} {arXiv:2303.01467 [hep-lat]} \BibitemShut
  {NoStop}%
\bibitem [{\citenamefont {Susskind}(1977)}]{Susskind1976}%
  \BibitemOpen
  \bibfield  {author} {\bibinfo {author} {\bibfnamefont {L.}~\bibnamefont
  {Susskind}},\ }\bibfield  {title} {\bibinfo {title} {{Lattice Fermions}},\
  }\href {https://doi.org/10.1103/PhysRevD.16.3031} {\bibfield  {journal}
  {\bibinfo  {journal} {Phys. Rev. D}\ }\textbf {\bibinfo {volume} {16}},\
  \bibinfo {pages} {3031} (\bibinfo {year} {1977})}\BibitemShut {NoStop}%
\bibitem [{\citenamefont {de~Forcrand}\ \emph {et~al.}(2018)\citenamefont
  {de~Forcrand}, \citenamefont {Unger},\ and\ \citenamefont
  {Vairinhos}}]{deForcrand2017}%
  \BibitemOpen
  \bibfield  {author} {\bibinfo {author} {\bibfnamefont {P.}~\bibnamefont
  {de~Forcrand}}, \bibinfo {author} {\bibfnamefont {W.}~\bibnamefont {Unger}},\
  and\ \bibinfo {author} {\bibfnamefont {H.}~\bibnamefont {Vairinhos}},\
  }\bibfield  {title} {\bibinfo {title} {{Strong-Coupling Lattice QCD on
  Anisotropic Lattices}},\ }\href {https://doi.org/10.1103/PhysRevD.97.034512}
  {\bibfield  {journal} {\bibinfo  {journal} {Phys. Rev.}\ }\textbf {\bibinfo
  {volume} {D97}},\ \bibinfo {pages} {034512} (\bibinfo {year} {2018})},\
  \Eprint {https://arxiv.org/abs/1710.00611} {arXiv:1710.00611 [hep-lat]}
  \BibitemShut {NoStop}%
%%CITATION = ARXIV:1710.00611;%%
\bibitem [{\citenamefont {Kim}\ and\ \citenamefont {Unger}(2016)}]{Kim2016}%
  \BibitemOpen
  \bibfield  {author} {\bibinfo {author} {\bibfnamefont {J.}~\bibnamefont
  {Kim}}\ and\ \bibinfo {author} {\bibfnamefont {W.}~\bibnamefont {Unger}},\
  }\bibfield  {title} {\bibinfo {title} {{Quark Mass Dependence of the QCD
  Critical End Point in the Strong Coupling Limit}},\ }\bibfield  {booktitle}
  {\emph {\bibinfo {booktitle} {{Proceedings, 34th International Symposium on
  Lattice Field Theory (Lattice 2016): Southampton, UK, July 24-30, 2016}}},\
  }\href {https://doi.org/10.22323/1.256.0035} {\bibfield  {journal} {\bibinfo
  {journal} {PoS}\ }\textbf {\bibinfo {volume} {LATTICE2016}},\ \bibinfo
  {pages} {035} (\bibinfo {year} {2016})},\ \Eprint
  {https://arxiv.org/abs/1611.09120} {arXiv:1611.09120 [hep-lat]} \BibitemShut
  {NoStop}%
%%CITATION = ARXIV:1611.09120;%%
\bibitem [{\citenamefont {de~Forcrand}\ \emph {et~al.}(2014)\citenamefont
  {de~Forcrand}, \citenamefont {Langelage}, \citenamefont {Philipsen},\ and\
  \citenamefont {Unger}}]{deForcrand2014}%
  \BibitemOpen
  \bibfield  {author} {\bibinfo {author} {\bibfnamefont {P.}~\bibnamefont
  {de~Forcrand}}, \bibinfo {author} {\bibfnamefont {J.}~\bibnamefont
  {Langelage}}, \bibinfo {author} {\bibfnamefont {O.}~\bibnamefont
  {Philipsen}},\ and\ \bibinfo {author} {\bibfnamefont {W.}~\bibnamefont
  {Unger}},\ }\bibfield  {title} {\bibinfo {title} {{Lattice QCD Phase Diagram
  In and Away from the Strong Coupling Limit}},\ }\href
  {https://doi.org/10.1103/PhysRevLett.113.152002} {\bibfield  {journal}
  {\bibinfo  {journal} {Phys. Rev. Lett.}\ }\textbf {\bibinfo {volume} {113}},\
  \bibinfo {pages} {152002} (\bibinfo {year} {2014})},\ \Eprint
  {https://arxiv.org/abs/1406.4397} {arXiv:1406.4397 [hep-lat]} \BibitemShut
  {NoStop}%
%%CITATION = ARXIV:1406.4397;%%
\bibitem [{\citenamefont {Unger}\ and\ \citenamefont
  {de~Forcrand}(2011)}]{Unger2011a}%
  \BibitemOpen
  \bibfield  {author} {\bibinfo {author} {\bibfnamefont {W.}~\bibnamefont
  {Unger}}\ and\ \bibinfo {author} {\bibfnamefont {P.}~\bibnamefont
  {de~Forcrand}},\ }\bibfield  {title} {\bibinfo {title} {{Continuous Time
  Monte Carlo for Lattice QCD in the Strong Coupling Limit}},\ }\href
  {https://doi.org/10.1088/0954-3899/38/12/124190} {\bibfield  {journal}
  {\bibinfo  {journal} {J. Phys. G}\ }\textbf {\bibinfo {volume} {38}},\
  \bibinfo {pages} {124190} (\bibinfo {year} {2011})},\ \Eprint
  {https://arxiv.org/abs/1107.1553} {arXiv:1107.1553 [hep-lat]} \BibitemShut
  {NoStop}%
\bibitem [{\citenamefont {Klegrewe}\ and\ \citenamefont
  {Unger}(2020)}]{Klegrewe2020}%
  \BibitemOpen
  \bibfield  {author} {\bibinfo {author} {\bibfnamefont {M.}~\bibnamefont
  {Klegrewe}}\ and\ \bibinfo {author} {\bibfnamefont {W.}~\bibnamefont
  {Unger}},\ }\bibfield  {title} {\bibinfo {title} {{Strong Coupling Lattice
  QCD in the Continuous Time Limit}},\ }\href
  {https://doi.org/10.1103/PhysRevD.102.034505} {\bibfield  {journal} {\bibinfo
   {journal} {Phys. Rev. D}\ }\textbf {\bibinfo {volume} {102}},\ \bibinfo
  {pages} {034505} (\bibinfo {year} {2020})},\ \Eprint
  {https://arxiv.org/abs/2005.10813} {arXiv:2005.10813 [hep-lat]} \BibitemShut
  {NoStop}%
\bibitem [{\citenamefont {Kim}\ \emph {et~al.}(2019{\natexlab{a}})\citenamefont
  {Kim}, \citenamefont {Klegrewe},\ and\ \citenamefont {Unger}}]{Kim2020a}%
  \BibitemOpen
  \bibfield  {author} {\bibinfo {author} {\bibfnamefont {J.}~\bibnamefont
  {Kim}}, \bibinfo {author} {\bibfnamefont {M.}~\bibnamefont {Klegrewe}},\ and\
  \bibinfo {author} {\bibfnamefont {W.}~\bibnamefont {Unger}},\ }\bibfield
  {title} {\bibinfo {title} {{Gauge Corrections to Strong Coupling Lattice QCD
  on Anisotropic Lattices}},\ }in\ \href@noop {} {\emph {\bibinfo {booktitle}
  {{37th International Symposium on Lattice Field Theory}}}}\ (\bibinfo {year}
  {2019})\ \Eprint {https://arxiv.org/abs/2001.06797} {arXiv:2001.06797
  [hep-lat]} \BibitemShut {NoStop}%
\bibitem [{\citenamefont {Kim}\ \emph {et~al.}(2019{\natexlab{b}})\citenamefont
  {Kim}, \citenamefont {Philipsen},\ and\ \citenamefont {Unger}}]{Kim2019}%
  \BibitemOpen
  \bibfield  {author} {\bibinfo {author} {\bibfnamefont {J.}~\bibnamefont
  {Kim}}, \bibinfo {author} {\bibfnamefont {O.}~\bibnamefont {Philipsen}},\
  and\ \bibinfo {author} {\bibfnamefont {W.}~\bibnamefont {Unger}},\ }\bibfield
   {title} {\bibinfo {title} {{On the $\beta$- and Quark Mass Dependence of the
  Nuclear Transition in the Strong Coupling Regime}},\ }\href
  {https://doi.org/10.22323/1.363.0064} {\bibfield  {journal} {\bibinfo
  {journal} {PoS}\ }\textbf {\bibinfo {volume} {LATTICE2019}},\ \bibinfo
  {pages} {064} (\bibinfo {year} {2019}{\natexlab{b}})},\ \Eprint
  {https://arxiv.org/abs/1912.00822} {arXiv:1912.00822 [hep-lat]} \BibitemShut
  {NoStop}%
\bibitem [{\citenamefont {Harris}\ \emph
  {et~al.}(2010{\natexlab{a}})\citenamefont {Harris}, \citenamefont
  {Johansson}, \citenamefont {Berkley}, \citenamefont {Johnson}, \citenamefont
  {Lanting}, \citenamefont {Han}, \citenamefont {Bunyk}, \citenamefont
  {Ladizinsky}, \citenamefont {Oh}, \citenamefont {Perminov}, \citenamefont
  {Tolkacheva}, \citenamefont {Uchaikin}, \citenamefont {Chapple},
  \citenamefont {Enderud}, \citenamefont {Rich}, \citenamefont {Thom},
  \citenamefont {Wang}, \citenamefont {Wilson},\ and\ \citenamefont
  {Rose}}]{PhysRevB.81.134510}%
  \BibitemOpen
  \bibfield  {author} {\bibinfo {author} {\bibfnamefont {R.}~\bibnamefont
  {Harris}}, \bibinfo {author} {\bibfnamefont {J.}~\bibnamefont {Johansson}},
  \bibinfo {author} {\bibfnamefont {A.~J.}\ \bibnamefont {Berkley}}, \bibinfo
  {author} {\bibfnamefont {M.~W.}\ \bibnamefont {Johnson}}, \bibinfo {author}
  {\bibfnamefont {T.}~\bibnamefont {Lanting}}, \bibinfo {author} {\bibfnamefont
  {S.}~\bibnamefont {Han}}, \bibinfo {author} {\bibfnamefont {P.}~\bibnamefont
  {Bunyk}}, \bibinfo {author} {\bibfnamefont {E.}~\bibnamefont {Ladizinsky}},
  \bibinfo {author} {\bibfnamefont {T.}~\bibnamefont {Oh}}, \bibinfo {author}
  {\bibfnamefont {I.}~\bibnamefont {Perminov}}, \bibinfo {author}
  {\bibfnamefont {E.}~\bibnamefont {Tolkacheva}}, \bibinfo {author}
  {\bibfnamefont {S.}~\bibnamefont {Uchaikin}}, \bibinfo {author}
  {\bibfnamefont {E.~M.}\ \bibnamefont {Chapple}}, \bibinfo {author}
  {\bibfnamefont {C.}~\bibnamefont {Enderud}}, \bibinfo {author} {\bibfnamefont
  {C.}~\bibnamefont {Rich}}, \bibinfo {author} {\bibfnamefont {M.}~\bibnamefont
  {Thom}}, \bibinfo {author} {\bibfnamefont {J.}~\bibnamefont {Wang}}, \bibinfo
  {author} {\bibfnamefont {B.}~\bibnamefont {Wilson}},\ and\ \bibinfo {author}
  {\bibfnamefont {G.}~\bibnamefont {Rose}},\ }\bibfield  {title} {\bibinfo
  {title} {Experimental demonstration of a robust and scalable flux qubit},\
  }\href {https://doi.org/10.1103/PhysRevB.81.134510} {\bibfield  {journal}
  {\bibinfo  {journal} {Phys. Rev. B}\ }\textbf {\bibinfo {volume} {81}},\
  \bibinfo {pages} {134510} (\bibinfo {year} {2010}{\natexlab{a}})}\BibitemShut
  {NoStop}%
\bibitem [{\citenamefont {Harris}\ \emph
  {et~al.}(2010{\natexlab{b}})\citenamefont {Harris}, \citenamefont {Johnson},
  \citenamefont {Lanting}, \citenamefont {Berkley}, \citenamefont {Johansson},
  \citenamefont {Bunyk}, \citenamefont {Tolkacheva}, \citenamefont
  {Ladizinsky}, \citenamefont {Ladizinsky}, \citenamefont {Oh}, \citenamefont
  {Cioata}, \citenamefont {Perminov}, \citenamefont {Spear}, \citenamefont
  {Enderud}, \citenamefont {Rich}, \citenamefont {Uchaikin}, \citenamefont
  {Thom}, \citenamefont {Chapple}, \citenamefont {Wang}, \citenamefont
  {Wilson}, \citenamefont {Amin}, \citenamefont {Dickson}, \citenamefont
  {Karimi}, \citenamefont {Macready}, \citenamefont {Truncik},\ and\
  \citenamefont {Rose}}]{PhysRevB.82.024511}%
  \BibitemOpen
  \bibfield  {author} {\bibinfo {author} {\bibfnamefont {R.}~\bibnamefont
  {Harris}}, \bibinfo {author} {\bibfnamefont {M.~W.}\ \bibnamefont {Johnson}},
  \bibinfo {author} {\bibfnamefont {T.}~\bibnamefont {Lanting}}, \bibinfo
  {author} {\bibfnamefont {A.~J.}\ \bibnamefont {Berkley}}, \bibinfo {author}
  {\bibfnamefont {J.}~\bibnamefont {Johansson}}, \bibinfo {author}
  {\bibfnamefont {P.}~\bibnamefont {Bunyk}}, \bibinfo {author} {\bibfnamefont
  {E.}~\bibnamefont {Tolkacheva}}, \bibinfo {author} {\bibfnamefont
  {E.}~\bibnamefont {Ladizinsky}}, \bibinfo {author} {\bibfnamefont
  {N.}~\bibnamefont {Ladizinsky}}, \bibinfo {author} {\bibfnamefont
  {T.}~\bibnamefont {Oh}}, \bibinfo {author} {\bibfnamefont {F.}~\bibnamefont
  {Cioata}}, \bibinfo {author} {\bibfnamefont {I.}~\bibnamefont {Perminov}},
  \bibinfo {author} {\bibfnamefont {P.}~\bibnamefont {Spear}}, \bibinfo
  {author} {\bibfnamefont {C.}~\bibnamefont {Enderud}}, \bibinfo {author}
  {\bibfnamefont {C.}~\bibnamefont {Rich}}, \bibinfo {author} {\bibfnamefont
  {S.}~\bibnamefont {Uchaikin}}, \bibinfo {author} {\bibfnamefont {M.~C.}\
  \bibnamefont {Thom}}, \bibinfo {author} {\bibfnamefont {E.~M.}\ \bibnamefont
  {Chapple}}, \bibinfo {author} {\bibfnamefont {J.}~\bibnamefont {Wang}},
  \bibinfo {author} {\bibfnamefont {B.}~\bibnamefont {Wilson}}, \bibinfo
  {author} {\bibfnamefont {M.~H.~S.}\ \bibnamefont {Amin}}, \bibinfo {author}
  {\bibfnamefont {N.}~\bibnamefont {Dickson}}, \bibinfo {author} {\bibfnamefont
  {K.}~\bibnamefont {Karimi}}, \bibinfo {author} {\bibfnamefont
  {B.}~\bibnamefont {Macready}}, \bibinfo {author} {\bibfnamefont {C.~J.~S.}\
  \bibnamefont {Truncik}},\ and\ \bibinfo {author} {\bibfnamefont
  {G.}~\bibnamefont {Rose}},\ }\bibfield  {title} {\bibinfo {title}
  {Experimental investigation of an eight-qubit unit cell in a superconducting
  optimization processor},\ }\href {https://doi.org/10.1103/PhysRevB.82.024511}
  {\bibfield  {journal} {\bibinfo  {journal} {Phys. Rev. B}\ }\textbf {\bibinfo
  {volume} {82}},\ \bibinfo {pages} {024511} (\bibinfo {year}
  {2010}{\natexlab{b}})}\BibitemShut {NoStop}%
\bibitem [{ann()}]{anneal}%
  \BibitemOpen
  \href@noop {} {\bibinfo {title} {Annealing implementation and controls}},\
  \bibinfo {note}
  {\url{https://docs.dwavesys.com/docs/latest/c_qpu_annealing.html}}\BibitemShut
  {NoStop}%
\bibitem [{cha()}]{chain}%
  \BibitemOpen
  \href@noop {} {\bibinfo {title} {Programming the d-wave qpu: Parameters for
  beginners, whitepaper}},\ \bibinfo {note}
  {\url{https://www.dwavesys.com/media/qvbjrzgg/guide-2.pdf}}\BibitemShut
  {NoStop}%
\bibitem [{\citenamefont {Ferrenberg}\ and\ \citenamefont
  {Swendsen}(1988)}]{Ferrenberg:1988yz}%
  \BibitemOpen
  \bibfield  {author} {\bibinfo {author} {\bibfnamefont {A.~M.}\ \bibnamefont
  {Ferrenberg}}\ and\ \bibinfo {author} {\bibfnamefont {R.~H.}\ \bibnamefont
  {Swendsen}},\ }\bibfield  {title} {\bibinfo {title} {{New Monte Carlo
  Technique for Studying Phase Transitions}},\ }\href
  {https://doi.org/10.1103/PhysRevLett.61.2635} {\bibfield  {journal} {\bibinfo
   {journal} {Phys. Rev. Lett.}\ }\textbf {\bibinfo {volume} {61}},\ \bibinfo
  {pages} {2635} (\bibinfo {year} {1988})}\BibitemShut {NoStop}%
\end{thebibliography}%

\end{document}